\documentclass[a4paper,12pt]{article}
 \usepackage[a4paper,left=2.5cm, right=2.5cm, top=2.5cm, bottom=2.5cm]{geometry}
\usepackage[latin1]{inputenc}   	
\usepackage[T1,TS1]{fontenc}        

\usepackage[colorinlistoftodos, textwidth=6.5cm, shadow]{todonotes}
\usepackage[doublespacing]{setspace} 
\usepackage{color}              
\usepackage{amsmath,amsthm,amsfonts,amssymb,amscd,mathrsfs}

\usepackage{bm}	
\usepackage{bbm}
\usepackage[ngerman, english]{babel}     %
\usepackage{ae}                 %
\usepackage{graphicx}           
\usepackage{epstopdf}
\usepackage{braket}
\usepackage{longtable}          
\usepackage{array}
\usepackage[inline]{enumitem}
\usepackage{multirow}           
\usepackage{url}
\usepackage{pdfpages}
\usepackage{epsfig}
\usepackage{fancybox}
\usepackage[font={sc},labelfont={sc}, hypcap=false,format=hang]{caption}
\usepackage{subcaption}
\usepackage{bold-extra}
\usepackage{titlesec}
\usepackage{xcolor}
\usepackage{fancyvrb}
\usepackage{arydshln}
\usepackage{lscape}
\usepackage{import}
\usepackage[titletoc]{appendix}
\usepackage{todonotes}
\usepackage[babel]{csquotes}
\usepackage{datetime}
\newdateformat{deformat}{\THEDAY{.} \monthname[\THEMONTH] \THEYEAR}
\definecolor{tuebingendarkred}{RGB}{165,30,55}
\usepackage{alphalph}
\usepackage{pifont}
\usepackage{microtype}
\usepackage{placeins}
\usepackage{MnSymbol}
\usepackage{tikz}
\usetikzlibrary{arrows,snakes,shapes,automata,backgrounds,petri,calc,matrix,decorations.markings,decorations.pathreplacing,patterns,arrows.meta}
\usepackage{pgf}
\usepackage{adjustbox}
\usepackage{booktabs}
\usepackage{multirow}

\makeatletter

\renewcommand\tableofcontents{%
  \null\hfill\textbf{\Large\contentsname}\hfill\null\par
  \@mkboth{\MakeUppercase\contentsname}{\MakeUppercase\contentsname}%
  \@starttoc{toc}%
}
\renewcommand\listoffigures{%
  \null\textbf{\large\listfigurename}\hfill\null\par
  \@mkboth{\MakeUppercase\listfigurename}{\MakeUppercase\listfigurename}%
  \@starttoc{lof}%
}
\renewcommand\listoftables{%
  \null\textbf{\large\listtablename}\hfill\null\par
  \@mkboth{\MakeUppercase\listtablename}{\MakeUppercase\listtablename}%
  \@starttoc{lot}%
}
\makeatother

\makeatletter
	\renewcommand\appendix{
		\addtocontents{toc}{\null\noindent\large \bfseries Appendices\normalsize\par} 
	\renewcommand\section{
			\newpage
			\thispagestyle{plain}
			\suppressfloats[t]
			\@afterindentfalse
			\secdef\Appendix\sAppendix
		}
		\setcounter{section}{0}%
		\setcounter{figure}{0}%
		\setcounter{equation}{0}%
		\renewcommand\thesection{\textsc{\Alph{section}}}%
		\renewcommand\theequation{\Alph{section}\arabic{equation}}%
		\renewcommand\thefigure{\Alph{section}\arabic{figure}}
	}
	
	\newcommand\Appendix[2][?]{
		\refstepcounter{section}
		\addcontentsline{toc}{section}%
		{\protect\numberline
		{\thesection}#1} {\centering \large \scshape \appendixname\,
		 \thesection \par#2\par}
		 \sectionmark{#1}
		 \@afterheading
		 \addvspace{\baselineskip}
	}
	\newcommand\sAppendix[1]{
		{\raggedleft\large\bfseries\appendixname\
		 \thesection\par \centering#1\par}
		 \@afterheading
		 \addvspace{\baselineskip}
	}
\makeatother

\RecustomVerbatimCommand{\VerbatimInput}{VerbatimInput}%
{fontsize=\footnotesize,
 frame=lines,  
 framesep=2em, 
 rulecolor=\color{Gray},
 label=\fbox{\color{Black}data.txt},
 labelposition=topline,
 commandchars=\|\(\), 
 commentchar=*        
}

\newtheoremstyle{LOBstyle}
  {2\topsep} 
  {\topsep} 
  {\itshape} 
  {} 
  {\bfseries} 
  {.} 
  {.5em} 
  {} 

\theoremstyle{LOBstyle}
\newtheorem{LOB}{Rule}

\theoremstyle{remark}
\newtheorem*{rem}{Remark}

\theoremstyle{plain}

\newcommand{\seq}[1]{\left<#1\right>}

\DeclareMathOperator*{\E}{\mathbb{E}} 	
\DeclareMathOperator*{\Var}{var}		

\DeclareMathOperator*{\diag}{diag}		

\makeatletter
\def\munderbar#1{\underline{\sbox\tw@{$#1$}\dp\tw@\z@\box\tw@}}
\let\@@pmod\pmod
\DeclareRobustCommand{\pmod}{\@ifstar\@pmods\@@pmod}
\def\@pmods#1{\mkern4mu({\operator@font mod}\mkern 6mu#1)}
\makeatother
\makeatletter 
\def\leqn{\tagsleft@true} 
\def\reqn{\tagsleft@false} 
\def\fleq{\@fleqntrue \let\mathindent\@mathmargin \@mathmargin=\leftmargini} 
\def\cneq{\@fleqnfalse} 
\makeatother 

\newcolumntype{C}[1]{>{\centering\arraybackslash}m{#1}}
\newcolumntype{R}[1]{>{\raggedleft\arraybackslash}m{#1}}  

\usepackage{simpler-wick}

\usepackage{forest}
\usepackage[colorlinks=true, linkcolor=blue, citecolor=blue]{hyperref}
\usepackage{cleveref}

\addto\extrasenglish{

}

\usepackage{booktabs}
\usepackage{environ}
\usepackage{expl3}
\ExplSyntaxOn
\seq_new:N \l_DT_contents_seq
\seq_new:N \l_DT_contents_A_seq
\tl_new:N \l_DT_header_tl
\cs_generate_variant:Nn \seq_set_split:Nnn { NnV }
\NewEnviron{dbltbl}[2]
  {
    \seq_set_split:NnV \l_DT_contents_seq { \\ } \BODY

    \seq_pop_left:NN \l_DT_contents_seq \l_DT_header_tl

    \seq_clear:N \l_DT_contents_A_seq
    \prg_replicate:nn
      { \int_div_truncate:nn { \seq_count:N \l_DT_contents_seq } {2} }
      {
        \seq_pop:NN \l_DT_contents_seq \l_tmpa_tl
        \seq_put_right:NV \l_DT_contents_A_seq \l_tmpa_tl
      }

    \begin{tabular}{#1p{#2}#1}
      \toprule
      \l_DT_header_tl && \l_DT_header_tl \\
      \midrule
      \seq_mapthread_function:NNN
        \l_DT_contents_A_seq
        \l_DT_contents_seq
        \DT_one_row:nn
      \bottomrule
    \end{tabular}
  }
\cs_new:Npn \DT_one_row:nn #1#2 { #1 && #2 \\ }
\ExplSyntaxOff

\usepackage[longnamesfirst]{natbib}

\begin{document}

\author{
Johannes Bleher\footnote{Corresponding author. Department of Econometrics, Statistics and Empirical Economics, University of T\"ubingen. Sigwartstr.\ 18, 
72074 T\"ubingen, Phone: +49
7071 29 78165. Email: \texttt{johannes.bleher@uni-tuebingen.de} }\\ \vspace*{-0.3cm} \small {\em University of T\"ubingen}, \and
Michael Bleher\footnote{Mathematical Institute, Heidelberg University. Mathematikon, Im Neuenheimer Feld 205, 69120 Heidelberg. Email: 
\texttt{mbleher@mathi.uni-heidelberg.de}}\\ \vspace*{-0.3cm} \small {\em Heidelberg University} \and
Thomas Dimpfl\footnote{Department of Econometrics, Statistics and Empirical Economics, University of T\"ubingen. Sigwartstr.\ 18, 72074 T\"ubingen, Tel: +49 
7071 29 78165. Email:
\texttt{thomas.dimpfl@uni-tuebingen.de}} \\  \small {\em University of T\"ubingen}  \\\vspace*{-0.4cm} }


\title{From orders to prices: \\
A stochastic description of the limit order book to forecast intraday returns
}
\date{\today}

\singlespace

{\let\newpage\relax\maketitle}

\thispagestyle{empty}
\vspace*{-0.5cm}
\begin{abstract}
	\noindent

    We propose a microscopic model to describe the dynamics of the fundamental events in the limit order book (LOB): order arrivals and cancellations.
    It is based on an operator algebra for individual orders and describes their effect on the LOB.
    The model inputs are arrival and cancellation rate distributions that emerge from individual behavior of traders, and we show how prices and liquidity arise from the LOB dynamics.
    In a simulation study we illustrate how the model works and highlight its sensitivity with respect to assumptions regarding the collective behavior of market participants.
    Empirically, we test the model on a LOB snapshot of XETRA, estimate several linearized model specifications, and conduct in- and out-of-sample forecasts.
    The in-sample results based on contemporaneous information suggest that our model describes returns very well, resulting in an adjusted $R^2$ of roughly 80\%.
    In the more realistic setting where only past information enters the model, we observe an adjusted $R^2$ around 15\%.
    The direction of the next return can be predicted (out-of-sample) with an accuracy above 75\% for time horizons below 10 minutes.
    On average, we obtain an $RMSPE$ that is 10 times lower than values documented in the literature.

\end{abstract}
\vspace{0.5cm}

\begin{tabular}{p{0.15\textwidth} p{0.7\textwidth}}
	  \textit{Key words:} &Limit Order Book, Master Equation, Continuous Markov Process, High Frequency, Market Microstructure\\
	  \\
	\textit{JEL:} & C58, D43, G12 \\
\end{tabular}

\newpage
\pagestyle{plain}
\pagenumbering{Roman}
\doublespacing
\newpage
\pagenumbering{arabic}
\setcounter{page}{1}
\section{Introduction}

Ever since \cite{Glosten94} raised the question whether 'the electronic open limit order book [was] inevitable', limit order books (LOBs) have become the most important way of trading, resulting in more than 75\% of exchanges  around the world (including the recently sprouting cryptocurrency exchanges) using order driven systems and relying (at least partially) on limit order books \cite[][]{Jain03}.
Nevertheless, 'no comprehensive and realistic models (either statistical or economic) exist' \cite[][Ch.~12, p.~118]{Hasbrouck07} which describe the deep-rooted mechanisms of limit order markets in their entirety.
While \citeauthor{Hasbrouck07}'s statement is more than 10 years old and the literature has made significant progress, a dynamic model which comprehensively describes the interaction of individual orders and is able to incorporate the (strategic) behavior of market participants has, to the best of our knowledge, not yet been developed.
\cite{GouldPWMFH13} provide an overview about research on the dynamics of the LOB and identify (roughly speaking) two branches of research.
One of them originates in the field of physics and focuses mainly on idealized models to describe statistical features of the LOB system, focusing on dynamic order flows.
The second branch is rooted in the economics literature which tends to treat order flows as static.
According to \cite{GouldPWMFH13}, economists primarily focus on the (strategic) behavior of traders, but neglect the dynamical structure.
Of course, this reduction is too simplistic and there are multiple attempts to combine strategic behavior and the order book dynamics.
For example, \cite{Parlour98} develops a stylized, dynamic model for the LOB and strategic order placement.
Nevertheless, only few models approach the subject by heuristically incorporating statistical regularities observed in market microstructure and incorporate trader interaction based on these statistical observations.
A notable exception is \cite{HautschH12} who use high-frequency cointegrated vector autoregressive models  to shift the spotlight on the order flow of incoming orders and therewith, based on
empirical analysis, draw attention to the intersection of LOB mechanics and strategic behavior of market participants.
They show that the revelation of trading intentions through limit order placements affects the LOB states.
\cite{Large07a} shows how trades of different size (marketable incoming orders) affect future states of the LOB.
\cite{Paulsen14} derives macroscopic limiting models for a microscopic LOB system in which bid, ask, and transaction prices drive the dynamics of the order flow.
These limiting models serve as first order approximations of the stochastic processes that describe the system.

This paper aims to link statistical models of dynamic order flow and theoretical models which consider strategic interaction.
Adapting ideas from \cite{Paulsen14} and inspired by \cite{Baez2018}, we relate the LOB dynamics to the modeling of reaction networks \cite[developed by][]{BaezP17} and present a simple but comprehensive description of the microscopic order book mechanisms.
Based on operator algebra, we construct the LOB dynamics bottom up from the elementary events of the book, namely the entry and exit of orders which enables us to model the LOB system by a Markov process.
Furthermore, the Hamiltonian, i.e., the operator which governs the time dynamics of the system, can be constructed from these elementary events.\footnote{In economics and for that matter as well in optimal control theory, the Hamiltonian is know as a function that describes optimality conditions with respect to some control variable for some dynamical system. A popular example is the Ramsey-Cass-Koopmanns model in economics. In our model, the Hamiltonian directly describes the LOB dynamics and is based on the mechanical algebra of the LOB. In principle, it is therefore similar to the Hamiltonian used in classical mechanics.}
Similar to \cite{Paulsen14}, our approach allows to incorporate both the dynamics of the fundamental LOB mechanisms as well as the (strategic) behavior of market participants.
Using the event log of the first quarter of 2004 from XETRA, we develop, simulate and empirically evaluate the implications of our operator formulation.
By describing the interaction of individual orders in a purely statistical fashion, the state space of the order book system is worked out.
It depends only on the rates of arriving and canceled orders as well as on the current state of the LOB.
Based on a limited set of key variables, we show that the return dynamics can be approximated rather accurately by a linear model.
This variable set also allows to forecast returns, arrival rates, and other measures such as order book imbalance and liquidity.
Compared to other prediction models in the literature such as \cite{ZhouPHTZ18}, we are able to reduce the root mean squared prediction error ($RMSPE$) decisively by a factor of 1/10.

Similar to our approach, \cite{ContST10} explore the idea of modeling the order book as a Markov process depending on the rates of arrivals and cancellations.
They work out closed form solutions for the probability of an increase in the midprice, execution of an order at the best bid price (before a change of the best ask price), and execution of both a buy and a sell order at the best quotes before the price moves.
\cite{ContD13} show that such a model can also be used to calculate the distribution of the duration between transactions.
Unfortunately, as \cite{ContD12} state, these models are based on several assumptions which empirical research has shown to be incorrect \citep{BouchaudMP02,Hasbrouck07}.
In particular, the time intervals between order arrivals and cancellations are neither independent nor exponentially distributed and orders are not equally sized.
Another model which is based on the Markov properties of the order book has been developed by \cite{DanielsFIS02} and extended by \cite{SmithFGK03}.
However, the model imposes broad restrictions on the functional structure, the parameters and the assumptions about the stochastic processes which govern the LOB dynamics.
Again, some of their assumptions like equal order size or balanced order flow, order placement with uniform probability, among others are 'too simple to be literally true' \cite[][]{SmithFGK03}, but
the resulting insights provide a useful foundation for LOB modeling.

In our model, we do not assume a specific functional structure for the stochastic processes of the order book.
Instead, our model realistically describes the dynamics for any order size, with an arbitrary
order placement probability and regardless of whether the order flow is balanced or not.
We also allow the time interval between LOB events to directly depend on the state of the order book.
While we maintain the assumption of an exponential distribution of the intervals, we only require them to be conditionally exponentially distributed.

The paper proceeds as follows.
In \Cref{sec:TheModel}, we introduce the description of a typical LOB by an algebra of operators.
Simultaneously, we present selected empirical characteristics of the order book which guide our model development.
A key insight is that the LOB dynamics depends on the rates of order arrivals and cancellations at different bid and ask prices.
\Cref{sec:TheData} presents the XETRA order book data and an analysis of observed arrival and cancellation rate distributions.
In \Cref{sec:TheSimulation}, we report the results of a simulation study where we identify key drivers of order flow and the statistical distribution of events in the order book on short time scales.
Finally, \Cref{sec:EmpiricalAnalysis} holds the empirical analysis of the XETRA LOB and \Cref{sec:Conclusion} concludes.


\section{The Model}
\label{sec:TheModel}
The limit order book is the place where traders' orders meet.
These orders carry information about a trader's willingness to accept a certain price, the limit price, in exchange for the chosen number of instruments, or vice-versa.
The price level at which two orders are matched is called \emph{the reference price}.
Within the LOB, we distinguish buy (ask) orders and sell (bid) orders.
Perceiving these two order types as species which populate price and order size levels inspired the use of the mathematical tools presented in \cite{Baez2018}.
At any point in time, the exchange keeps track of all orders within the LOB.
Aside from the market side, the location within the LOB is defined by three key components: the limit price and the number of securities, both determined by the trader, and the time when the order
arrives in the LOB.

If traders require immediacy, they rely on \emph{market orders} which can be thought to have an infinite (bid order) or zero (bid order) limit price depending on the market side they were issued from. These orders are matched immediately and, normally, do not reside in the order book for an extended amount of time.
They enter the LOB at the best price level of all limit orders currently residing in the LOB on the opposite market side.

The majority of orders, however, are designed to remain in the book for some time at their specified limit price level.
These are called \emph{limit orders} which (as we will show in \Cref{sec:TheData}) make up for roughly 90\% of all orders in the XETRA LOB while only 3\% are market orders.
Limit orders have a well defined location in the price dimension.
If their designated price location is behind the best price level of all limit orders which currently reside in the LOB on the other side of the market, they are
matched (partially) before they can reach their designated limit price level.
The smallest populated price level on the sell side of the market is the \emph{best ask} and the highest price level on the bid side is the \emph{best bid}.
We will refer to one or the other as the \emph{best quote}.

There are generally further order types that are only submitted to the market if certain conditions are met.
For example, \emph{stop orders} are inserted in the LOB once the reference price hits a certain threshold price.
As soon as such an order enters the LOB, however, it is effectively equivalent to a market or a limit order.
Hence, conditional orders can be perceived as more sophisticated versions of limit or market orders and can in principle be incorporated in the framework presented below.
For the purpose of this paper, we restrict our considerations, thus, to plain market and limit orders.

\subsection{The LOB Algebra} \label{sec:TheLOBAlgebra}
In the following, we describe order creation and cancellation in the LOB as determined by the rules of a typical order book.
For this purpose, we borrow the so-called \emph{Dirac} or \emph{Bra-Ket Notation} from physics where the state of a system is denoted by a \emph{ket} $\ket{\psi}$.
In \Cref{sec:StateSpace}, we will discuss the underlying notion of a state in detail.
For now, we can refer to any possible configuration of the order book with $\ket{\psi}$.
Furthermore, we can assign weights (probabilities) to each possible configuration and refer to such a weighted bundle of pure states by $\ket{\psi}$.
To begin with, however, we consider the empty order book (or vacuum) $\ket{0}$.
From this vacuum state, more complicated order book states are created by successively acting on it with creation and annihilation operators.
As we will see below, the rules of the LOB induce certain commutation relations in the algebra of these operators.
It is convenient to also introduce the notation
\begin{align}
	\ket{0} = |0|
\end{align}
which represents an empty ledger.

\let\theLOBsafehaven\theLOB
\renewcommand{\theLOB}{1a}
\begin{LOB}[Ask Order Submission] \label{rule:askSubmit}
Traders can submit a limit ask order of quantity~$q$ at a specified price level~$k$.
The order is represented by a creation operator $a^+_{k,q}$ that acts on the order book state \emph{from the right}\footnote{This choice will become relevant in the context of price-time priority, see Rule~\ref{rule:PriceTimePriority}.}.
\end{LOB}
For example, if $n$ ask orders are residing in the book, each with its associated limit price $k_i$ and size $q_i$, $i \in {1,\ldots,n}$, the order book state is given by
\begin{align}
	|0| a^+_{k_1,q_1} \ldots  a^+_{k_n,q_n}  .
\end{align}

\renewcommand{\theLOB}{1b}
\begin{LOB}[Bid Order Submission] \label{rule:bidSubmit}
Traders can submit a limit bid order of quantity~$q$ at a specified price level~$k$.
The order is represented by a creation operator $b^+_{k,q}$ that acts on the order book state \emph{from the left}.
\end{LOB}
Analogously, for $m$ bid orders residing in the book with specified limit prices $k_j$  and sizes $q_j$  with $j \in {1,\ldots,m}$, the string of operators
\begin{align}
	b^+_{k_m,q_m} \ldots b^+_{k_1,q_1} |0|  .
\end{align}
describes the current state.
Note that, so far, the rules only describe the successive submission of orders.
In particular, we do not yet have a rule that would allow us to reorder the queue of creation operators.
Put differently, creation operators generally do not commute: $a^+_{k,q} a^+_{s,p} \neq a^+_{s,p} a^+_{k,q}$.
As a result, the strings of creation operators of ask and bid type are time-ordered.

\renewcommand{\theLOB}{2a}
\begin{LOB}[Ask Order Cancellation] \label{rule:askCancel}
Traders can cancel a previously submitted ask order.
An ask order cancellation is represented by an annihilation operator $a^-_{k,q}$ which acts from the right and satisfies
\begin{align} \label{eq:DefAnnihilationOperator}
	|0| a^+_{k,q} a^-_{k,q} = |0|  .
\end{align}
\end{LOB}
Clearly, the probability of a cancellation must be zero if there is no order in the book.
This means that when an annihilation operator acts on the empty order book, it generates a state with probability mass zero:
\begin{align} \label{eq:defAnnihilationOperator2}
	|0| a^-_{k,q} = 0 .
\end{align}

There is a standard argument that there is always one more possibility to create and then delete an object than deleting and then creating one.
In terms of the operators this argument is represented by the commutation relation\footnote{
In physics, these commutation relations are known as \emph{canonical commutation relations}.}
\begin{align*}
	[ a^+_{k,q},a^-_{k,q} ] = 1
\end{align*}
where $[A,B]:= AB - BA$ denotes the commutator of two operators.
In fact this relation directly follows from \eqref{eq:DefAnnihilationOperator} and \eqref{eq:defAnnihilationOperator2}
\begin{align*}
	|0| [ a^+_{k,q},a^-_{k,q} ] = |0| (a^+_{k,q}a^-_{k,q} -a^-_{k,q}a^+_{k,q}) = |0| a^+_{k,q} a^-_{k,q} - |0| a^-_{k,q} a^+_{k,q} = |0|
\end{align*}

Furthermore, since the cancellation of an order $a^+_{k,q}$ does not influence other orders $a^+_{s,p}$, we also have
\begin{align*}
	[ a^+_{s,p} , a^-_{k,q}] = 0
\end{align*}
whenever $s\neq k$ and $p\neq q$.
We can summarize these algebraic relations as follows:
\begin{align}
	[a^+_{k,q},  a^-_{s,p}]  = \delta_{sk} \delta_{pq}
\end{align}
where $\delta_{ij}$ is the Kronecker-Delta, defined on an index set $\mathcal{I}\ni i,j $ by
\begin{align*}
  \delta_{ij} = \begin{cases}
              1 & \text{if }i=j\\
              0 & \text{else.}
             \end{cases}
\end{align*}
In fact, these commutation relations are usually viewed as the defining properties of creation and annihilation operators.

\renewcommand{\theLOB}{2b}
\begin{LOB}[Bid Order Cancellation] \label{rule:bidCancel}
Traders can cancel a previously submitted bid order.
A bid order cancellation is an annihilation operator $b^-_{k,q}$ that acts from the left.
By analogy with the ask cancellations, it satisfies
\begin{align}
	b^-_{k,q} |0| &= 0 \\
	[b^-_{k,q} , b^+_{j,p}] &= \delta_{kj} \delta_{qp}  .
\end{align}
\end{LOB}
In comparison to the commutation relation of ask orders, the order of annihilation and creation operators is reversed since bid orders act from the left.

When there are several identical limit orders, i.e., orders with the same price level and quantity, we can distinguish their position in the order queue by means of their time stamp.
In contrast, for cancellation orders, an observer cannot predict which of the identical limit orders is supposed to be canceled.
The algebraic formalism captures this uncertainty: up to normalization, the commutation relations lead to a stochastically mixed state that contains each possible cancellation,
for example
\begin{align*}
	(|0|a^+_{k,q} a^+_{r,s} a^+_{k,q}) a^-_{k,q} = |0| a^+_{k,q} a^+_{r,s} + |0|a^+_{r,s} a^+_{k,q}. \label{eq:mixed_state_split}
\end{align*}

\begin{rem}
 We also introduce the convention that arrivals and cancellations with size $q=0$ are equivalent to the identity operator.
 This is motivated by the fact that such arrivals and cancellations in practice do not exist.
 However, if they would exist, they would render the current LOB state unchanged:
 \begin{align*}
  a^+_{k,0}  = a^-_{k,0} = b^+_{k,0}  = b^-_{k,0} = 1.
 \end{align*}
\end{rem}

\renewcommand{\theLOB}{\theLOBsafehaven}
\setcounter{LOB}{2}
\begin{LOB}[Price-Time Priority] \label{rule:PriceTimePriority}
Orders are organized according to price-time priority.
\end{LOB}
The order book state is the result of successive order submissions and the corresponding string of operators is strictly ordered by time.
Hence, we get a \emph{price-time ordering} by rearranging the operators into groups with identical price level whilst maintaining the time ordering within each group.
This can be achieved by letting ask and bid orders commute whenever they have different price levels $k\neq s$:
\begin{align}
	[a^+_{k,q}, a^+_{s,p}] &= 0, \\
	[b^+_{k,q}, b^+_{s,p}] &= 0  .
\end{align}
Using these relations, the order book state can always be written in the price-time ordered form
\begin{align}
	\ket{\psi} =\ b^+_{k_1,q_1}\ \cdots\  b^+_{k_n,q_n}\ |0|\ a^+_{k_{n+1},q_{n+1}}\ \cdots\ a^+_{k_{n+m},q_{n+m}}  ,
\end{align}
where $k_{i}\leq k_{i+1}$.
Whenever $k_i = k_{i+1}$, the order nearer to $|0|$ was submitted first.

Given a LOB state in price-time ordered form, the priority of an order is encoded by its distance to $|0|$, where orders closer to $|0|$ have higher priority.


\begin{LOB}[Order Matching]\label{rule:MatchingRule}\label{rule:4}
Two orders from different market sides \emph{permit a transaction} if they have highest priority and the  bid price is bigger or equal to the ask price.
When the LOB executes orders that permit a transaction, the quantities are matched up as far as possible and unmatched quantities remain in the book.
We write $\wick{\c{b}_{k,q}|0|\c{a}_{s,p}}$ for a pair of executed orders, such that for $k\geq s$ the matching procedure is captured by
\begin{align} \label{eq:MatchingRule}
\wick{\c{b}^+_{k,q}|0|\c{a}^+_{s,p}} =
\begin{cases}
|0|a^+_{s,p-q}		&\text{if }q>p\\
|0|				 				&\text{if }q=p\\
b^+_{k,q-p}|0|		&\text{if }q<p\\
\end{cases}
\end{align}
or as an algebraic relation of creation operators
\begin{align*}
 \wick{\c{b}^+_{k,q} \c{a}^+_{s,p}}  =
\theta(q-p)\  b^+_{k,q-p}  +
\theta(p-q)\  a^+_{s,p-q}
\end{align*}
where $\theta(x)$ is the Heaviside step function
\begin{align*}
 \theta(x) = \begin{cases}
              1 & \text{if }x>0\\
							\frac{1}{2} &\text{if } x=0 \\
              0 & \text{else.} \\
             \end{cases}
\end{align*}
\end{LOB}
Recall for the case $q=p$ that orders of size $0$ are equivalent to the identity operator.

Since incoming bid (ask) orders always act on a state from the left (right), we need to commute them through older orders until they reach their designated position in the price-time ordered queue.
This means orders automatically 'walk the book'\footnote{
In the LOB literature, 'walking the book' usually refers to an arriving, marketable order that is executed against several orders on the opposite market side.
We borrow this notion of the walking order and extend it.
In our case, every order 'walks through the book', however, only marketable orders encounter orders on their way to their destined limit price level.
} until they reach their destined price level $k$.
Along its walk, an order may encounter orders of the other market side and will then be executed as described by Rule~\ref{rule:4}.
It follows that \emph{market orders} are described by creation operators $a^+_{k=0,q}$ and $b^+_{k=\infty,q}$, which will walk all the way through the book until they are completely executed.

Let us stress that the price level $k$ of an order $a^+_{k,q}$ is not necessarily the \emph{transaction price} at which the order will be executed.
Instead, the transaction price is usually determined by the price level of the 'settled order' that is encountered by the 'walking order'.
The 'settled order', however, may depend on the trading mode (see \Cref{sec:Transactions}).
Also note that we did not yet specify \emph{when} orders are executed and when a transaction will take place.
The reason is that such rules do not add further algebraic relations.
The question when orders are matched is not relevant for the description of the current state of the book.
However, it is relevant for the time dynamics, i.e., if we examine time series of order book states.
The question is whether matching occurs after each event, as continuous trading dictates, or whether matching is only conducted hypothetically after each event to produce indicative prices like in the pre-auction phase or at the end of an auction.
These different modes matter for the evolution of the book.
A detailed discussion of transactions and related issues follows in \Cref{sec:Transactions}.

\subsection{State Space and the Probability Generating Function}
\label{sec:StateSpace}

As we have seen in \Cref{sec:TheLOBAlgebra}, at any given time the state of the order book is given by a price-time ordered string of creation operators
\begin{align*}
	\ket{\psi} =\ b^+_{k_1,q_1}\ \cdots\  b^+_{k_j,q_j}\ |0|\ a^+_{k_{j+1},q_{j+1}}\ \cdots\ a^+_{k_{n},q_{n}},
\end{align*}
where $k_i \leq k_{i+1}$ for all $i\in S=\{1,\ldots, n\}$.
Each operator in this string is specified by its market side $m\in \{a,b\} = \mathcal{M}$, the price level $k\in \mathcal{K}\subset \mathbb{R}$, and the order size $q\in \mathcal{Q}\subset \mathbb{R}$.
Here $\mathcal{K}$ and $\mathcal{Q}$ are the price and quantity levels at which orders can be submitted to the LOB.
These typically are discrete subsets of $\mathbb{R}$.
In principle, price and quantity levels can become arbitrarily large, but in practice one can introduce a cutoff for both  prices and quantities at a large enough value.
As a result, $\mathcal{K}$ and $\mathcal{Q}$, can be thought of as finite sets.

It is convenient to introduce a partial ordering $\leq$ on creation operators via the following set of relations:
\begin{align*}
	b^+_{k_1,q_1} \leq b^+_{k_2,q_2} \leq a^+_{k_3,q_3} \leq a^+_{k_4,q_4}  \ \iff\  k_1\leq k_2\leq k_3 \leq k_4  .
\end{align*}
Then any price-time ordered string of creation operators is equivalent to a monotonically increasing map
\begin{align*}
	z: \ S &\rightarrow \mathcal{M}\times\mathcal{K}\times\mathcal{Q} \\
		i &\mapsto z_i = {(m_i^+)}_{k_i,q_i} \qquad \text{s.t. }z_i \leq z_{i+1}
\end{align*}
from some finite set $S\subset \mathbb{N}$ to the set of creation operators.
We denote the associated states by
\begin{align*}
	\ket{z} :=  z_1\cdots z_j |0| z_{j+1} \cdots z_{n} .
\end{align*}
The collection $\{\ket{z}\}$ fully describes the possible configurations of the order book at any given moment.
We may refer to these states as \emph{pure states}.
Note that $S$, $\mathcal{M}$, $\mathcal{K}$, and $\mathcal{Q}$ are countable sets, so the set $\{\ket{z}\}$ is countable as well.

Clearly, any prediction of the future state of the order book must be probabilistic.
So, while pure states are potentially observable, \emph{mixed states} are not.
Mixed states are composed of several pure states in which each pure state is weighted with some probability mass.
For this reason, mixed and pure states are elements of the vector space spanned by the pure states $\ket{z}$:
\begin{align*}
	\mathcal{H} = \bigg\{ \ket{\psi} = \sum_{\ket{z}\in\mathcal{H}} p(z) \ket{z} \quad\bigg|\  p(z)\in\mathbb{R}\  \bigg\}  .
\end{align*}
In reality, the order book must be in a \emph{stochastically normalized} state, i.e., a state ${\ket{\psi}=\sum_{\ket{z}} p(z) \ket{z}}$ where $0\leq p(z)\leq 1$ for all $\ket{z}$ and $\sum_{\ket{z}\in
\mathcal{H}} p(z) = 1$.
In this case, (the coefficient) $p(z)$ is the probability that the state $\ket{z}$ will be realized.

\begin{rem}
The normalized states are a closed subset of the full vector space $\mathcal{H}$.
In particular, they do not form a vector space themselves.
It is often easier to work with unnormalized vectors and rescale the result to a normalized state at the end.
\end{rem}

In \Cref{sec:time_evolution}, we describe the time evolution of an initial state $\ket{z_0}$ at time $t_0$.
We will see that (by construction) time evolution produces a state
\begin{align}
 \ket{\psi(t); z_0, t_0} = \sum_{z \in \mathcal{H}} p(z,t|z_0,t_0) \ket{z} \label{eq:gen_fun}  .
\end{align}
In the literature, this object is usually referred to as \emph{generalized probability generating function}  \cite[see for example][Section~2]{Weber2017}.
For brevity, we often write $\ket{\psi(t)}$ and drop the reference to the conditional nature of the generalized probability generating function.

It is customary to denote linear functionals by 'bra' vectors $\bra{\psi}\in\mathcal{H}^\ast$ and introduce the dual basis $\bra{z}$, which satisfies
\begin{align*}
	\braket{z^\prime|z} = \delta_{z^\prime,z} .
\end{align*}
In this notation, the conditional probability to find a state $\ket{z}$ at time $t$ in $\ket{\psi(t); z_0,t_0}$ is given by
\begin{align} \label{eq:cond_prob}
 p(z,t|z_0,t_0) = \braket{z | \psi(t)} .
\end{align}

%

\subsection{Time Evolution}
\label{sec:time_evolution}
In this section, we introduce dynamics to the order book, i.e., we explain how an order book state evolves over time.
Throughout this section, we closely follow \cite{Baez2018}, where the general theory of stochastic time evolution is laid out in great detail.

The future state of the order book arises from acting on an initial state with the order operators introduced in \Cref{sec:TheLOBAlgebra}.
This means that we are automatically in the situation of a Markov process.

The only issue is that the rate (probability) of incoming orders can depend on the history of the order book.
It is, however, not sensible to assume that the entire history of the order book affects the properties of arrival and cancellation rates as old configurations of the LOB are usually not relevant for the decision process of market participants.
They usually seek to maximize the probability of order execution based on the current state of the order book and possibly a very narrow history of preceding order book configurations.
This only implies that arrival rates may be dependent on several preceding states of the LOB, which is not in contradiction to the Markov property per se.
It would only mean that the process governing the LOB dynamics might be of Markov order higher than 1.
However, theoretically, by appropriately extending the state space, every Markov process of finite order can be expressed as a Markov process of order one.
Thus, we assume that the dynamics of the order book follow a Markov process of order 1.

As a continuous Markov process, the order book satisfies the Master equation \cite[cp.][]{vanKampen1992,Weber2017}.
In our notation, the Master equation is given by
\begin{align} \label{eq:MasterEquation}
\frac{\partial}{\partial t} \ket{\psi(t)} = {H} \ket{\psi(t)} ,
\end{align}
where the so-called Hamiltonian operator ${H}$ encodes all information on the transition probabilities between order book states.

A solution of the Master equation is provided by a \emph{stochastic time evolution operator} $U(t,t_0)$ via
\begin{align*}
	\ket{\psi(t)} = U (t,t_0) \ket{\psi(t_0)}  .
\end{align*}
If the Hamiltonian is time-independent, the time evolution operator is remarkably easy:
\begin{align}\label{eq:Sol_MasterEQ_timeindep}
U (t,t_0) = e^{H (t-t_0)}.
\end{align}
If the Hamiltonian is time dependent, the time evolution operator can similarly be written as
\begin{align}\label{eq:Sol_MasterEQ_timedep}
{U}(t,t_0) = \exp\big(\;\int_{t_0}^t H (\tau) d\tau \big)   ,
\end{align}
but the evaluation of this expression is typically more involved.

We assume that other variables, like news from outside the order book, may impact the rates of incoming orders.
However, these variables are pre-determined outside the mechanism of the order book.
This may lead to time dependent arrival and cancellation rates.
In the system description, this would mean that the Hamiltonian is time dependent.
Nevertheless, a time-independent approximation of such a system may still serve as a good approximation if the time intervals are small enough.
In \Cref{sec:EmpiricalAnalysis}, we will take an indeterministic approach towards those other variables and regard them as predetermined outside the book.
Again, this is not in contradiction to the Markov property of the LOB system which is the key assumption for the Master equation \eqref{eq:MasterEquation} to hold.
At this point, it is also interesting to note that beyond the model presented in this paper, the LOB may be an open Markov process, which can be described by relying on a compositional model framework -- in the sense of \cite{BaezP17} -- and would allow to incorporate trader behavior.

Given the above considerations, the dynamics of the order book are fully described by the choice of a Hamiltonian $H$.
\cite{Baez2018} show how $H$ can be constructed from \emph{infinitesimal stochastic operators} which describe the elementary transitions that can take place in a system.

In the LOB, there are four possible transitions for each price level $k$ and each quantity $q$:
\begin{align*}
\text{entry of an ask order }\quad E^A_{k,q}&=\left(a^+_{k,q}-1\right) \\
\text{entry of a bid order }\quad E^B_{k,q}&=\left(b^+_{k,q}-1\right)\\
\text{cancellation of an ask order }\quad C^A_{k,q}&=\left(a^-_{k,q}- N_{k,q}^A\right)\\
\text{cancellation of a bid order }\quad C^B_{k,q}&=\left(b^-_{k,q}- N_{k,q}^B\right)
\end{align*}
where $a^\pm_{k,q}$, $b^\pm_{k,q}$ are the creation and annihilation operators of \Cref{sec:TheLOBAlgebra}.
The number operators $N_{k,q}^A = a^+_{k,q}a^-_{k,q}$ and $N_{k,q}^B = b^+_{k,q}b^-_{k,q}$ return the number of active bid and ask orders on price level $k$ and of quantity $q$ when they act on a pure LOB state (see \Cref{sec:Observables}).
\begin{rem}
Creation and annihilation operators are not infinitesimal stochastic operators.
This is why there are additional terms $(-1,-N)$ in the operators corresponding to order entry and cancellation.
\end{rem}

As mentioned above, the Hamiltonian of the LOB is a combination of elementary transitions
\begin{align}
H = \sum_k \sum_q
E^A_{k,q}\alpha_A(k,q)+E^B_{k,q}\alpha_B(k,q)+C^A_{k,q}\omega_A(k,q)+C^B_{k,q}\omega_B(k,q),
\label{eq:H_k}
\end{align}
where each transition is weighted by its arrival rate $\alpha$ or cancellation rate $\omega$, respectively.
Also note that the arrival rates need to be scaled such that the time evolution operator ${U}(t,t_0)$ is indeed stochastic and maps one stochastically normalized state to another.

Generally, the arrival and cancellation rates in a LOB are observed to be time dependent.
Intraday patterns of order flow have been documented for example by \cite{BiaisHS95}.
Even for the very recent development of international Bitcoin markets, in which trading is possible 24/7, \cite{ErossMU19} document activity patterns related to the opening and closing of major
markets.
The observed clustering of transactions in time can be conceived as the result of time dependent arrival and cancellation rates.
These are usually modeled using Autoregressive Conditional Duration (ACD) models \cite[see][]{EngleR98,FernandesG06}.
We may treat the arrival and cancellation rates, especially on small and intermediate time scales, as mainly determined by the state of the order book, in the sense that the distributions of the rates across $k$ and $q$ depend on the current state of the order book -- for example via current best bid/ask or the spread.
With this dependence on the current state, we allow for a quite general feedback mechanism between the current state of the order book and arrival and cancellation rates.
If the state of the LOB changes by an event, the arrival and cancellation rates may change subsequently, as well.
We investigate both static and dynamic specifications of arrival rates.
In \Cref{sec:TheSimulation}, we will investigate static distributions using empirical unconditional frequencies and a uniform as well as a theoretical discrete Gaussian exponential (DGX) distribution for arrival and cancellation rates across price levels.
The latter can be justified heuristically by the characteristics found in our data as described in \Cref{sec:TheData}, in particular \Cref{fig:bin_freq}.
We will, in one simulation scenario, also allow the parameters of the assumed DGX distribution, for arrival and cancellation rates across relative price levels, to depend on the spread.
In \Cref{sec:EmpiricalAnalysis}, we measure the arrival rates during fixed non-overlapping time intervals and therewith allow them to vary over time.
We also incorporate in our empirical analysis in \Cref{sec:EmpiricalAnalysis} the idea of conditional autoregressive arrival and cancellation rates and include lagged terms of arrival rates, moments of the spread 
and the distance to the opposite best quote.
Sampling the LOB data on different time intervals, i.e., taking snapshots of the current state at different frequencies (for example 1, 2, and 5 minutes), allows to relate the moments of the relative
integer distance $d_l$ (as defined in \Cref{eq:logintegerdistance} in \Cref{sec:TheData})  and the quantity of incoming and canceled orders $q$ to price changes and other observables of the system.
Empirical tests of these implications can be found in \Cref{sec:EmpiricalAnalysis}.

For now, we focus on the conceptual implications of these empirical findings and on how they affect the set up of the time evolution of the LOB system.
Thus, we denote the arrival rate of an order at price $k$ and quantity $q$ as $\alpha_M(k,q)$, $M\in \mathcal{M}$.
Since the distance to the opposite market side $d$ and the prevalent spread $\Delta$ depend on the current state of the order book, the arrival rates must be considered to be operators.
When $\alpha_M(k,q)$ acts on a pure state $\ket{z}$, it returns an arrival rate which depends on the values of $d$ and $\Delta$ that are realized in the state $\ket{z}$:
\begin{align*}
	\alpha_M(k,q;z) = \braket{z|\alpha_M(k,q)|z}
\qquad M \in \{A,B\}.
\end{align*}
A similar operator yields the distribution of cancellation rates corresponding to the current state of the order book:
\begin{align*}
	\omega_M(k,q;z) = \braket{z|\omega_M(k,q)|z}
\qquad M \in \{A,B\}.
\end{align*}
In the Hamiltonian given in \Cref{eq:H_k} the operators $\alpha_M(k,q)$ and $\omega_M(k,q)$ act on the state first, thus determining the rate of the corresponding transition $E^M_{k,q}$ that acts on the state subsequently.

\begin{rem}
Since the Hamiltonian $H$ is linear in the transition operators, it can be decomposed into smaller pieces that describe a subsystem of the LOB.
For example, we can split up the Hamiltonian into ask and bid Hamiltonians
\begin{align*}
{H} &= {H}^A+{H}^B \\
H^M &= \sum_{k,q} E^M_{k,q}\alpha_M(k,q)+C^M_{k,q}\omega_M(k,q) \quad,\ M\in\mathcal{M} .
\end{align*}
Similarly, we could decompose $H$ into the Hamiltonians for all price and quantity levels:
\begin{align*}
{H} &= \sum_{k,q} {H}_{k,q} \\
H_{k,q} &= E^A_{k,q}\alpha_A(k,q)+ E^B_{k,q}\alpha_B(k,q)+ C^A_{k,q}\omega_A(k,q) +C^B_{k,q}\omega_B(k,q).
\end{align*}
While these decompositions are convenient in calculations, they also allow a different view on the evolution of the book:
In principle, one could argue that the time evolution should be based on (groups of) traders, whose order submissions and cancellations can be described by Hamiltonians $H_g$ where the index $g$ may indicate a group of traders or individual traders.
The notion of particular groups can be found quite often in the literature.
For example, \cite{FoucaultST11} group traders into institutional and individual traders, whereas \cite{FoucaultKK05}
distinguish
patient and impatient traders.
These subsystems sum up to an \emph{effective Hamiltonian} $H_\text{eff}=\sum H_g$ which will necessarily be of the form \eqref{eq:H_k}.
The only difference is that now the rates $\alpha$ and $\omega$ become population parameters in a fundamental model about traders.
In this paper, we refrain from modeling traders and instead estimate effective arrival rate distributions from LOB data.

However, there is surely a trader induced clustering or autocorrelation in arrival rates which we cannot ignore.
There are also patterns induced by general business activity throughout the day.
Additionally, when submitting orders to the LOB, traders often care about the probability that their submitted orders are executed in due time.
There is a trade-off between immediacy and a slightly delayed order execution.
The probability that an order is executed is directly linked to the arrival rates of orders in the LOB.
Thus, traders may incorporate the history in their decision process, i.e., when, at which limit price, and with which quantity they want to submit their orders to the LOB and again induce
autocorrelation into arrival and cancellation rates.
The decomposition of the Hamiltonian, as discussed above, would allow for an explicit model and cover such a scenario.
In general, the model above does not exclude the notion of autocorrelation in the arrival rates.
Especially in \Cref{sec:EmpiricalAnalysis}, however, we take a more indeterministic view in that we allow prior arrival and cancellation rates of ask or bid orders to proxy current arrival and cancellation rates.
The idea that prior arrival rates determine current rates is also the guiding notion for the ACD literature mentioned above.
\end{rem}

\subsection{Observables} \label{sec:Observables}
A specific configuration $\ket{\psi}$ of the order book contains an enormous amount of information.
Usually, the focus lies on selected descriptive quantities which can be extracted from the order book at any state.
We will call these quantities \emph{observables} and describe them by the action of an operator~$O$ on pure order book states $\ket{z}$.
The value of $O$ for a given state $\ket{z}$ can be calculated as
\begin{align*}
    O(z) = \braket{z| O |z}  .
\end{align*}
More generally, given a state $\ket{\psi}$, the $\nu$th conditional moment of the observable $O$ is given by\footnote{In quantum mechanics, a similar relation holds, known as the Born rule $\bra{\Psi} \hat{O}^\nu \ket{\Psi}$. Since we work with stochastic probabilities (and not with quantum mechanical amplitudes), $\bra{\Psi}$ needs to be replaced by the sum over all dual basis vectors $\bra{z}$.}
\begin{align}
	\E[O^\nu;\psi] = \sum_{\ket{z} \in \mathcal{H}} \bra{z} O^\nu \ket{\psi} . \label{eq:cond_moments}
\end{align}
Similarly, we can calculate the expected value of sums and products of distinct operators.
This gives rise to covariance and correlation measures, e.g.,
\begin{align*}
\operatorname{Cov}(O_1,O_2) = \sum_{\ket{z} \in \mathcal{H}} \bra{z} (O_1 -\E[O_1])(O_2-\E(O_2)) \ket{\psi}.
\end{align*}

Combined with the time evolution of an initial state $\ket{\psi_0}$, we obtain the moments of an observable's probability distribution at time $t$ ($t>t_0$) as
\begin{align} \label{eq:MomentProjection}
	\E[O^\nu;\psi(t) ] = \sum_{\ket{z}} \bra{z} O^\nu e^{H(t -t_0)} \ket{\psi_0}
			= \sum_{\ket{z}}\bra{z} O^\nu \big(1 + H(t-t_0) + \frac{1}{2} H^2(t-t_0)^2 + \ldots \big) \ket{\psi_0}.
\end{align}
Note that the expected value in \Cref{eq:MomentProjection} is a conditional expectation.
It is conditional on the state $\ket{\psi_0}$ at time $t_0$.
Later on, in \Cref{sec:EmpiricalAnalysis}, to make this conditioning clear, we will denote conditional moments with ${\E}_{t_0}[O^\nu]$.
The following example illustrates the rationale behind the formula in \Cref{eq:MomentProjection}.
Consider the operator $\beta_B$ which extracts the value of the best bid order in a state $\ket{z}$ (cf.\ \Cref{sec:Observables}).
Furthermore, assume that at $t_0 = 0$, the initial state is given in price-time ordered form by
\begin{align*}
	\ket{\psi_0} =  b^+_{k_1,q_2} b^+_{k_2,q_3} |0|  a^+_{k_3,q_1}.
\end{align*}
Clearly, since $k_1 < k_2 <k_3$, the best bid is currently at price level $k_2$.
According to \eqref{eq:MomentProjection}, the expected value of $\beta_B$ at time $t$ is given by an infinite sum.
To begin with, we consider terms for which the state does not change during the time period $\Delta t = t-t_0$.
These include of course the identity operator $1$ in Equation~\eqref{eq:MomentProjection}.
But $H$ is build from terms of the form $\alpha(p,q) (b^+_{p,q} -1)$, so there are additional contributions at any order in $H^k$.
Together, they contribute the following term to the expected value of the best bid price only for price level $k_2$:
\begin{align*}
	k_2 \bigg(1 - \sum_{k,q} \alpha(k,q) \Delta t  - \sum_{k,q} \omega(k,q) \Delta t - \ldots \bigg).
\end{align*}
The expression in parenthesis is of course nothing but the probability that the state will not change within $\Delta t$.

For other price levels, these contributions are different.
Therefore, we next investigate the linear terms in $H$ which describe the entry of a single order.
There are only three cases in which the best bid changes.
First, we may observe an entry of a limit bid order with a price level in between best ask and best bid.
It's contribution to the expected value is $\sum_{k_2 < k < k_3, q} k \alpha_B(k,q) \Delta t$.
Second, the order residing on $k_2$, the current best bid price level, may be canceled.
In this case the contribution to the expected value is $k_1 \omega(k_2,q_2) \Delta t$.
Third, an arrival of an ask order above the best bid which exhausts the best bid order's quantity may arrive.
In this case we get $\sum_{k\geq k_2,q\geq q_2} k_3 \alpha_A(k,q) \Delta t$.
In all other cases of incoming limit orders the value of $\beta_B$ remains at $k_2$.

A similar analysis is possible at order $H^2$.
This would entail interaction terms of two orders entering the book: $a^+_{k,q} \alpha_A(k,q) b^+_{\ell,r} \alpha_B(\ell,r)$.
There are again several different cases which depend on the type, price level and quantity of the incoming orders, each contributing differently to the expected value.
More generally, at order $H^n$ one encounters the probabilities that $n$ orders enter the book and influence the best bid during time period $\Delta t$.
For short time periods, such higher order contributions are negligible compared to the linear contributions because they depend on products of arrival rates, which are typically very small.
However, for long time periods the powers ${\Delta t}^n$ will eventually dominate.

The example above illustrates an important property of the model.
According to \Cref{eq:MomentProjection}, the moments of observables depend solely on the (current) distributions $\alpha_M(k,q)$ and $\omega_M(k,q)$.
These distributions, if they vary sufficiently slowly, may be measured or modeled from the event stream of the book.
Therefore, a testable hypothesis implied by our model is whether the distributional moments of $k$ (or equivalently $d_l$ for that matter) and $q$ (calculated by perceiving $\alpha_M(k,q)$ and
$\omega_M(k,q)$ as their underlying probability functions) can be used to predict the expected value of observables, including price changes and inter-transaction duration.

In the following, we present a selection of observables, which are important for our analysis.
\subsubsection{Number and Volume Operators}
A basic observable is the number of active orders on price level $k$ with size $q$.
It can be described for the bid and ask side by the \emph{number operators}
\begin{align*}
	N_{k,q}^B &= b^+_{k,q} b^-_{k,q}, \\
	N_{k,q}^A &= a^+_{k,q} a^-_{k,q}  .
\end{align*}
These operators can be utilized to extract several other observables.
In particular, the total number of active orders on price level $k$ of ask or bid type $M\in \mathcal{M} =\{A,B\}$
			\begin{align*}
			 {N}_k^M = \sum_{q} {N}_{k,q}^M  ,
			\end{align*}
the \emph{quantity} of active orders on price level $k$ and the \emph{total quantity} on each market side $M \in\mathcal{M}$
			\begin{align*}
				Q_k^M &= \sum_{q} q N_{k,q}^M  , \\
				Q^M &= \sum_k Q_k^M   ,
			\end{align*}
or the \emph{volume} of active orders at price level $k$ and the \emph{total volume} on each market side
			\begin{align*}
				V_k^M &= k Q_k^M,\\
				V^M &= \sum_k V_k^M  .
			\end{align*}

There are also operators that describe a global aspect of the configuration of an order book state $\ket{z}$, e.g. the best bid and best ask prices $\beta_M$, $M\in\mathcal{M}$.
In the following, let the state of the LOB be
\begin{align*}
	\ket{z} =\  b^+_{k_1,q_1} \  \ldots\ b^+_{k_j,q_j}\ |0|\ a^+_{k_{j+1},q_{j+1}}\ \ldots\ a^+_{k_n,q_n} .
\end{align*}
Then the \emph{best bid} and \emph{best ask} operators act on $\ket{z}$ as follows:
\begin{align*}
	\beta_B \ket{z} &= k_j \ket{z}, \\
	\beta_A \ket{z} &= k_{j+1} \ket{z} .
\end{align*}
Note that $k$ on the right hand side is not an operator but the price level associated with the best quote.
Combining the two, one obtains the \emph{spread} operator $\Delta$ and \emph{mid price operator} $\beta_\text{mid}$ as
\begin{align*}
	\Delta &= \beta_A - \beta_B , \\
	\beta_\text{mid} &= \tfrac{1}{2} \big( \beta_B + \beta_A \big)  .
\end{align*}


\subsubsection{Liquidity}
\cite{Harris2003} defines liquidity as "the ability to trade large size quickly, at low cost, when you want to trade."
According to the same source, the notion of liquidity incorporates four dimensions: immediacy of trade execution for a given size, depth, width, and resilience of the market.
Therefore, the spread itself is used frequently as a liquidity measure in the literature.

There are multiple approaches to measure liquidity and we rely on the exchange liquidity measure (${XLM}$) which is based on the concept of implementation shortfall, introduced by \cite{Gomber2002}.
It covers three dimensions of liquidity: depth, width, and immediacy.
The ${XLM}$ (also known as XETRA Liquidity Measure) is composed of liquidity measures for the ask side (${XLM}_A$) and the bid side of the market (${XLM}_B$),
\begin{align}
 {XLM} = {XLM}_A+{XLM}_B, \label{eq:XLM_all}
\end{align}
where
\begin{align}
 {XLM}_A &= 10,000 \frac{\frac{\sum^{\infty}_{k}V^A_k}{\sum_{k} Q^A_k}-\beta_\text{mid}}{
 \frac{\sum^{\infty}_{k}V^A_k}{\sum_{k} Q^A_k}}, \label{eq:XLMA} \\
 {XLM}_B &= 10,000 \frac{\beta_\text{mid}-\frac{\sum^{\infty}_{k}V^B_k}{\sum_{k} Q^B_k}
 }{
 \frac{\sum^{\infty}_{k}V^B_k}{\sum_{k} Q^B_k}
 }.\label{eq:XLMV}
\end{align}

The ${XLM}$ depends on the volume weighted price which can be realized immediately on each side of the market for a round trip order with a certain volume $\bar{V}$, i.e., simultaneously submitting
marketable ask and bid orders with a total volume of $\bar{V}$.
In other words, the ${XLM}$ measures the cost of a round trip order (in basis points).

\subsection{Transactions} \label{sec:Transactions}
Up to now, we have deferred the discussion of transactions since, strictly speaking, they are not necessary to set up the order book states.
In this section we first discuss the trading modes of the XETRA order book and explain how one can augment the LOB states to also record information about transactions.
This will allow us to study the transaction price and transaction rates, which were so far not available in the order book state.

The XETRA order book is organized as continuous trading augmented by opening-, intraday-, and closing-auctions.
Before stating the rules for these modes, we make a small change in notation:
Instead of the symbol $|0|$ for the empty book, we record via $|T_{k,q;t}|$ the last price $k$, quantity $q$, and time $t$ at which a transaction occurred.
\let\theLOBsafehaven\theLOB
\renewcommand{\theLOB}{5a}
\begin{LOB}[Continuous Trading]
Assume an incoming order is assigned highest priority and is such that it permits a transaction with its partner on the opposite market side.
Then the orders will be executed at the price of the partner that was already residing in the market and a transaction of the matched-up quantity will be issued at this price.
For an arriving ask order, this results in
\begin{align}
	\big(\ \cdots b^+_{k,q}\ |0|\ \cdots \big)\, a^+_{s,p} &= \cdots \wick{\c{b}^+_{k,q}\ |\, T_{k, \mathrm{min}(q,p);t}\, |\ \c{a}^+_{s,p}} \cdots \ , \label{eq:Auction1}
\end{align}
while for an arriving bid order, we have
\begin{align}
    b^+_{k,q}\, \big(\ \cdots |0|\ a^+_{s,p}\ \cdots \big) &= \cdots \wick{\c{b}^+_{k,q}\ |\, T_{s, \mathrm{min}(q,p);t}\, |\ \c{a}^+_{s,p}} \cdots \ .\label{eq:Auction2}
\end{align}
\end{LOB}
\renewcommand{\theLOB}{5b}
\begin{LOB}[Auction]
Auctions consist of an outcry/call phase, during which incoming orders are collected and ordered by price-time priority as usual, but are \emph{not executed}.
The exchange may provide an indicative pricing to market participants i.e., the price level at which the current order book state would settle if the call phase were to end immediately.
Upon closing of the call phase the transaction price is determined according to the principle of highest traded volume.
Subsequently, orders of highest priority are executed iteratively at the previously determined transaction price.
The transaction is recorded at the transaction price and with the total traded quantity.
A description of the matching procedure like in \Cref{eq:Auction1}~and~\Cref{eq:Auction2} is possible for concrete situations.
The principle of highest traded volume makes a general formulation exhausting and is not particularly illuminating.
Therefore, we omit a general formulation at this point.
\end{LOB}
\renewcommand{\theLOB}{\theLOBsafehaven}
\setcounter{LOB}{5}

The rules above are illustrated in \Cref{fig:TransactionMatching}.
We can now introduce the \emph{transaction price}, \emph{transaction quantity}, and \emph{transaction volume operators}, which extract the corresponding numbers from the last recorded transaction.
Let the state of the book be
\begin{align*}
	\ket{z} = z_1 \ldots z_i\, |\, T_{k,q;t}\, |\, z_{i+1} \ldots z_n \ .
\end{align*}
The operators are then defined as follows
\begin{align*}
	T_K \ket{z}  &= k \ket{z}, \\
	T_Q \ket{z} &= q \ket{z}, \\
	T_V \ket{z} &= kq \ket{z}.
\end{align*}
Furthermore, we can extract the time at which the last transaction occurred via $T_t \ket{z}= t \ket{z}$.
This is the basis for an important observable, the \emph{inter-trade duration} $T_{\Delta t} = t_2 -t_1$, i.e., the time between two transaction.
In our current setup, $T_{\Delta t}$ cannot be expressed as an operator which only acts on the current state, while in practice we can calculate the time intervals from remembering earlier
transactions' time stamps.
\begin{figure}
\centering
\caption{Transaction matching}
\label{fig:TransactionMatching}
\begin{minipage}{0.9\linewidth}
 The figure illustrates the matching  $\wick{\c{b}_{k,q}|0|\c{a}_{s,p}}$ for $s\leq k$ and $q>p$.
 \vspace{0.2cm}
\end{minipage}

 \begin{tikzpicture}[node distance=2cm,fill=gray,auto,>=stealth',scale=0.9]
	\draw[->,thick] (0,0) -- (1,0) -- (1,0.2) -- (1,-0.2) node[below,inner sep=0.25cm] at (1,-0.2) {$s$} -- (1,0) -- (3,0) -- (3,0.2) -- (3,-0.2) node[below] at (3,-0.2) {$k$}   -- (3,0) --  (5,0) -- (5,0) -- (5,0) -- (5,0) -- (7,0) node[below] at (7,0) {price};
    \node[circle,draw=black,fill=green!60!black!20,very thick,dashed] (askl) at (1,1) {$a^+_{s,p}$};
    \node[circle,draw,fill=red!70!black!40,thick,text width=1.5cm,align=center] (bidm) at (3,1.2) {$b^+_{k,q}$};
    \node[circle,draw,fill=green!30!black!20,thick] (askr) at (5,1) {$a^+_{s,p}$};
    \draw[-,very thick] (askl) -- ++(0,1.5cm) -- ++ (2,0) -| (bidm);
    \draw[->,semithick] (askr) to (askl);
    \draw[snake=coil,->,very thick] (3,-1) -- (3,-2);

	\draw[->,thick] (0,-4.5) -- (1,-4.5) -- (1,-4.3) -- (1,-4.7) node[below,inner sep=0.25cm] at (1,-4.7) {$s$} -- (1,-4.5) -- (3,-4.5) -- (3,-4.3) -- (3,-4.7) node[below] at (3,-4.7) {$k$}   -- (3,-4.5) --  (5,-4.5) -- (5,-4.5) -- (5,-4.5) -- (5,-4.5) -- (7,-4.5) node[below] at (7,-4.5) {price};
	\node[circle,draw,fill=red!70!black!40,thick,text width=1cm,align=center] (bidm) at (3,-3.3) {$b^+_{k,q-p}$};
	\end{tikzpicture}
\end{figure}

\section{Data} \label{sec:TheData}

The data used in the present paper were provided by Deutsche B\"orse in 2004 and have been previously used by \cite{GrammigHR04}.
They consist of all recorded order book events of the XETRA system\footnote{XETRA Release 7.0} for trading of the 30 stocks that constituted the German stock market index, DAX between January 2 and March 26, 2004.
Additionally,\ Deutsche B\"orse provided the open order positions in their books as of January 1, 2004, 12 pm.
The data allow for the full recovery of the order book.
Over the three months period, 228,275,832 events were recorded.
Additionally, 2,282 initial positions are available at the beginning of the period.
The data cover order arrivals, (partial) matches, changes and cancellations.
The XETRA trading system allows for limit and market orders.
It is also possible to mix the two standard order types (market and limit orders) with a market-to-limit order (MTL).
An MTL order is filled on the best limit price in the book, either fully or partially.
If an MTL order is matched only partially on the best ask or bid price, it enters the LOB on the best limit price with the remaining order size.
Additionally, iceberg orders (ICE) are allowed for which only a fraction of the total volume chosen by the issuer is displayed to market participants.

Market and limit orders can also have a specified stop price.
They are then called stop orders.
Different from the limit price, i.e., the price upon which the trader is willing to trade, the stop price specifies a price level from which onwards the trader is willing to submit an order.
Hence, if the reference price exceeds (in case of bid order) or undercuts (ask order) the stop price, the order is inserted in the book.

Furthermore, the XETRA system also allowed so called XETRA BEST execution orders during the sample period at hand.
The BEST execution orders are matched against incoming market or crossing limit orders at a price level just before the currently prevailing best ask and bid prices.
In that way they introduce an extra hidden layer of occupied price levels in front of the prevailing best ask and bid prices in the LOB.
XETRA BEST orders can be market or limit orders.
Market-to-limit orders, iceberg orders and stop-orders cannot be submitted as XETRA BEST orders.
Also, if XETRA BEST orders are not executed they do not enter the LOB event log at all.
This happens if the associated price level is better than the current best quote.
If they are matched in a trade, they are recorded as the counterparty of a transaction.
If they are submitted as limit orders and the associated price level is worse than the best quote, the enter as regular limit orders.
In the latter case, we cannot distinguish regular limit orders from XETRA BEST orders in our data.

For all orders validity constraints can be set.
Users may specify a termination date up to which the order is valid.
Without such a restriction, orders are valid for 90 days.
Iceberg orders, however, are only good for the day.
Traders may also specify whether orders are valid only for specific trading phases such as auctions, or during which auction the order shall be valid.
Orders with such a restricted validity reside in the book and become active during the trading phase for which they are valid.

With regard to execution restrictions, the XETRA system allows for the following two specifications for market, limit or MTL orders.
First, the Fill-or-Kill order (FOK) is either filled entirely or canceled.
FOK orders are only recorded as entries to the book if successfully filled.
If no immediate filling is possible, FOK orders are canceled without notification within the LOB event log.
Second the Immediate-or-Cancel order (IOC) is filled as far as possible upon entry, or canceled.
Similar to the FOK order, a record of the order is only entered in the LOB event log in case of a successful (partial) filling.

All these different order types and restrictions can be incorporated in the time evolution set out in \Cref{sec:TheModel} by introducing different types of arrival and cancellation rates for the related events as elements of the Hamiltonian.
These order types, however, do not affect the generality of the algebra set up in \Cref{sec:TheLOBAlgebra}.

\Cref{tab:OrderNumbers} in \Cref{appendix:dist_of_events} provides an overview of the distribution of the events in the LOB log.
It presents the total number of submitted limit, market, iceberg, and MTL orders along with their relative occurrences on the bid and ask side.

\begin{figure}
\begin{center}
\caption{Frequency of order arrivals}
\label{fig:bin_freq}
\begin{minipage}{0.8\linewidth}
 The graphs show the logarithmic frequency of arriving (a and b) or canceled (c and d) limit orders (including stop and
iceberg orders) for the stock VOW, respectively, on their logarithmic relative integer distance to the best bid or ask prices.
 The logarithmic relative integer distance is defined as $d_l = \log\left(100\max(d,0)+1\right)$.
 The red line is the logarithmic probability of a truncated discrete Gaussian exponential (DGX) distribution for $d_l>0$, i.e., $d>1$ (as described in \Cref{appendix:DGX}).
 The theoretical value for $d_l = 0$ or $d=1$ is intentionally ignored for the fitting of the parameters of the DGX distribution.
 At $d_l=0$ the logarithmic frequencies of several types of marketable orders are displayed.
 The blue point represents the logarithmic frequency of market orders.
 Also, the log frequency of marketable limit orders at (red) and behind (blue) the best quote is shown, as well as marketable iceberg orders in purple, and marketable stop orders in orange.
 Crossing cancellations occur in the XETRA event log when the orders are deleted before they are matched.
\end{minipage}
\begin{subfigure}[b]{0.45\linewidth}
 \includegraphics[width=\linewidth]{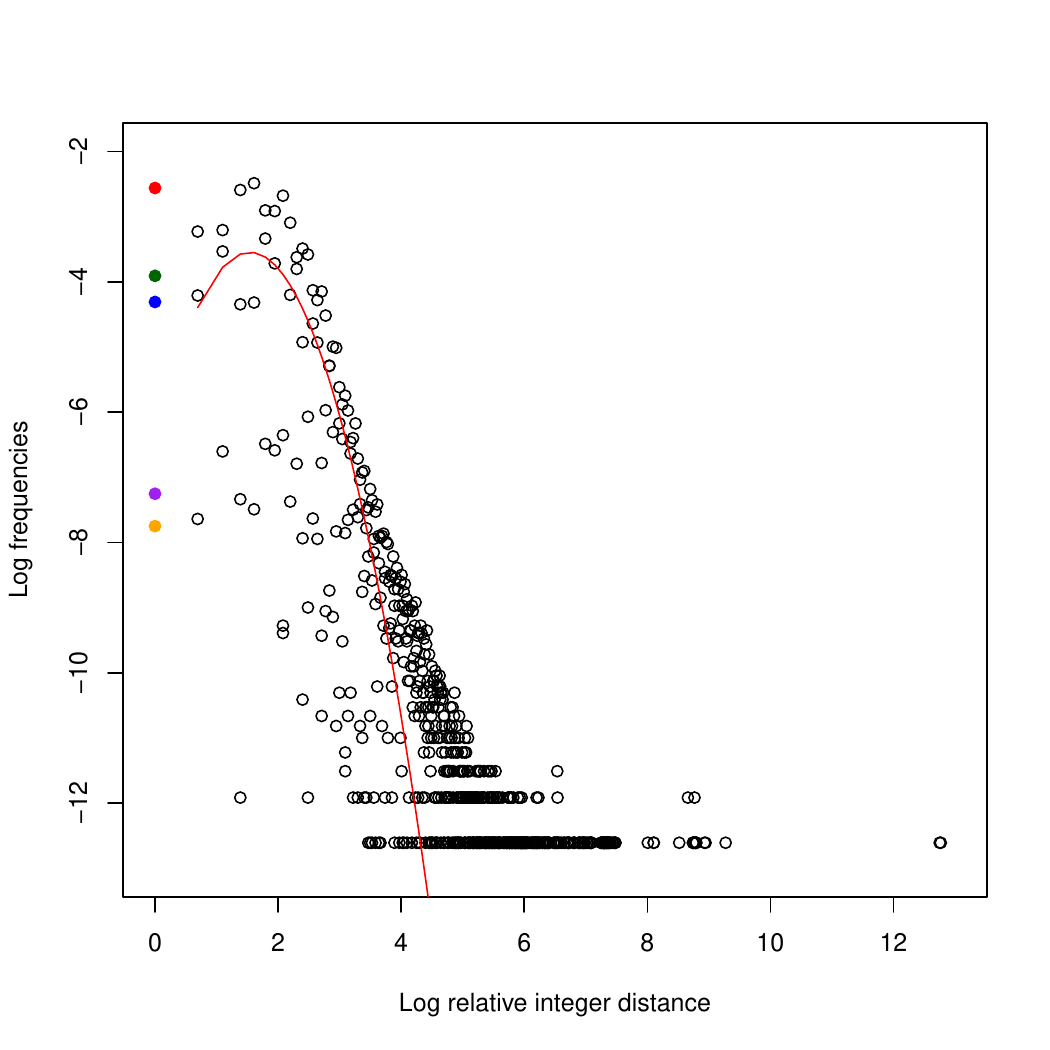}
 \caption{Ask arrivals}
 \label{fig:bin_freq_askarrival}
\end{subfigure}
\begin{subfigure}[b]{0.45\linewidth}
 \includegraphics[width=\linewidth]{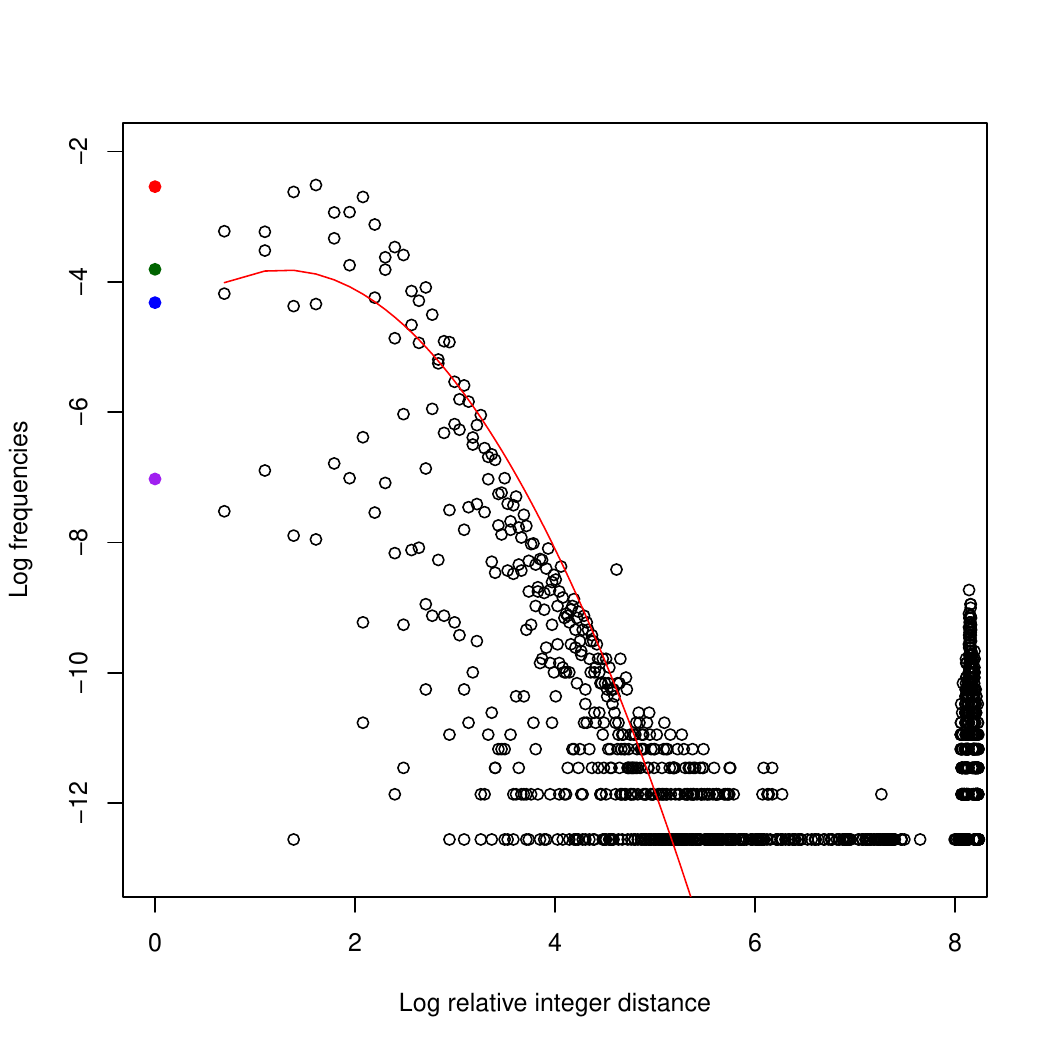}
 \caption{Bid arrivals}
 \label{fig:bin_freq_bidarrival}
\end{subfigure}
\begin{subfigure}[b]{0.45\linewidth}
 \includegraphics[width=\linewidth]{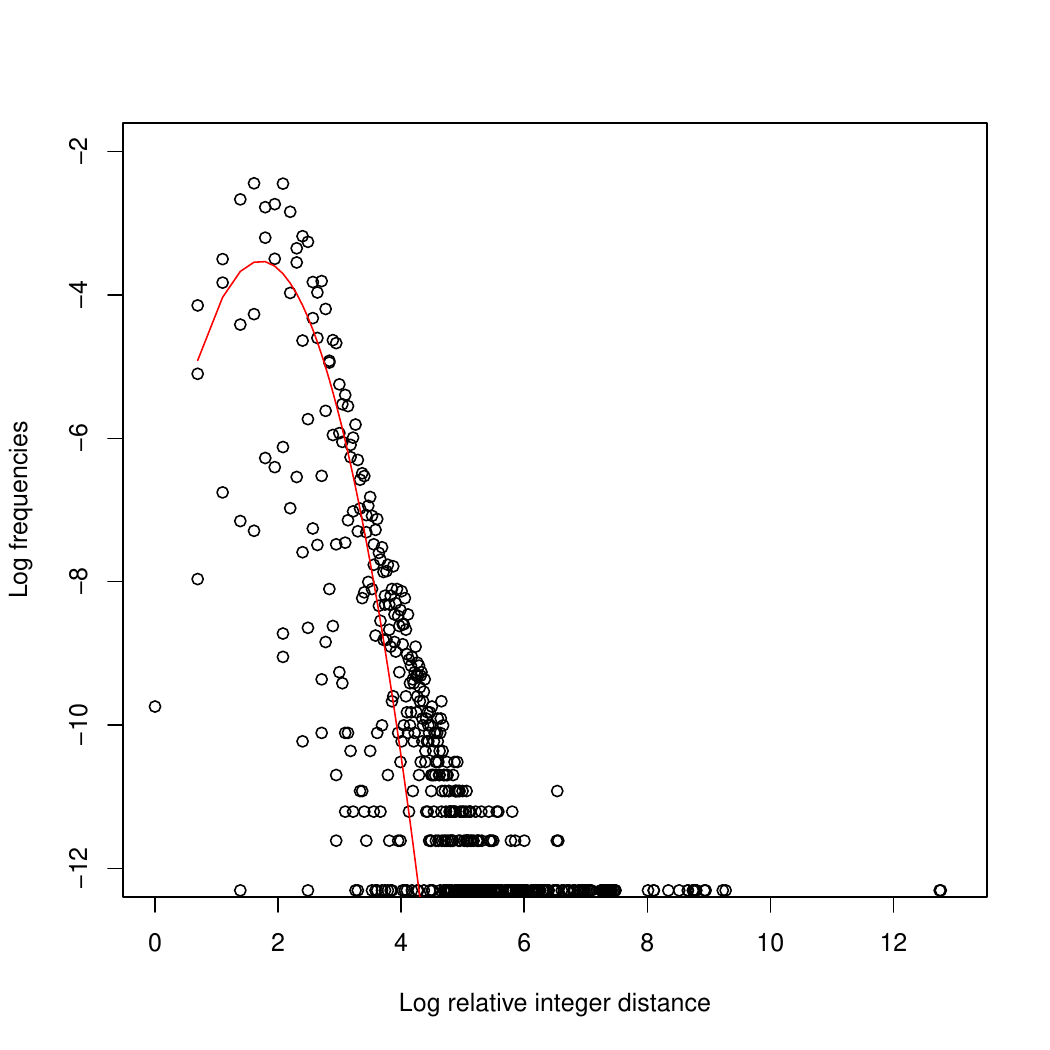}
 \caption{Ask cancellations}
 \label{fig:bin_freq_askcancellations}
\end{subfigure}
\begin{subfigure}[b]{0.45\linewidth}
 \includegraphics[width=\linewidth]{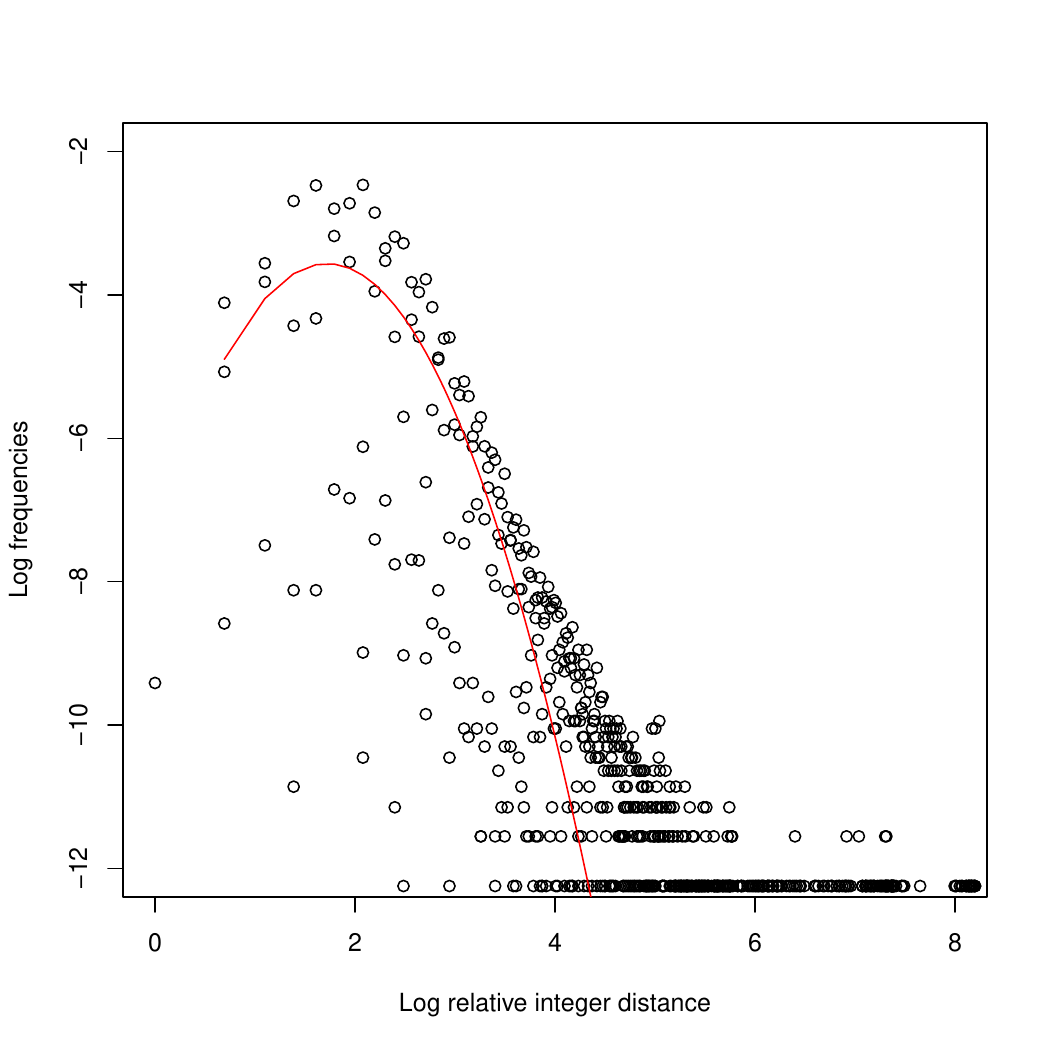}
 \caption{Bid cancellations}
 \label{fig:bin_freq_bidcancellations}
\end{subfigure}
\end{center}
\end{figure}

In empirical investigations of LOB data, it is frequently observed that the distributions of arrival and cancellation rates show a relatively stable connection with the current distance $d = |\beta_M - k|$ of the respective price level $k$ to the best active price level on the opposite market side $\beta_M$, $M\in\mathcal{M} = \{A,B\}$ \cite[cp.][]{BouchaudMP02}.
We find a similar phenomenon in the XETRA data.
We set the distance $d$ of all crossing, arriving orders to 0 and define the logarithmic relative integer distance as
\begin{align}
 d_l = \log \left(100\max(d,0)+1\right). \label{eq:logintegerdistance}
\end{align}
Taking logarithms exposes the heavy tails of the distribution across price levels and its similarities with the DGX-distribution (see \Cref{appendix:DGX} for a short description and \cite{BiFCK01} for further details).
For our simulation study in \Cref{sec:TheSimulation}, we use the DGX distribution to model order arrivals across $d_l$.
\Cref{fig:bin_freq} displays the empirical logarithmic frequencies of order arrivals and cancellations at the logarithmic relative integer distance $d_l$ of their limit order price to the best ask or best bid price, respectively.
The red line indicates a DGX distribution (truncated at 1).
The logarithmic frequencies of different types of marketable orders (limit, stop, iceberg and market orders) are displayed separately at $d_{l0}=0$ in \Cref{fig:bin_freq}.

\begin{figure}
\begin{center}
\caption{Relation between spread and relative price distance}
\label{fig:bin_vs_spread}
\begin{minipage}{0.8\linewidth}
 The graphs present the logarithmic relative integer distance to the best bid or ask price of orders arriving (a and b) or being canceled (c and d) related to the stock ALT
together with the prevailing logarithmic integer spread.
 The logarithmic relative integer distance is defined as $d_l = \log\left(100\max(d,0)+1\right)$ whereas the logarithmic integer is $\log(100\Delta)$ with $\Delta$ being the prevailing spread at arrival or cancellation.
 Even though cancellations smaller than the spread seem counter intuitive, they occur when orders are immediately canceled right after their insertion into the event log.
 The red line indicates the bisecting line.
\end{minipage}
\begin{subfigure}[b]{0.45\textwidth}
 \includegraphics[width=\linewidth]{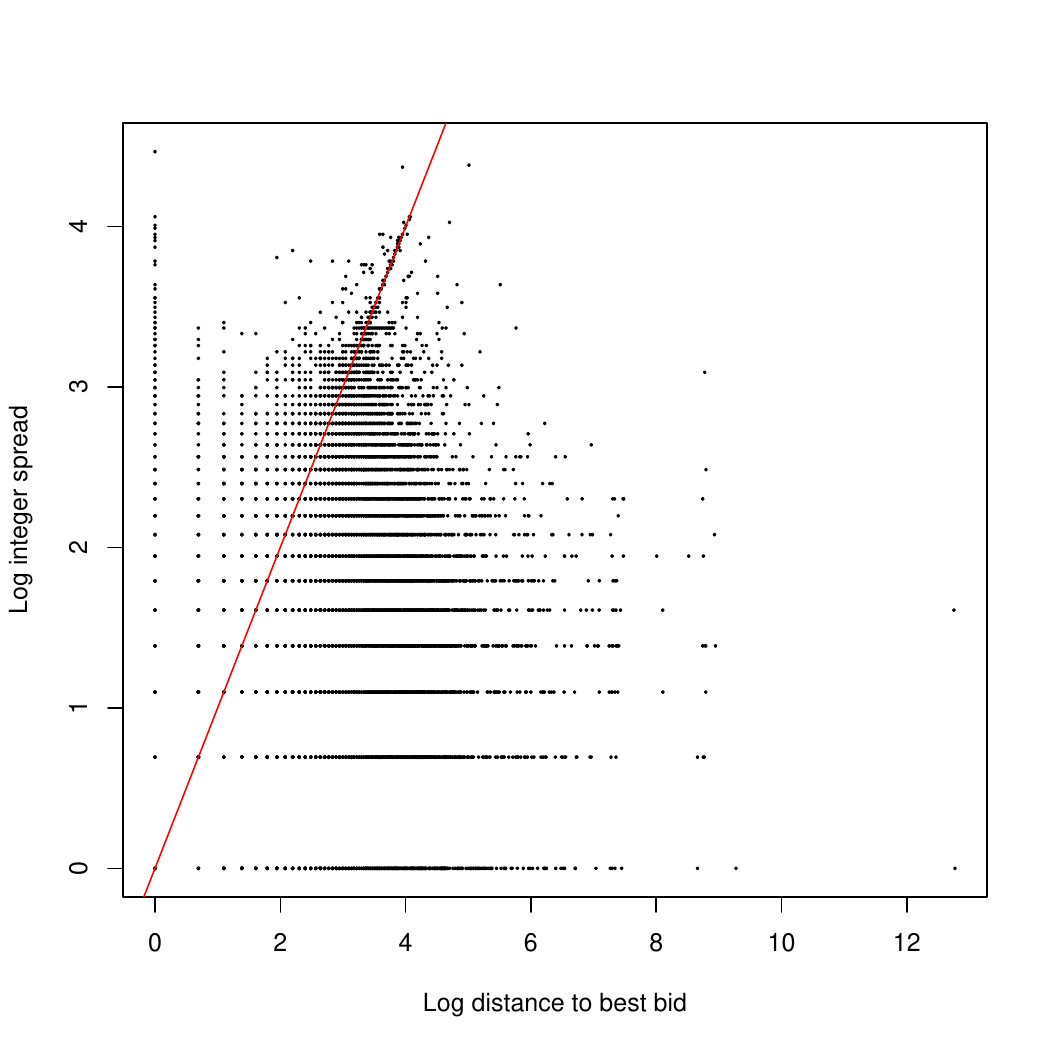}
 \caption{Ask arrivals}
 \label{fig:bin_vs_spread_askarrival}
\end{subfigure}
\begin{subfigure}[b]{0.45\textwidth}
 \includegraphics[width=\linewidth]{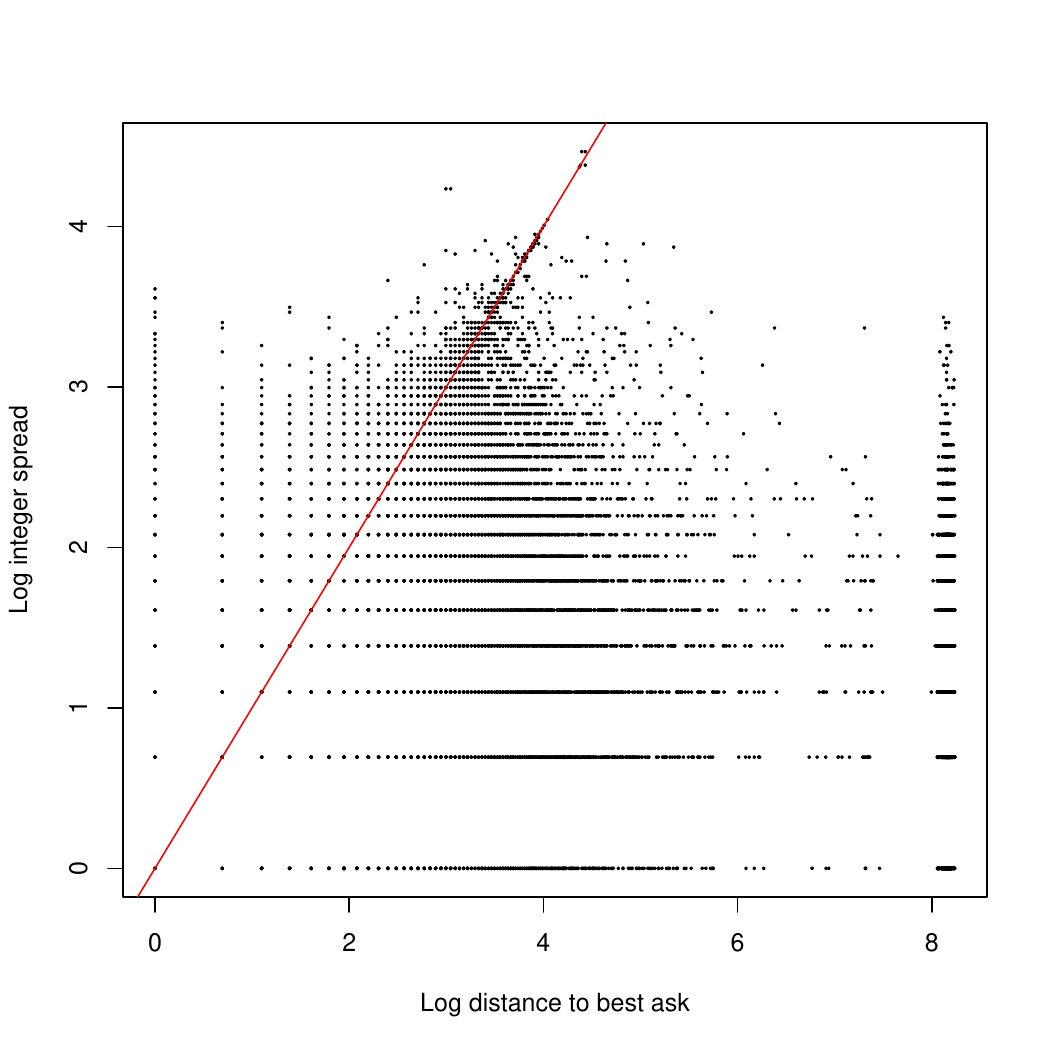}
 \caption{Bid arrivals}
 \label{fig:bin_vs_spread_bidarrival}
\end{subfigure}
\begin{subfigure}[b]{0.45\textwidth}
 \includegraphics[width=\linewidth]{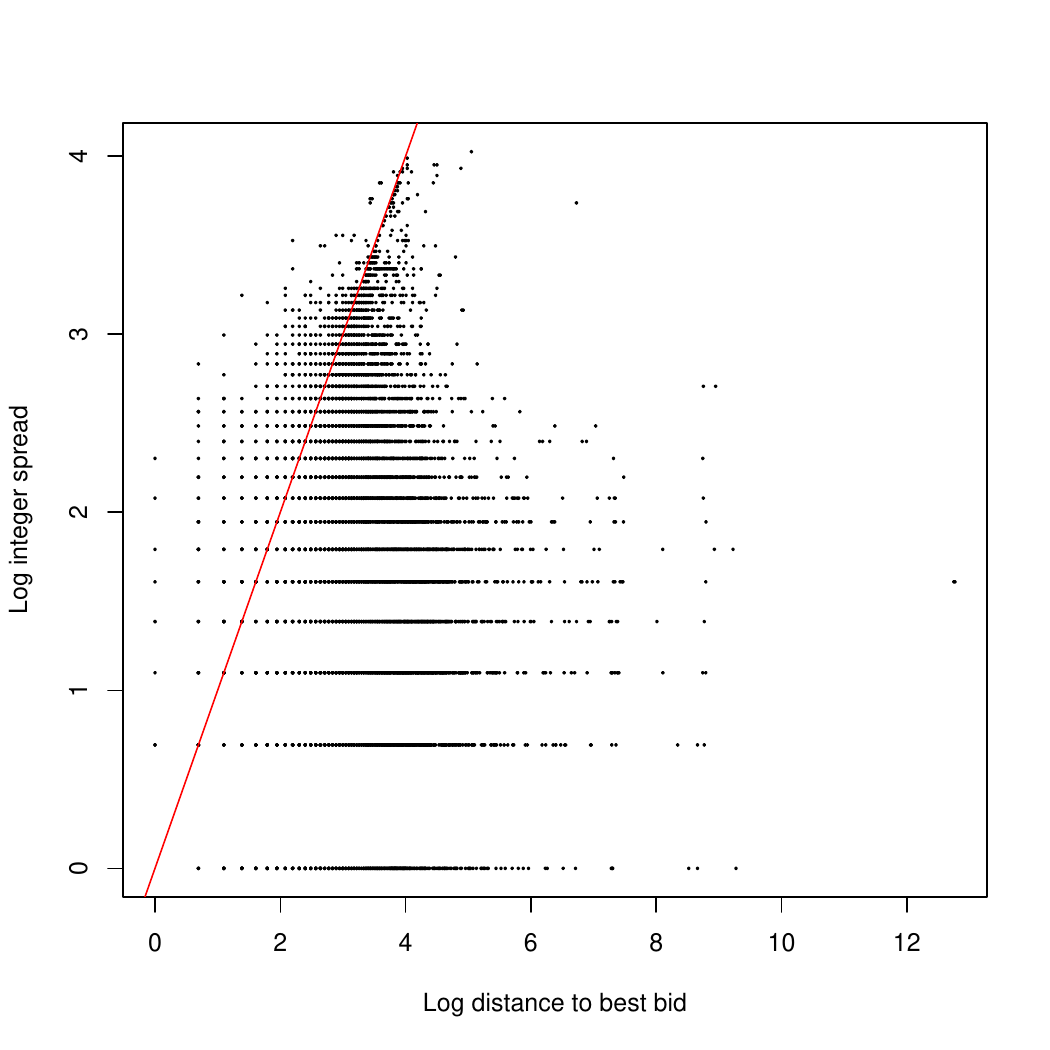}
 \caption{Ask cancellations}
 \label{fig:bin_vs_spread_askcancellations}
\end{subfigure}
\begin{subfigure}[b]{0.45\textwidth}
 \includegraphics[width=\linewidth]{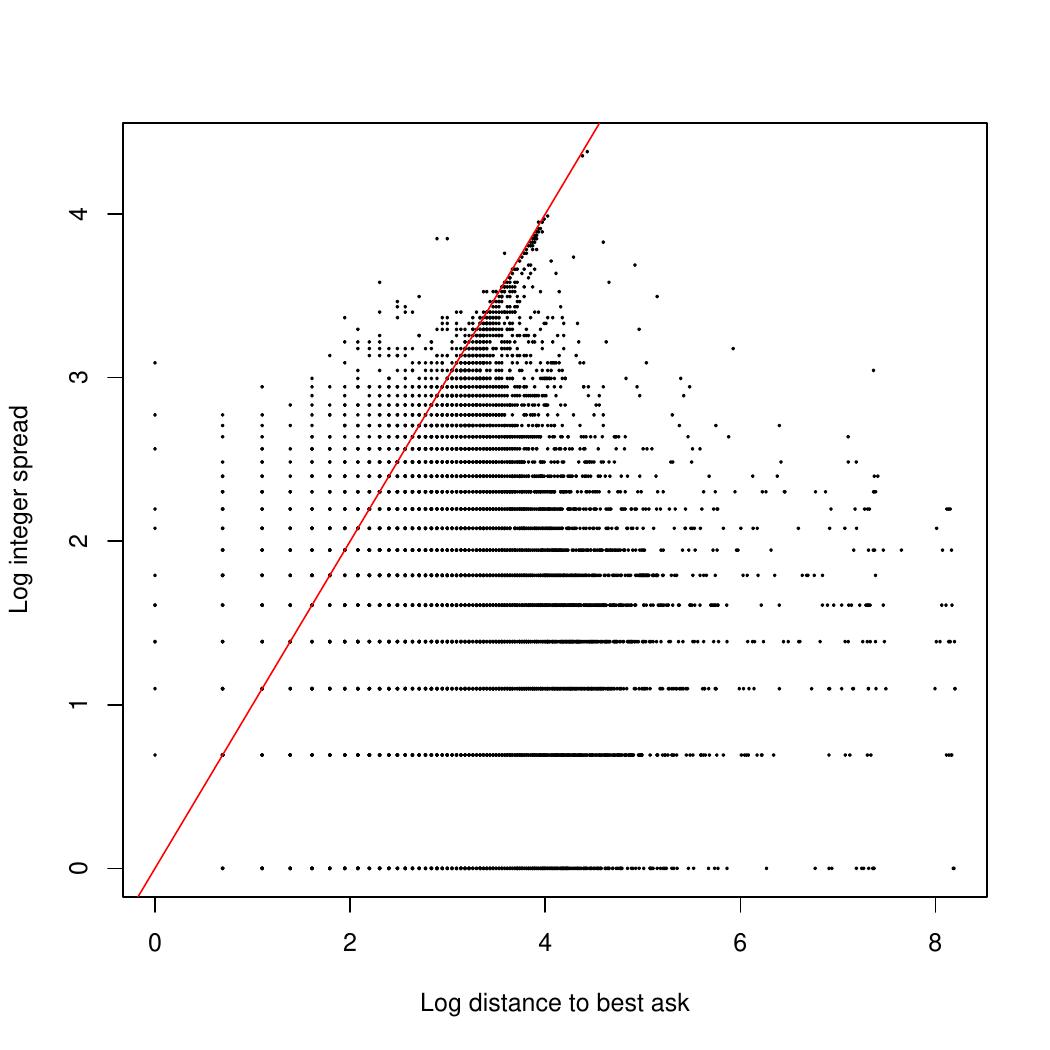}
 \caption{Bid cancellations}
 \label{fig:bin_vs_spread_bidcancellations}
\end{subfigure}
\end{center}
\end{figure}

In our data, we also observe a small correlation between $d_l$ and the prevailing spread $\Delta = v_A - v_B$ in logarithms.
\Cref{fig:bin_vs_spread} presents scatter plots of $d_l$ against the logarithmic integer spread $\log(100\Delta)$.
For large spreads there is a stronger correlation between events that are mainly concerned with price levels around the best limit price of the same market side.
When the spread is small, it seems that events occur more evenly spread out up to as much away as EUR 4 from the best limit price of the opposite market side.
We also note that events on the bid side are less dispersed across price levels than events on the ask side as the natural limit price for a bid order is a price level of~0.
For ask orders, theoretically, no such limit exists.

With respect to order size, we observe no clear pattern between order size and the logarithmic relative integer distance in the data, only a very small negative correlation can be noted.
Budget restrictions of market participants would suggest that order sizes further away from the opposite best quote on the ask side bring order sizes down with growing price levels.
For the bid side, the inverse argument should hold, nevertheless, a small negative correlation can also be observed for the bid side.
However, as the correlations are low, the price level and quantity, at least in logs, may justify the approximating assumption of stochastic independence used in several scenarios of the simulation study in Section~\ref{sec:TheSimulation}.
\Cref{tab:Correlations_k_q} lists the correlations between the order volume $q$ associated with a certain event and the logarithmic integer distance to the best opposite quote for the MEO stock.
As can be seen, the approximating assumption that the two variables $q$ and $d_l$ are independent does not mirror reality exactly.
Nevertheless, we will make the assumption several times in this paper as it keeps the estimation and simulation manageable.

\begin{table}
\caption{Correlations between $d_l$ and $q$}
\label{tab:Correlations_k_q}
\centering
\begin{minipage}{0.8\linewidth}
The table lists the correlation coefficients $r^2$ between the logarithmic integer distance to the opposite best quote $d_l$ as defined in \Cref{eq:logintegerdistance} together with
standard errors calculated as $\text{s.e.}=\sqrt{(1-r^2)/(n-2)}$ where $n$ is the number of observations.
The approximately standard normally distributed $z$-score$=(r^2)/(\text{s.e.})$ is reported as well, complemented by its $p$-value.
All values reported concern the MEO stock.
\vspace{0.2cm}
\end{minipage}
 \begin{tabular}{lrrrr}
 \toprule
  \multicolumn{1}{c}{Event}& \multicolumn{1}{c}{$r^2$} &\multicolumn{1}{c}{ s.e.} &\multicolumn{1}{c}{ $z$-score} &\multicolumn{1}{c}{$p$-value}\\
  \midrule
  ask arrivals&-0.1517&0.0060&-25.2&<0.001\\

  bid arrivals&-0.2090&0.0058&-35.9&<0.001\\

  ask cancellations&0.0291&0.0062&4.7&1\\

  bid cancellations&0.0359&0.0061&5.87&1\\

  \bottomrule
 \end{tabular}
\end{table}

\begin{figure}
\begin{center}
\caption{Relation between order size and relative price distance}
\label{fig:bin_vs_size}
\begin{minipage}{0.8\linewidth}
 The graphs show the decadic logarithmic relative integer distance to the best bid or ask price of arriving (a and b) or canceled (c and d) orders related to the stock BAS against the decadic
logarithm of the size.
 The decadic, logarithmic, relative integer distance is defined as $d_l = \log_{10}\left(100\max(d,0)+1\right)$.
 For the arriving orders, the order size depicted is the original order size, while for the canceled orders, the order size depicted is the actually canceled order size, not the original size at
entry.
\end{minipage}
\begin{subfigure}[b]{0.45\textwidth}
  \includegraphics[width=\linewidth]{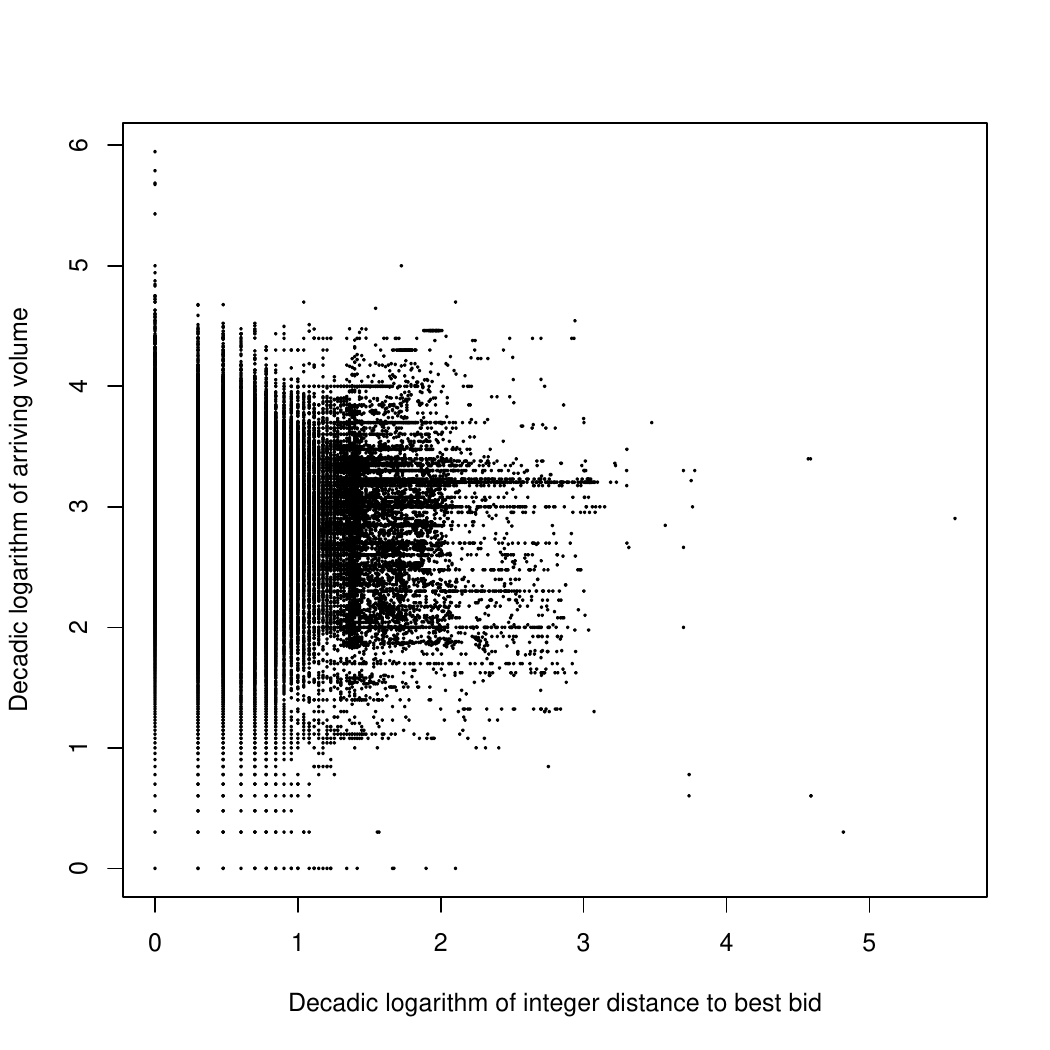}
 \caption{Ask arrivals}
 \label{fig:bin_vs_size_askarrival}
\end{subfigure}
\begin{subfigure}[b]{0.45\textwidth}
  \includegraphics[width=\linewidth]{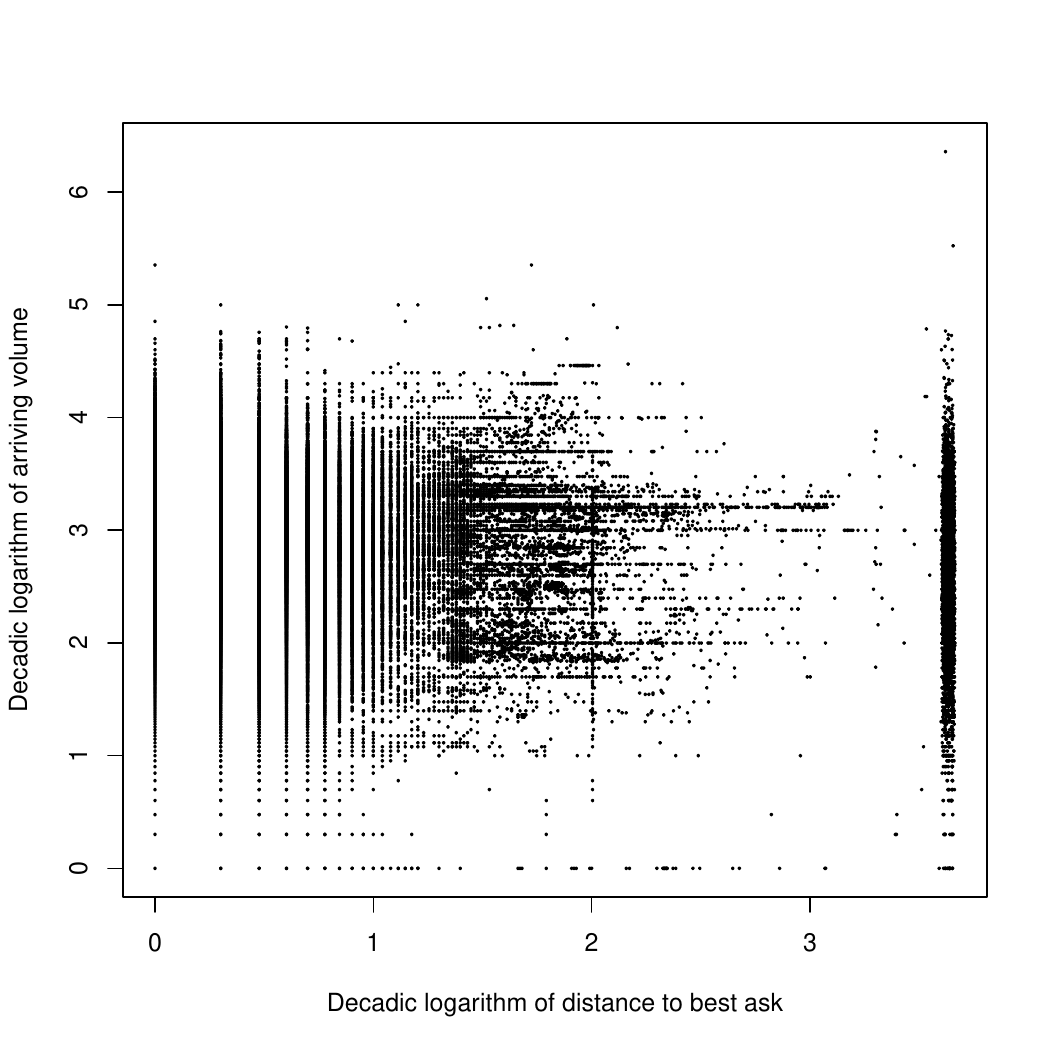}
 \caption{Bid arrivals}
 \label{fig:bin_vs_size_bidarrival}
\end{subfigure}
\begin{subfigure}[b]{0.45\textwidth}
 \includegraphics[width=\linewidth]{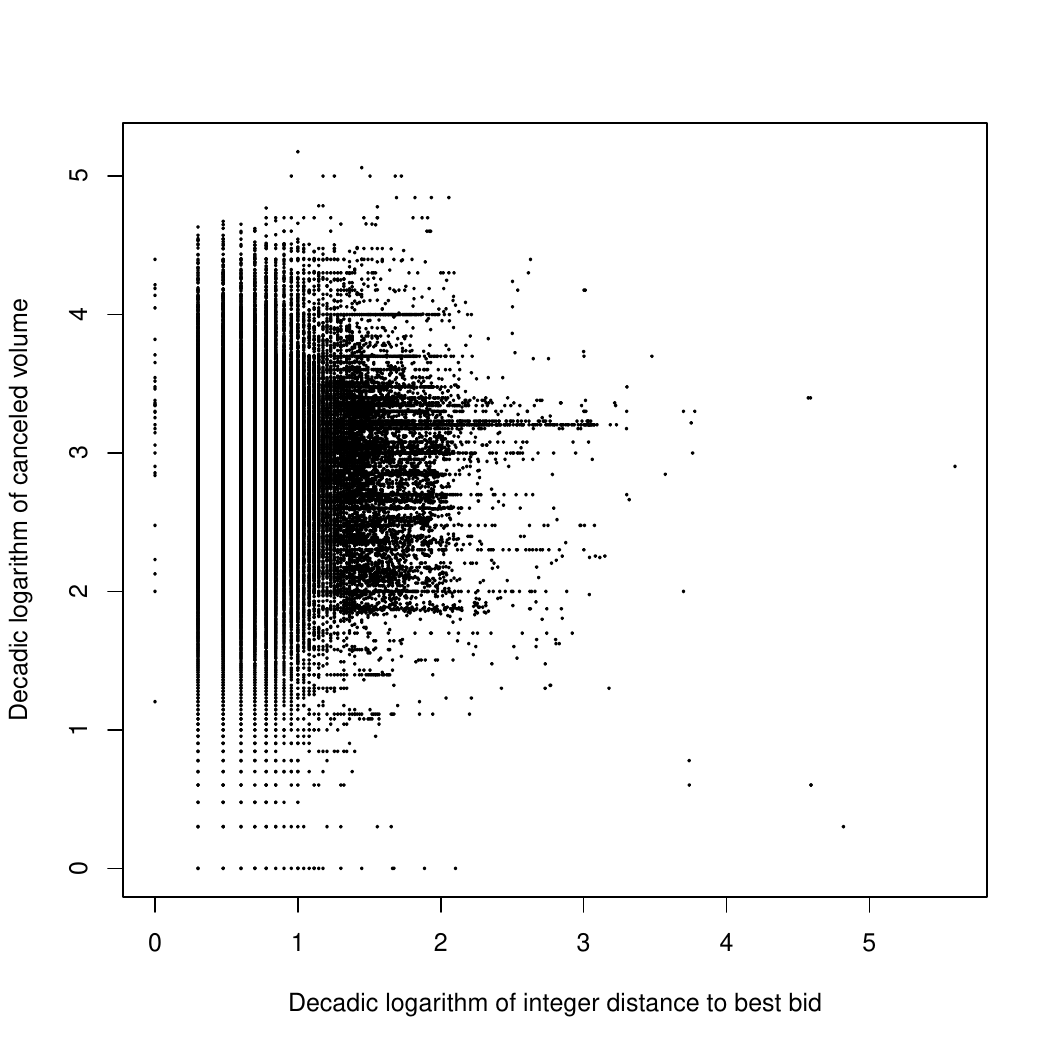}
 \caption{Ask cancellations}
 \label{fig:bin_vs_size_askcancellations}
\end{subfigure}
\begin{subfigure}[b]{0.45\textwidth}
  \includegraphics[width=\linewidth]{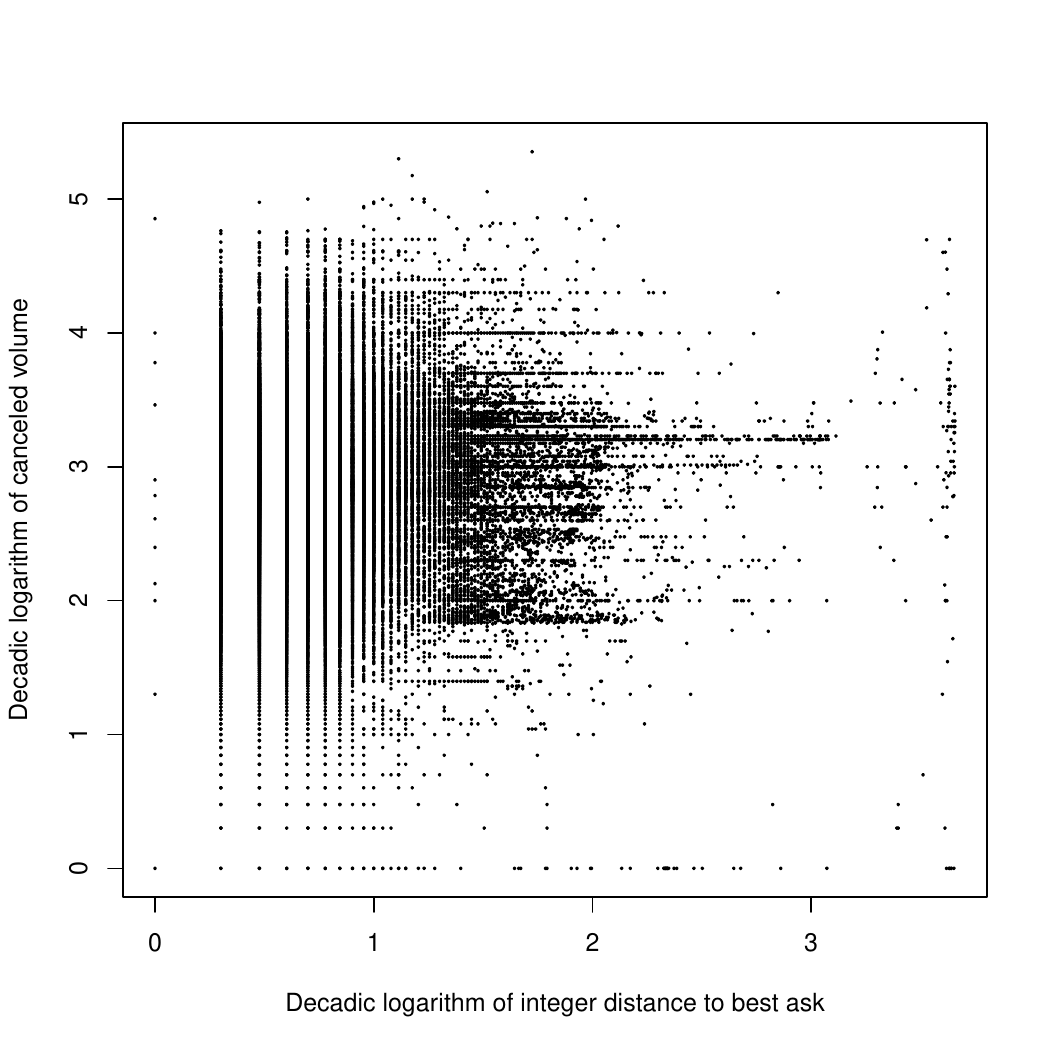}
 \caption{Bid cancellations}
 \label{fig:bin_vs_size_bidcancellations}
\end{subfigure}
\end{center}
\end{figure}

\section{Simulation Study}
\label{sec:TheSimulation}

\subsection{The Simulation Algorithm}
In order to simulate the LOB, several assumptions have to be imposed on the functional structure of the arrival and cancellation rates included in the Hamiltonian $H$ of the model developed in \Cref{sec:TheModel}.
The following section gives a short overview on the calibration of the stochastic simulation algorithm (SSA) that we use, which was developed by \cite{Gillespie1977}.
With the algorithm, we simulate an artificial history of the order book.
\Cref{appendix:Simulation} presents the theory and the successive steps of the algorithm in more detail.
The possibility to exactly simulate the system with the SSA is a direct consequence of the model.
Note that the assumptions made about the immanent functional structure of the arrival and cancellation rates are the crucial ingredients of the model.
We therefore explore possible calibrations of our model in the subsequent simulation study.
Our goal is not to fit the simulation results to an observed LOB history as closely as possible.
Nonetheless, our choice of parameters will often be guided by empirical observations.
The simulation, however, has the purpose to offer insights into the sensitivity of the order book dynamics, especially with respect to transaction price dynamics, when the structure of arrival and cancellation rates are changed.

\begin{figure}
\begin{center}
\caption{Distribution of size}
\label{fig:size_dist}
\begin{minipage}{0.8\linewidth}
 The figure presents the logarithmic frequencies of logarithmic order sizes of the MEO stock for arriving (a and b) or canceled (c and d) orders.
 For incoming orders, the logarithm of the original order size is used, whereas for order cancellations, the actually canceled remaining order size is utilized.
\end{minipage}
\begin{subfigure}[b]{0.45\textwidth}
 \includegraphics[width=\linewidth]{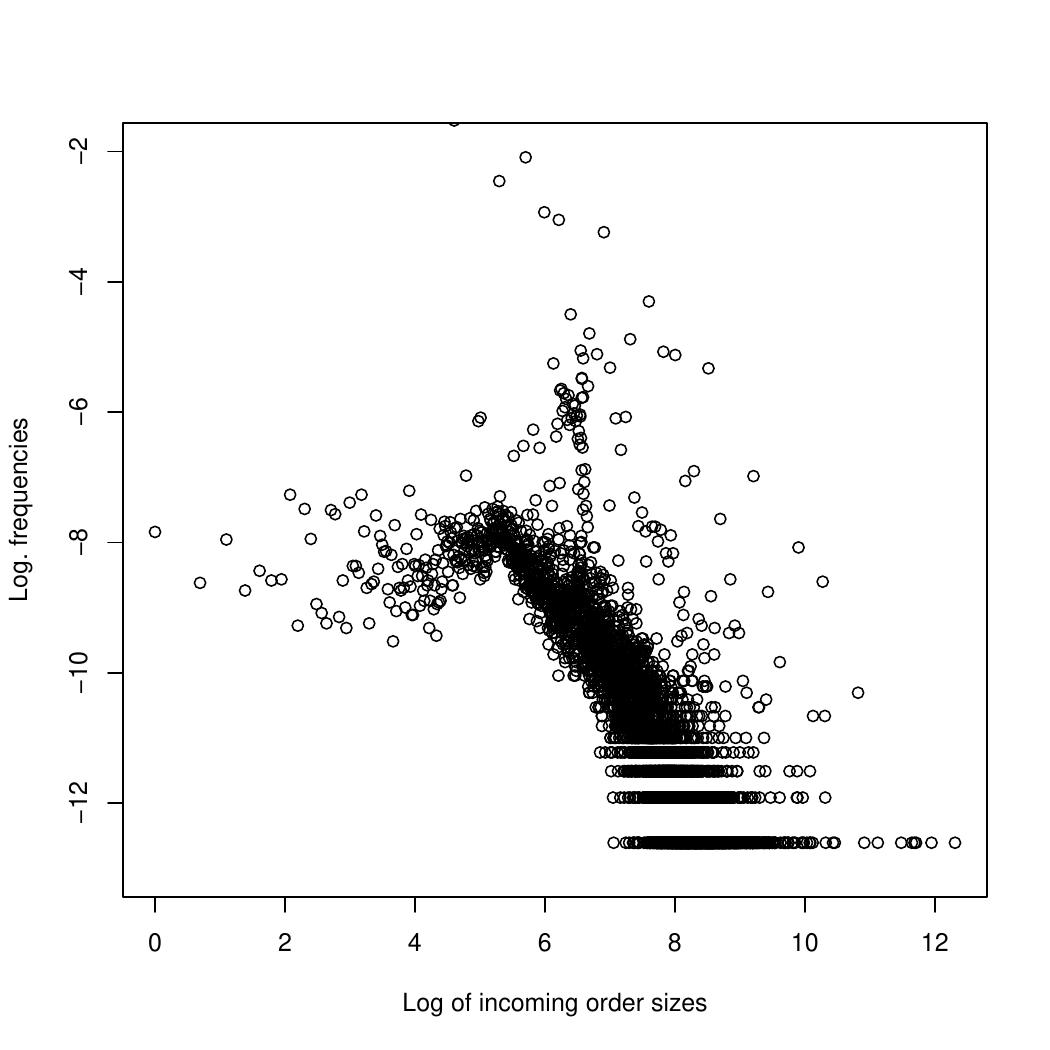}
 \caption{Ask arrivals}
 \label{fig:size_askarrival}
\end{subfigure}
\begin{subfigure}[b]{0.45\textwidth}
 \includegraphics[width=\linewidth]{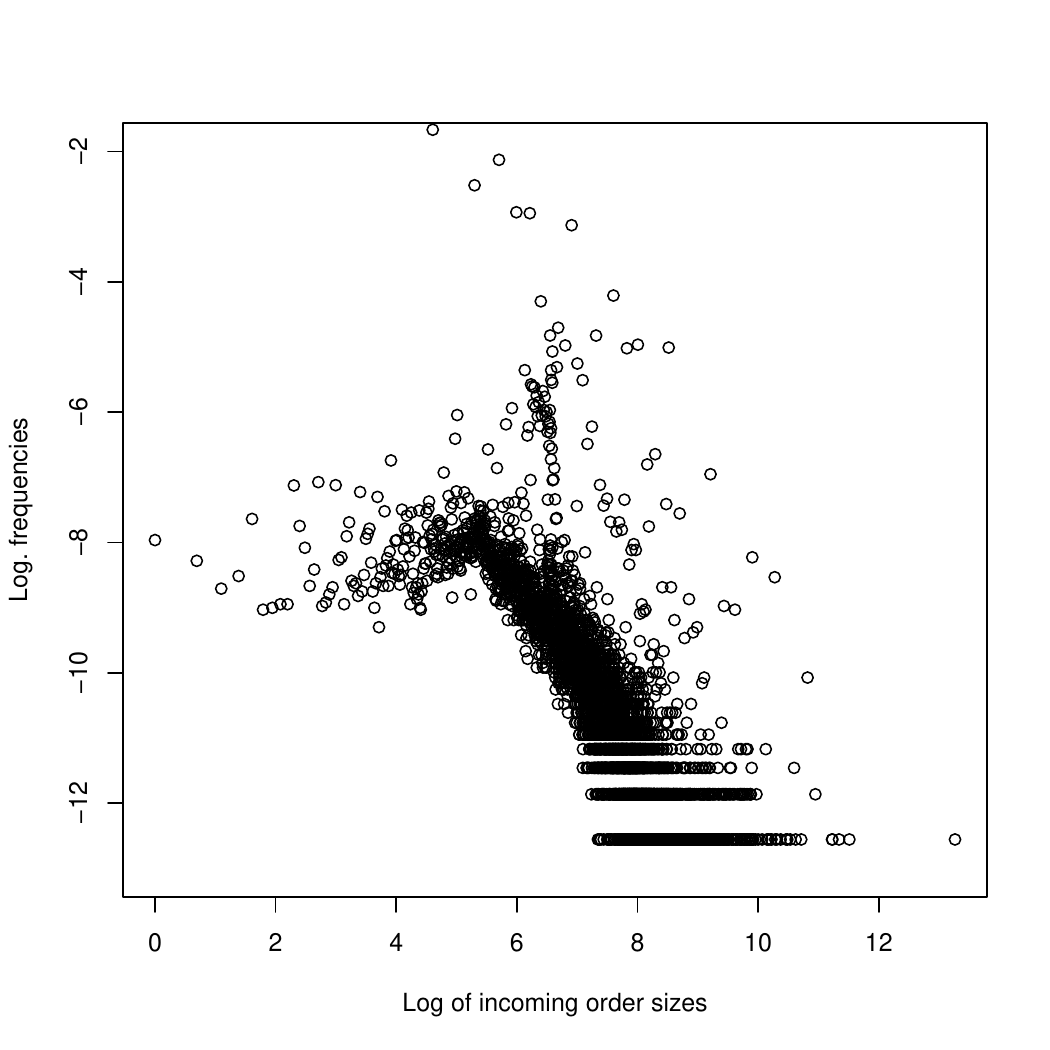}
 \caption{Bid arrivals}
 \label{fig:size_bidarrival}
\end{subfigure}
\begin{subfigure}[b]{0.45\textwidth}
 \includegraphics[width=\linewidth]{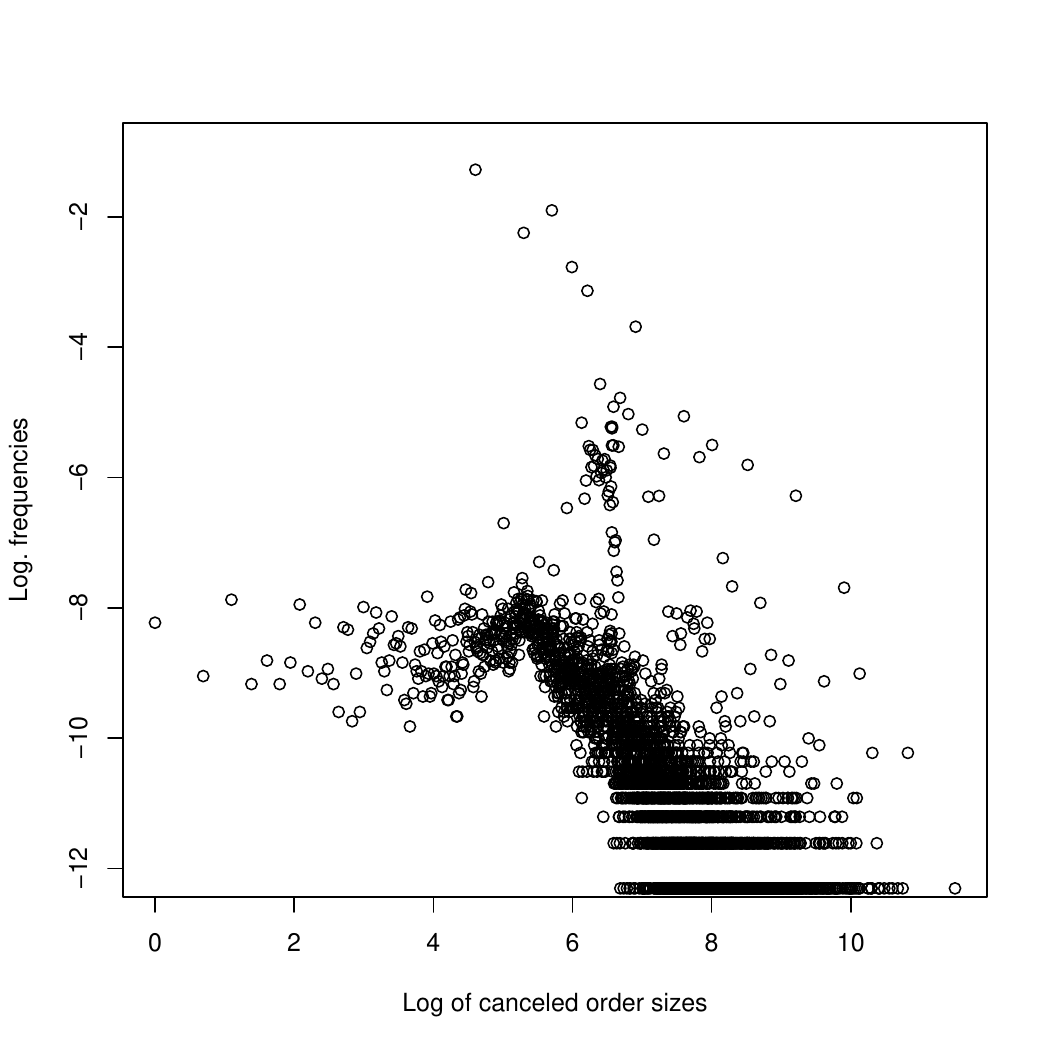}
 \caption{Ask cancellations}
 \label{fig:size_askcancellations}
\end{subfigure}
\begin{subfigure}[b]{0.45\textwidth}
 \includegraphics[width=\linewidth]{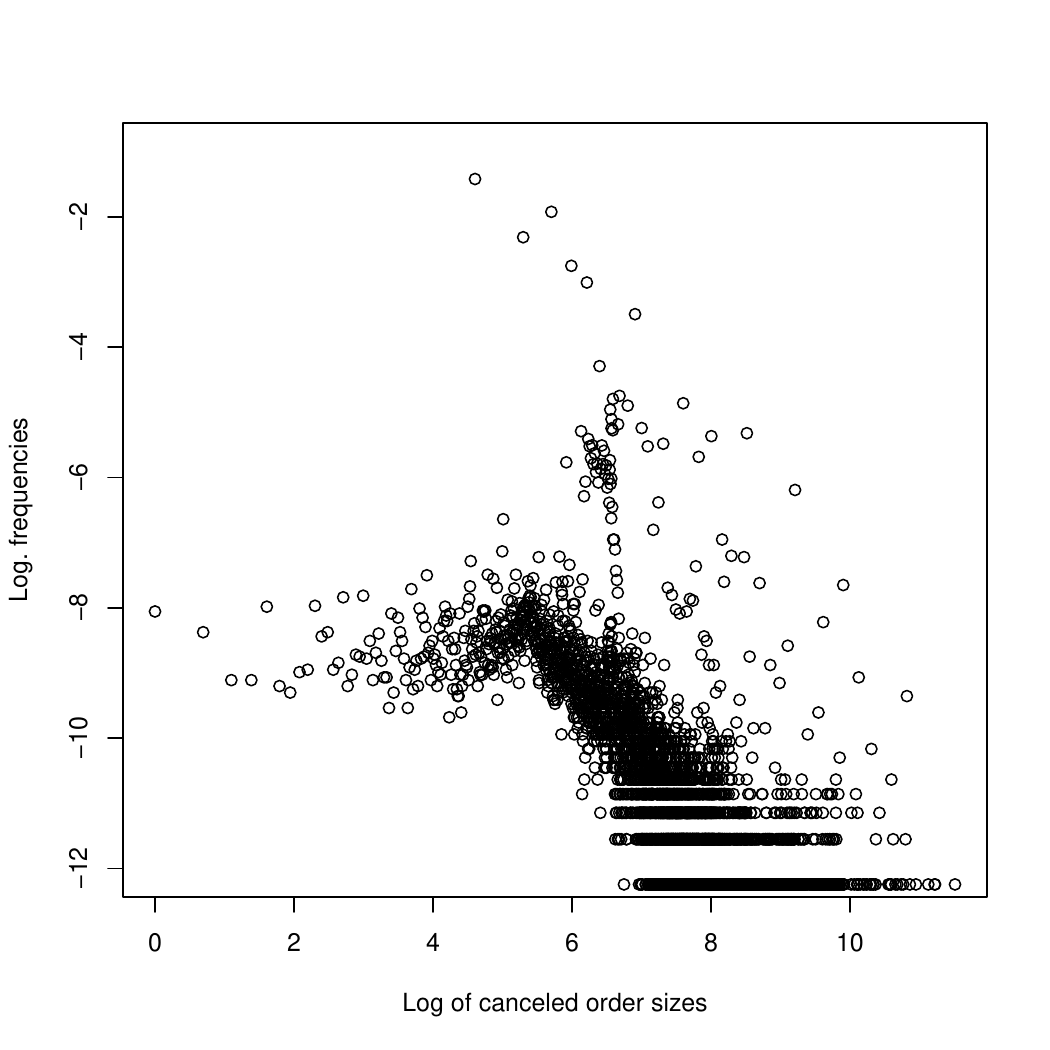}
 \caption{Bid cancellations}
 \label{fig:size_bidcancellations}
\end{subfigure}
\end{center}
\end{figure}

We consider three theoretical probability mass functions for the arrival and cancellation rates across price levels (denoted by $p_{K,M}(\cdot)$):
The uniform distribution (uni), a discrete log-normal distribution with fixed parameters (fix), and a discrete log-normal distribution which depends on the prevailing spread (dyn).
For the probability distribution across volume levels (denoted by $p_{Q,M}(\cdot)$), we only consider a power law distribution (pow).
The power-law distribution is empirically motivated to capture the heavy tails of the volume distribution.
The distribution of order size and the heavy tails can be seen in \Cref{fig:size_dist} which depicts the frequencies of order arrivals and cancellations for an 
exemplary stock in our dataset, the retail company Metro (MEO).

\Cref{appendix:Simulation} lists all the functional specifications as well as a description on how market orders are incorporated in the distributional setup.
Iceberg orders, stop orders, or fill-or-kill restrictions are neglected in the simulation study, as the events marked by these order types only make up for less than 1\% of all events in our data set.

Additionally, we also investigate cases in which $p_{K,M}(\cdot)$ and $p_{Q,M}(\cdot)$ are described by the empirical univariate frequency distributions in our sample across $k$ and $q$, respectively (emp).
We also utilize the joint frequency distribution of the observed pairs $(k,q)$ in one scenario (emp,emp).
Note that although we use the empirical frequencies, the rates are fixed over the entire simulation run.
Thus, in the scenario 'emp', no dynamic feedback between the state of the book and the arrival and cancellation rates is introduced.

For all combinations of these distributional specifications (in total 8 scenarios\footnote{The scenarios are (uni,pow), (uni,emp), (fix,pow), (fix,emp), (dyn,pow), (dyn,emp), (emp,pow) and (emp,emp)
where the list of pairs utilizes the introduced abbreviations and states the distribution across $k$ in the first coordinate and the one across $q$ in the second coordinate.}) for each stock, we
simulate 200 realizations of LOB evolutions over half a trading day (4 hours).
The state at the end of our sample, i.e., after the opening auction on March 31, 2004 at 9h00 CET, serves as a starting point for the simulation.
We also ran the simulations for a different starting point, namely the positions after the opening auction at 9h00 CET of January 2, 2004.
The results are not qualitatively different for the two simulation setups.

\subsection{Discussion of the Simulation Results}

\begin{figure}
\caption{Scenario: Uniformly distributed arrival and cancellation rates}
\label{fig:simul_uni}
\centering
\begin{minipage}{0.9\linewidth}
The graphs show 200 simulated paths of transaction prices (in red) using the scenario in which the arrivals and cancellations of orders follow a uniform distribution.
The starting point of each simulation is the LOB position of the MEO stock on March 31, 2004 after the midday auction at 13h00.
The true history of transaction prices during the first half of that day is depicted in black.
In \ref{fig:simul_uni_emp}, the empirical order size distribution is taken to generate the samples.
In \ref{fig:simul_uni_pow}, a power law is assumed to generate order sizes.
Paths that end earlier than 12h00 result in an empty order book.
\end{minipage}
\begin{subfigure}[b]{0.45\textwidth}
 \includegraphics[width=\linewidth]{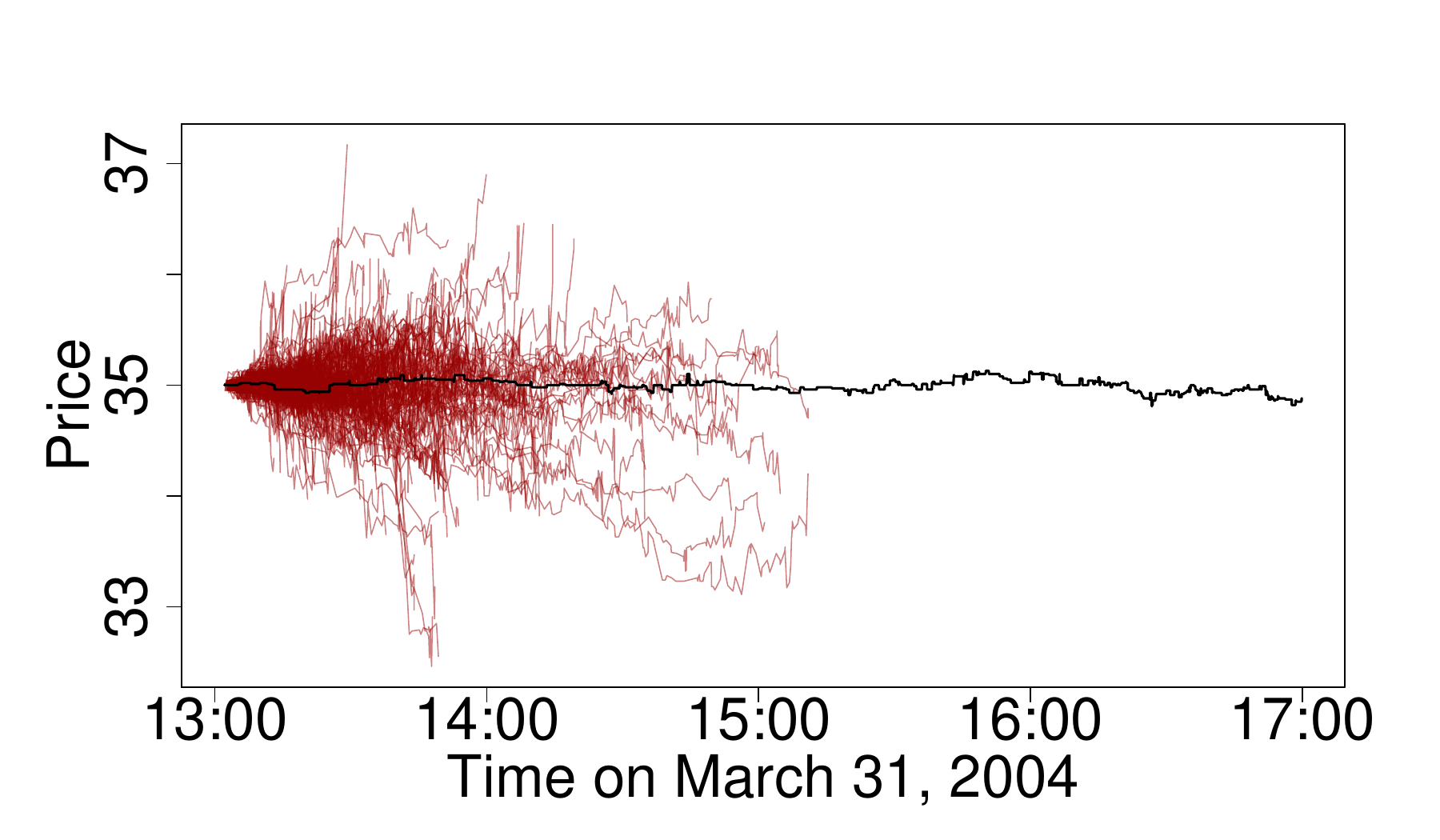}
 \caption{Empirical}
 \label{fig:simul_uni_emp}
\end{subfigure}
\begin{subfigure}[b]{0.45\textwidth}
 \includegraphics[width=\linewidth]{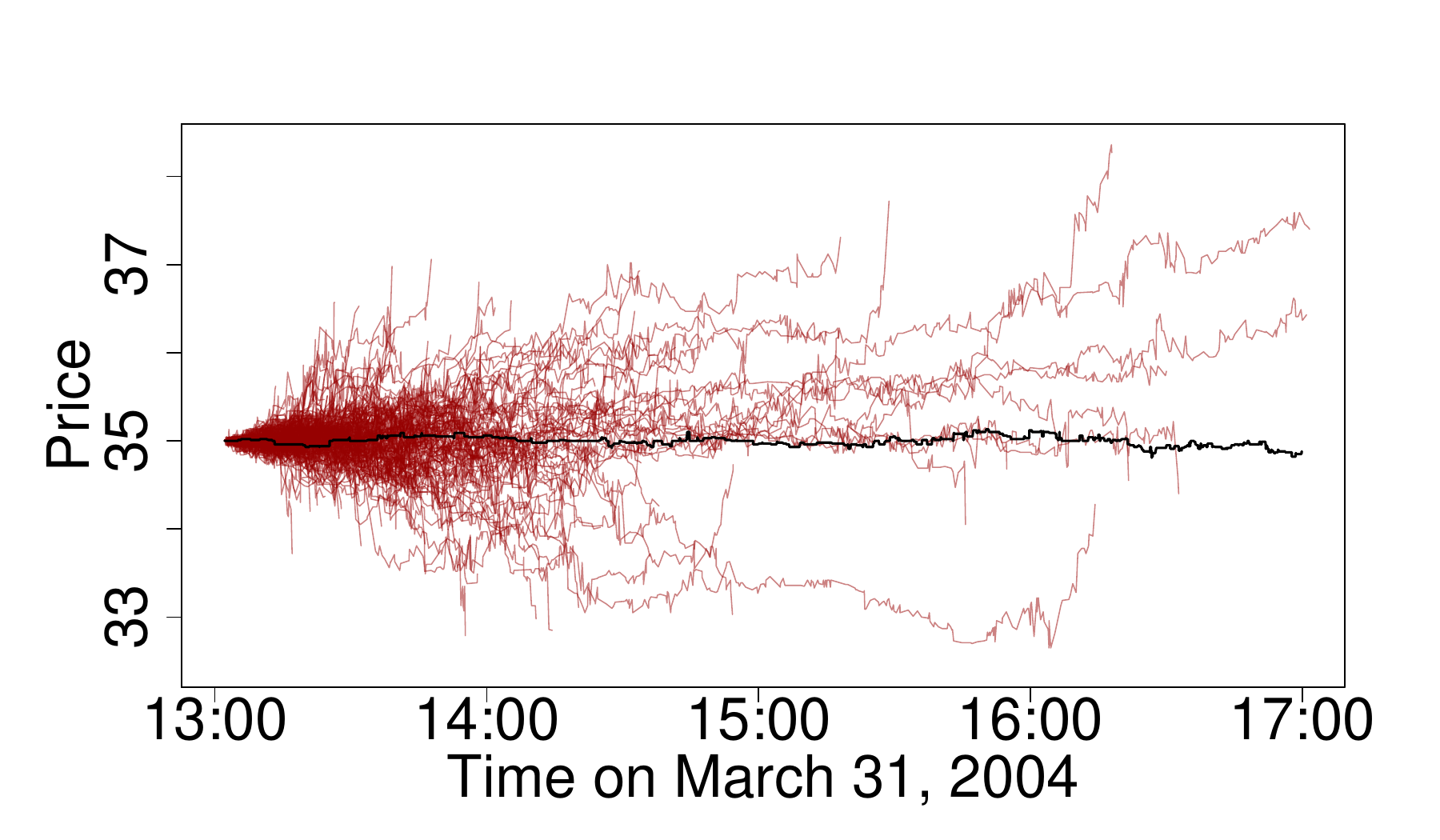}
 \caption{Power law}
 \label{fig:simul_uni_pow}
\end{subfigure}
\end{figure}

We first turn to the results from the uni scenario which are depicted in \Cref{fig:simul_uni}.
In this scenario, a lot of simulation runs ended with an empty order book, indicated by the sudden stop of the individual (red) price paths.
Furthermore, the variance of transaction prices, which is induced by the uniform distribution, is rather high, especially, when using the empirical volume distribution.
This can be seen in Table \ref{tab:MeanStdSimulated_Tprices} which shows the average mean and standard deviation of the simulated logarithmic transaction price changes across all simulation runs for all the different scenarios.
The 'emp/pow' as well as the two 'uni' scenarios result in a high standard deviation for the simulated returns.
Compared to the observed empirical values, the simulated values in these scenarios are implausibly high.
The average of the time series means and standard deviations of the observed logarithmic transaction price changes across all thirty stocks for the same time period are displayed in the last row in \Cref{tab:MeanStdSimulated_Tprices}. For the time after the opening auction (9h00 - 12h00) on January 2, 2020 the values are $\bar{\mu}_{\text{emp}} = 0.02\cdot 10^3$ and $\bar{\sigma}_{\text{emp}} = 0.67\cdot 10^3$.

\begin{table}
\caption{Mean and standard deviation of simulated price changes}
\label{tab:MeanStdSimulated_Tprices}
\centering
\begin{minipage}{0.9\linewidth}
For each of the 30 stocks, the time series mean and standard deviation of the logarithmic transaction price changes (in event time) across the 200 simulations has been calculated.
 The table reports the average across the 30 means $\bar{\mu}$ and standard deviations $\bar{\sigma}$ multiplied by $10^3$. In the last row, the average across the observed time series means and standard deviations of the logarithmic transaction price changes are reported.
 \vspace{0.2cm}
\end{minipage}
\resizebox{\textwidth}{!}{
 \begin{tabular}{ccccc cc cccc cc cccc cc cccc}
 \toprule
 \multicolumn{2}{c}{Initial Position:}&&
 \multicolumn{6}{c}{January 2, 2004} &&&
 \multicolumn{6}{c}{March 31, 2004}\\
 &&&
 \multicolumn{2}{c}{Opening Auction} &&&
 \multicolumn{2}{c}{Midday Auction}
 &&&
 \multicolumn{2}{c}{Opening Auction} &&&
 \multicolumn{2}{c}{Midday Auction}\\
 \cmidrule{4-9}
 \cmidrule{12-18}
\multicolumn{2}{c}{Scenario}&&$\bar{\mu}$&$\bar{\sigma}$&&&$\bar{\mu}$&$\bar{\sigma}$
&&&$\bar{\mu}$&$\bar{\sigma}$&&&$\bar{\mu}$&$\bar{\sigma}$\\
  \midrule
  uni & emp && -0.12 & 3.82    &&& -0.02 & 3.05   &&& -0.06 & 3.82  &&& -0.04 & 3.17\\
  uni & pow && -0.11 & 3.17    &&& 0.00 & 2.57    &&& -0.03 & 3.29  &&& -0.01 & 2.67\\
  fix & emp && 0.01 & 0.70       &&& 0.01 & 0.69    &&&0.01 & 0.71  &&& 0.01 & 0.70\\
  fix & pow && 0.01 & 0.70     &&& 0.01 & 0.68   &&& 0.01 & 0.69    &&& 0.01 & 0.69\\
  dyn & emp && 0.00 & 1.21       &&& 0.00 & 1.17     &&& -0.00 & 1.21  &&& 0.00 & 1.19 \\
  dyn & pow && 0.00 & 0.87      &&& -0.00 & 0.85    &&& 0.00 & 0.87 &&& -0.00 & 0.86\\
  emp & emp && -0.01 & 3.43     &&& -0.01 & 3.43    &&& -0.04 & 2.67&&& -0.04 & 2.89\\
  emp & pow && -0.04 & 2.37    &&& -0.02 & 1.48   &&& -0.03 & 1.53  &&& -0.03 & 1.56 \\

  \midrule 
 \multicolumn{2}{c}{observed} &&0.02&0.67&&&0.01 &0.58 &&& 0.00 &0.67&&&-0.00&0.49\\
   \bottomrule
\end{tabular}
}
\end{table}

Note that for these simulations, the average event rates on each market side (which is denoted $\bar{r}_{0,M,i,j,\cdot}$ in \Cref{eq:rate_decomp} in \Cref{appendix:Simulation}) are the same as in the case of the fixed and dynamic arrival and cancellation rates.
We may associate the uniform distribution across price levels with somewhat uninformed traders who, regardless of the price, randomly submit orders in the vicinity of the current best quote.
With the uniform distribution, the mean and variance of the price level is rather high compared to the DGX specifications in other simulation scenarios as well as the empirically observed equivalents.
Throughout all simulation scenarios, we see that a higher mean and variance in the distribution across price levels are related to a higher variance in transaction prices.
This result is interesting since, as noted before, arrival rates are linked to trader's behavior.
So, if traders are uniformed on how the asset should be valued and constantly shift their valuation with no clear tendency and/or if traders are indifferent between immediate execution and delayed execution, transaction prices become highly volatile.

\begin{figure}
\caption{Special Case: Fixed DGX Distribution with an Imbalance in Arrival Rates}
\label{fig:simul_meo_fix}
\centering
\begin{minipage}{0.9\linewidth}
The graphs show 200 simulated paths of transaction prices (in red) using the scenario in which the arrival and cancellations of orders follow a fixed DGX distribution across price levels.
However, an imbalance in the distribution of arrival rates is inserted as bid orders arrive densely in vicinity to the best ask price level.
The starting point of each simulation is the LOB position for the MEO stock on March 31, 2004 after the midday auction at 13h00.
The true history of transaction prices for the first half of that day is depicted in black.
In \ref{fig:simul_fix_emp}, the empirical order size distribution is taken to generate the samples.
In \ref{fig:simul_fix_pow}, a power law is assumed to generate order sizes.
\end{minipage}
\begin{subfigure}[b]{0.45\textwidth}
 \includegraphics[width=\linewidth]{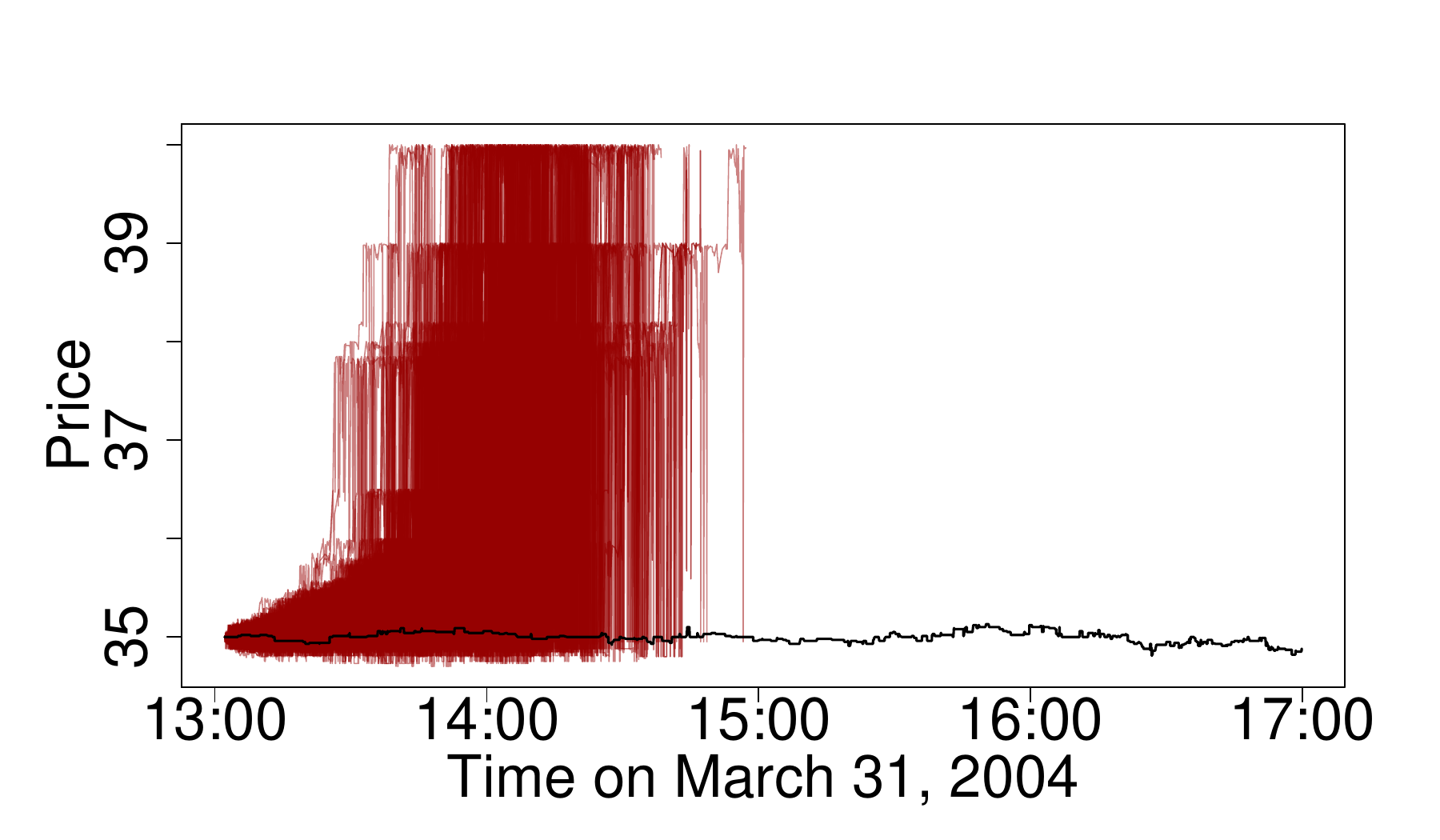}
 \caption{Empirical}
 \label{fig:simul_fix_emp}
\end{subfigure}
\begin{subfigure}[b]{0.45\textwidth}
 \includegraphics[width=\linewidth]{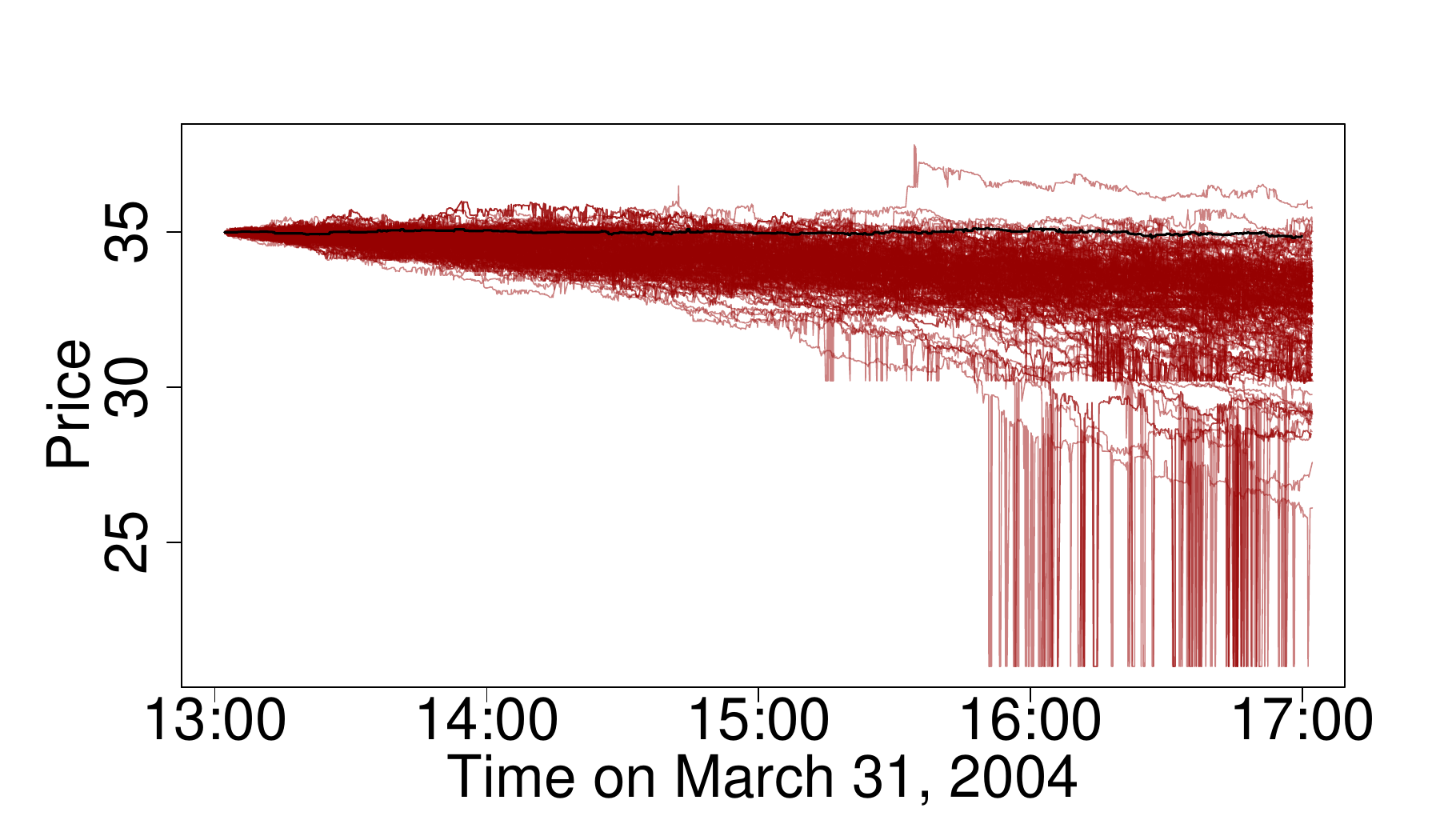}
 \caption{Power law}
 \label{fig:simul_fix_pow}
\end{subfigure}
\end{figure}

\begin{figure}
\caption{Scenario: Fixed DGX distribution for arrival and cancellation rates}
\label{fig:simul_fix}
\centering
\begin{minipage}{0.9\linewidth}
The graphs show 200 simulated paths of transaction prices (in red) using the scenario in which the arrival and cancellations of orders follow a fixed DGX distribution across price levels.
The starting point of each simulation is the LOB position for the MEO stock on March 31, 2004 after the midday auction at 13h00.
The true history of transaction prices for the first half of that day is depicted in black.
In \ref{fig:simul_fix_emp}, the empirical order size distribution is taken to generate the samples.
In \ref{fig:simul_fix_pow}, a power law is assumed to generate order sizes.
\end{minipage}
\begin{subfigure}[b]{0.45\textwidth}
 \includegraphics[width=\linewidth]{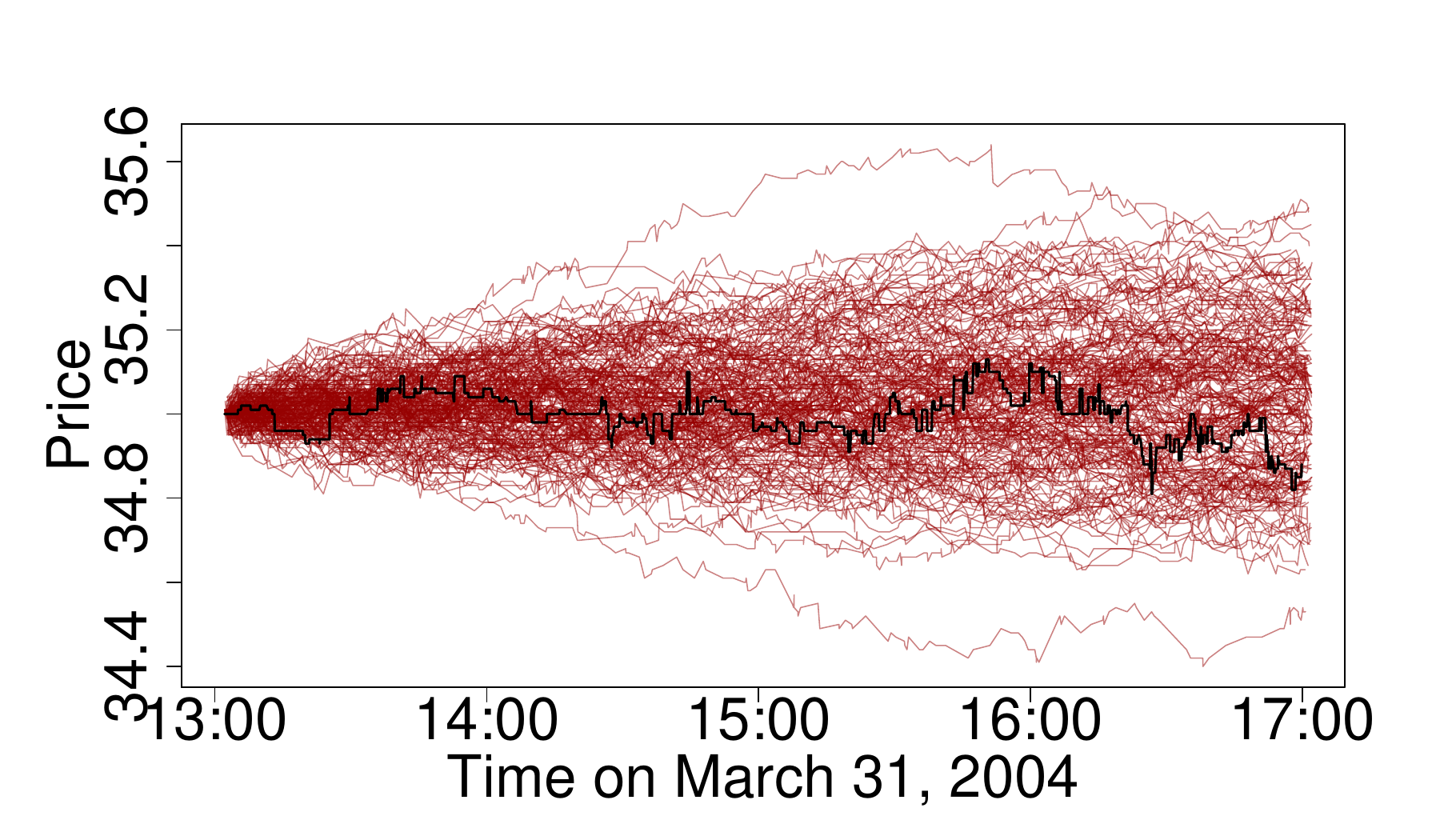}
 \caption{Empirical}
 \label{fig:simul_fix_emp}
\end{subfigure}
\begin{subfigure}[b]{0.45\textwidth}
 \includegraphics[width=\linewidth]{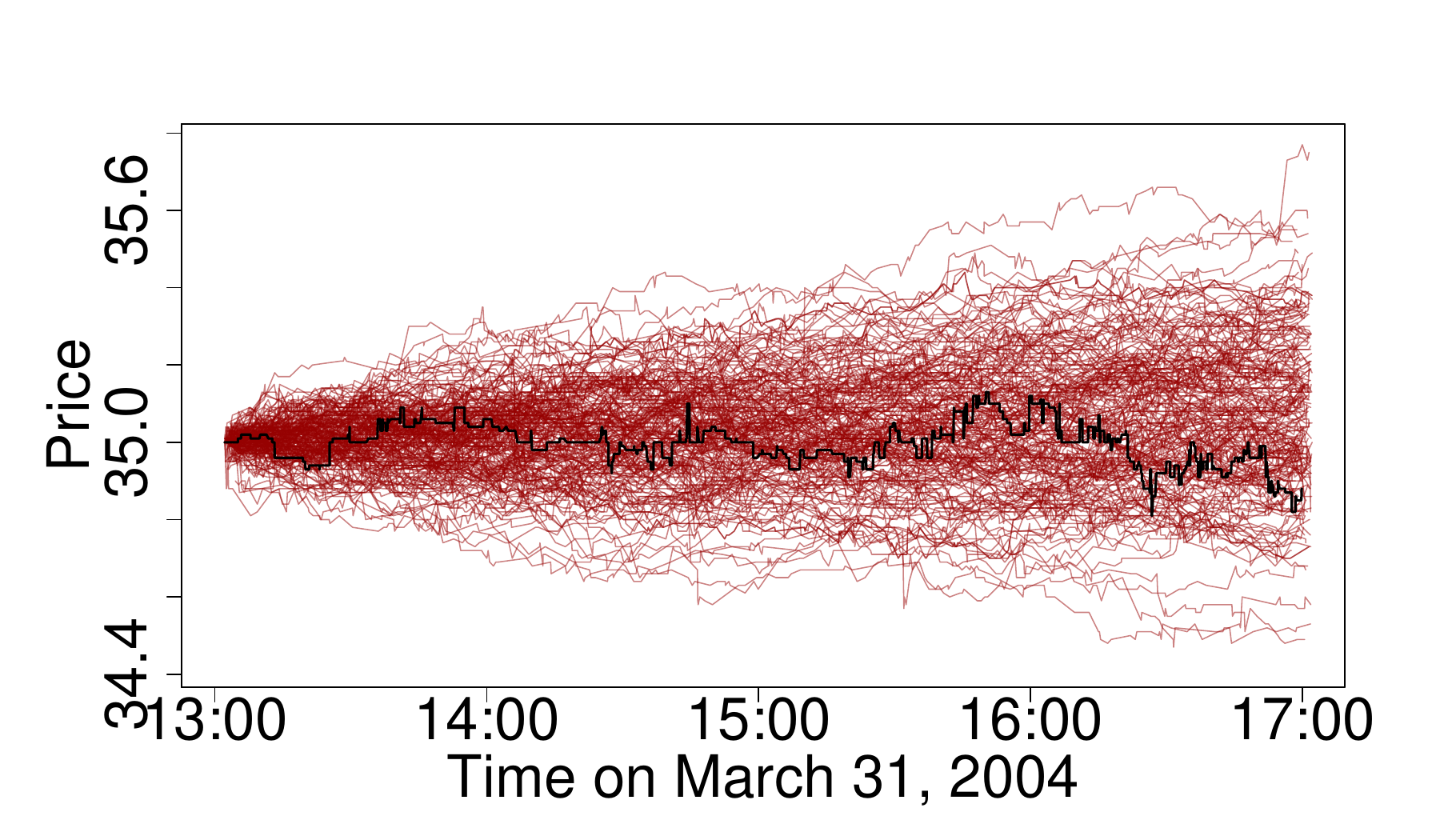}
 \caption{Power law}
 \label{fig:simul_fix_pow}
\end{subfigure}
\end{figure}

Second, the results for the fixed DGX distribution across price levels are presented in \Cref{fig:simul_fix}.
We can see that the power law very rarely induces large jumps in transaction prices due to the extremely large order sizes that are possible under this distributional scheme.
In general, however, differences between the volume distributions are not obvious, neither in the 'uni' scenario nor in the 'fix' scenario.
The average of the time series means and standard deviations for the simulations with the fixed DGX distributions as provided in 
\Cref{tab:MeanStdSimulated_Tprices} are close to the empirical ones.
One very interesting result for the 'fix' scenario concerns the parameters of the DGX distribution.
The distributional parameters $\mu$ and $\sigma$ are almost identically defined: The parameter $\mu$ of the DGX distribution for incoming and canceled orders on 
the bid side has been specified slightly higher (at $\mu_{B,a} = 1{.}766$ and $\mu_{B,c}= 1{.}674$) than the one for the ask side ($\mu_{A,a}= 1{.}726$ and 
$\mu_{A,c}= 1{.}620$) to match estimated parameters from empirically observed frequencies.
However, this small difference, does not seem to have any effect.
In order to analyse the effect, we ran the simulation of the 'fix' scenario only for the stock MEO again with adjusted parameters:
Leaving the distribution of the cancellation rates untouched, only the parameters of DGX distribution across arrival rates are altered to $\mu_{B,a} = 0{.}1$ 
and $\mu_{A,a}= 2$ as well as 
$\sigma_{B,a} = 0{.}2$ and $\sigma_{A,a}= 1$.
Therewith, the distribution of bid order arrivals is much more dense around the best ask price.
The result can be seen in \Cref{fig:simul_meo_fix}.

The behavior of the transaction prices also depends crucially on the cancellation rates as well as the volume distribution. 
In the case where order size is distributed according to a power law distribution, order sizes on both market sides are identically distributed.
Furthermore, the power law distribution generates a lot of small orders while  large orders are very rare.
At the same time, the initial position on the bid side contains several large orders close to the best bid price (which are more likely to be canceled).
The ask side consists of several medium sized orders close to the best ask, while the large orders rest deep in the book.
So, the frequent small orders inserted at or close to the best ask price are not able to move the market upwards permanently due to the medium sized orders sitting in the book on the ask side.
At the same time large orders at the front of the bid side (from the initial position) are canceled frequently.
One rare large ask order generated by the power law, thus, is able to move the bid price quite a lot.
The longer the simulation is running, the more likely it is for a large ask order to occur and the more likely it is that the large orders at the top of the bid side are already canceled.
This makes it easier for the ask side to move the best ask down and therefore transaction prices deteriorate.

In the empirical distribution, the volume distribution on the bid side dominates the volume distribution of the ask side.
Thus, bid orders inserted into the book are larger in size than inserted ask orders.
Since bid orders are inserted close to or at the best ask price level, the bid side moves the market soon after the simulation starts upwards.
This leads to a hefty upward drift of transaction prices in each simulated transaction path and empty order books on the ask side.
Hence, we can conclude that limit order distributions with a high probability mass in the vicinity of the best quote in one market side push transaction prices 
in the direction of the opposite market side if the order size distribution allows for frequent medium sized orders.



\begin{figure}
\caption{Scenario: Dynamical DGX distribution for arrival and cancellation rates}
\label{fig:simul_dyn}
\centering
\begin{minipage}{0.9\linewidth}
The graphs show 200 simulation paths of transaction prices (in red) using the scenario in which the arrival and cancellations of orders follow a dynamical DGX distribution across price levels.
In the dynamical DGX distributions the parameters $\mu$ and $\sigma$ are functions of the prevailing integer spread.
The starting point of each simulation were the LOB positions for the MEO stock on March 31, 2004 after the midday auction at 13h00.
The true history of transaction prices for the first half of that day is depicted in black.
In \ref{fig:simul_dyn_emp}, the empirical order size distribution is taken to generate the samples.
In \ref{fig:simul_dyn_pow}, a power law is assumed to generate order sizes.
\end{minipage}
\begin{subfigure}[b]{0.45\textwidth}
 \includegraphics[width=\linewidth]{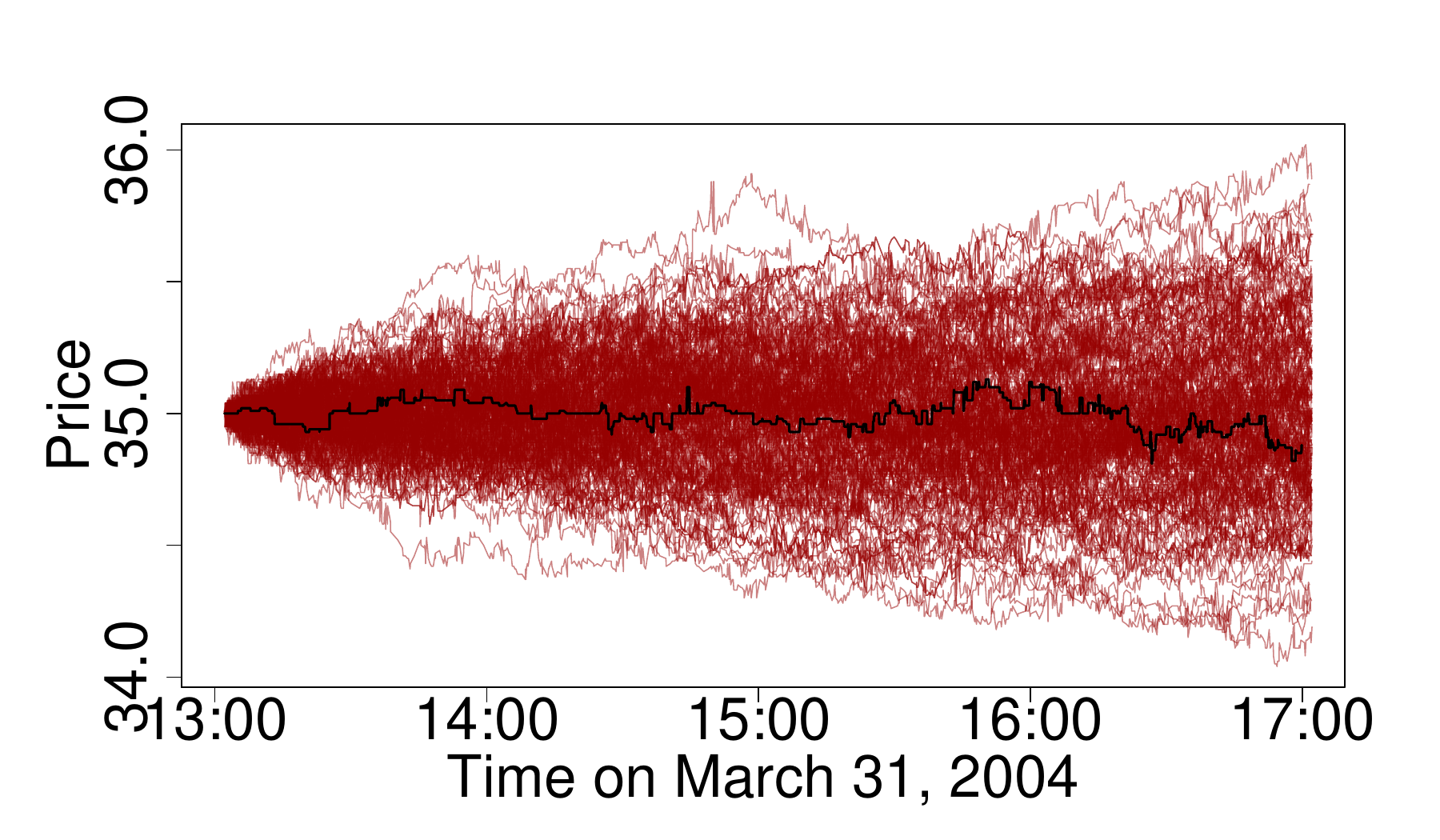}
 \caption{Empirical}
 \label{fig:simul_dyn_emp}
\end{subfigure}
\begin{subfigure}[b]{0.45\textwidth}
 \includegraphics[width=\linewidth]{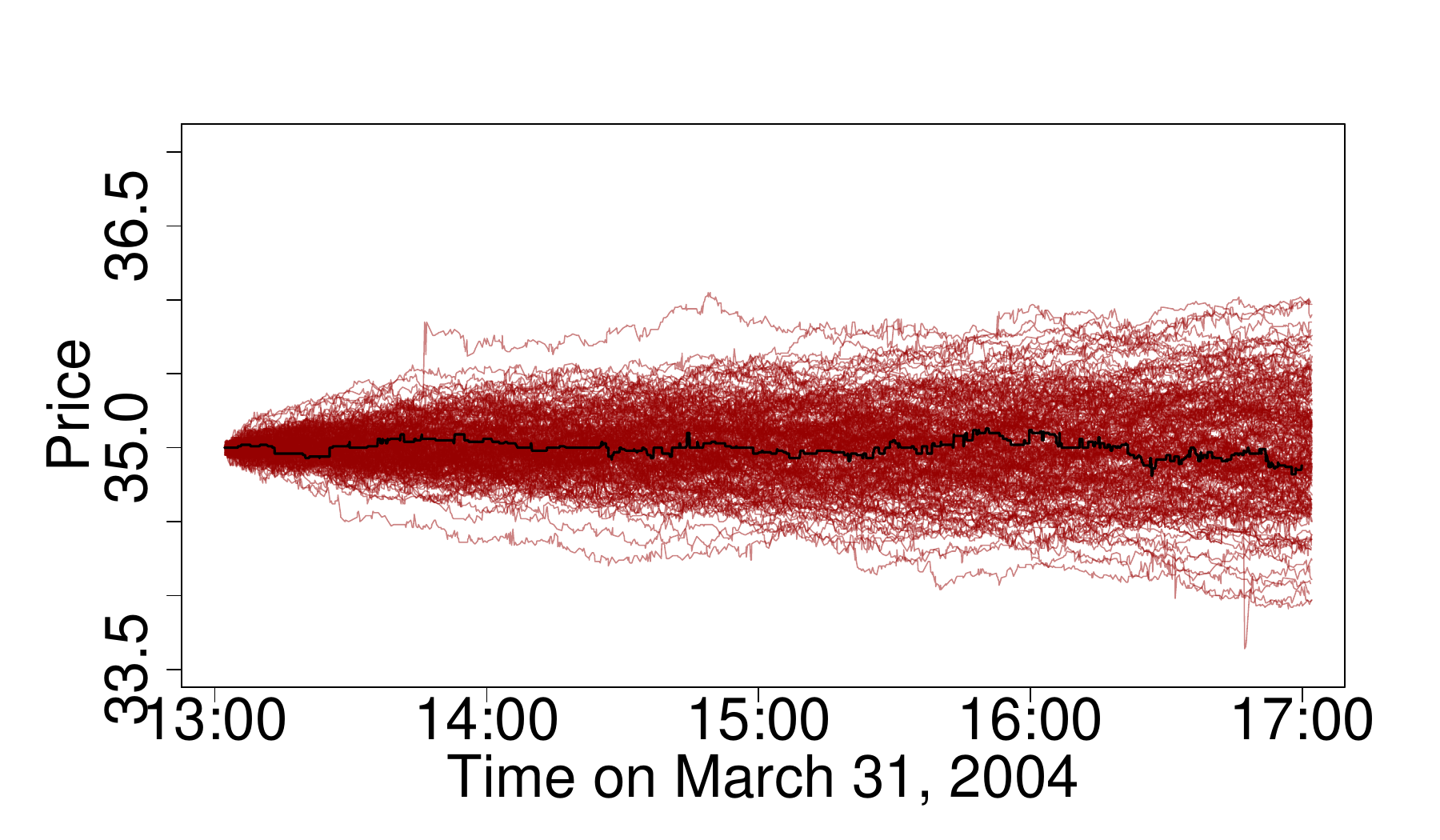}
 \caption{Power law}
 \label{fig:simul_dyn_pow}
\end{subfigure}
\end{figure}

Third, the simulated transaction prices resulting from the scenario with dynamical shifting and scaling DGX distributions which depend on the prevailing spread, are shown in \Cref{fig:simul_dyn}.
In this scenario the moments of the DGX distribution depend on the prevailing spread.
The dynamical adjustment of the arrival and cancellation rates across price levels is balanced, So, the mean and standard deviation of the DGX distribution across price levels is the same on both market sides.
Furthermore, the large jumps induced by the power law distribution are still rare, but rather pronounced.
These jumps are, however, not sufficient to cause an increase in volatility.
In fact, the scenarios with a power law distribution exhibit on average a slightly smaller volatility which might be due to the fact that the order size is 
rather small.
For the power law distributed order size, the time series mean and standard deviation of the dynamically adjusting simulation scenario are close to the time 
series mean and standard deviation of real observed logarithmic transaction changes as presented in \Cref{tab:MeanStdSimulated_Tprices}.
The scenario with the empirical order size distribution is too volatile.

\begin{figure}
\caption{Scenario: Empirical frequency distribution for arrival and cancellation rates}
\label{fig:simul_emp}
\centering
\begin{minipage}{0.9\linewidth}
The graphs show 200 simulation paths of transaction prices (in red) using the scenario in which the arrival and cancellations of orders follow a dynamical DGX distribution across price levels.
In the dynamical DGX distributions the parameters $\mu$ and $\sigma$ are functions of the prevailing integer spread.
The starting point of each simulation were the LOB position for the MEO stock on March 31, 2004 after the midday auction at 13h00.
The true history of transaction prices for the first half of that day is depicted in black.
In \ref{fig:simul_emp_emp}, the empirical order size distribution is taken to generate the samples.
In \ref{fig:simul_emp_pow}, a power law is assumed to generate order sizes.
\end{minipage}
\begin{subfigure}[b]{0.45\textwidth}
 \includegraphics[width=\linewidth]{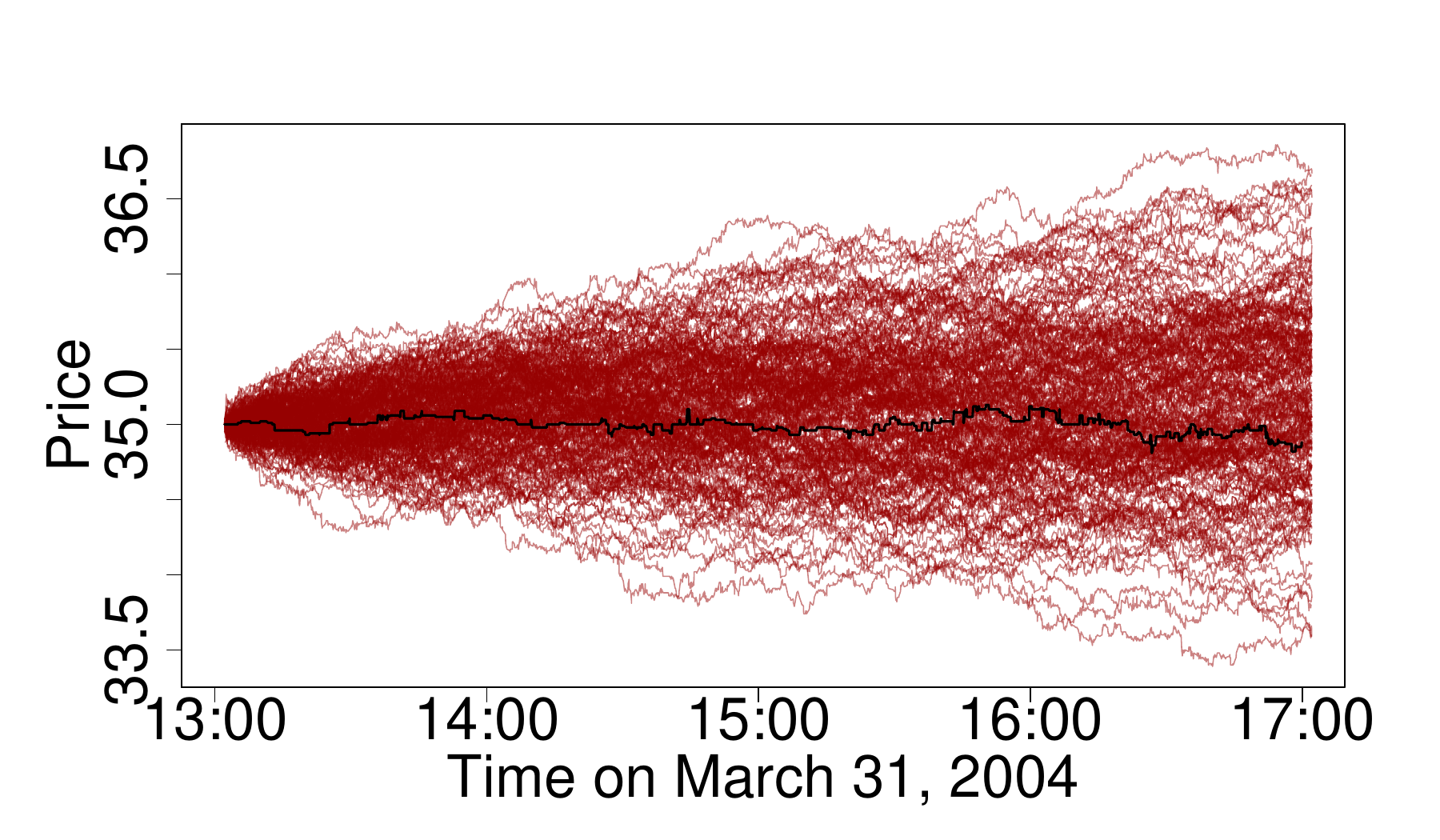}
 \caption{Empirical}
 \label{fig:simul_emp_emp}
\end{subfigure}
\begin{subfigure}[b]{0.45\textwidth}
 \includegraphics[width=\linewidth]{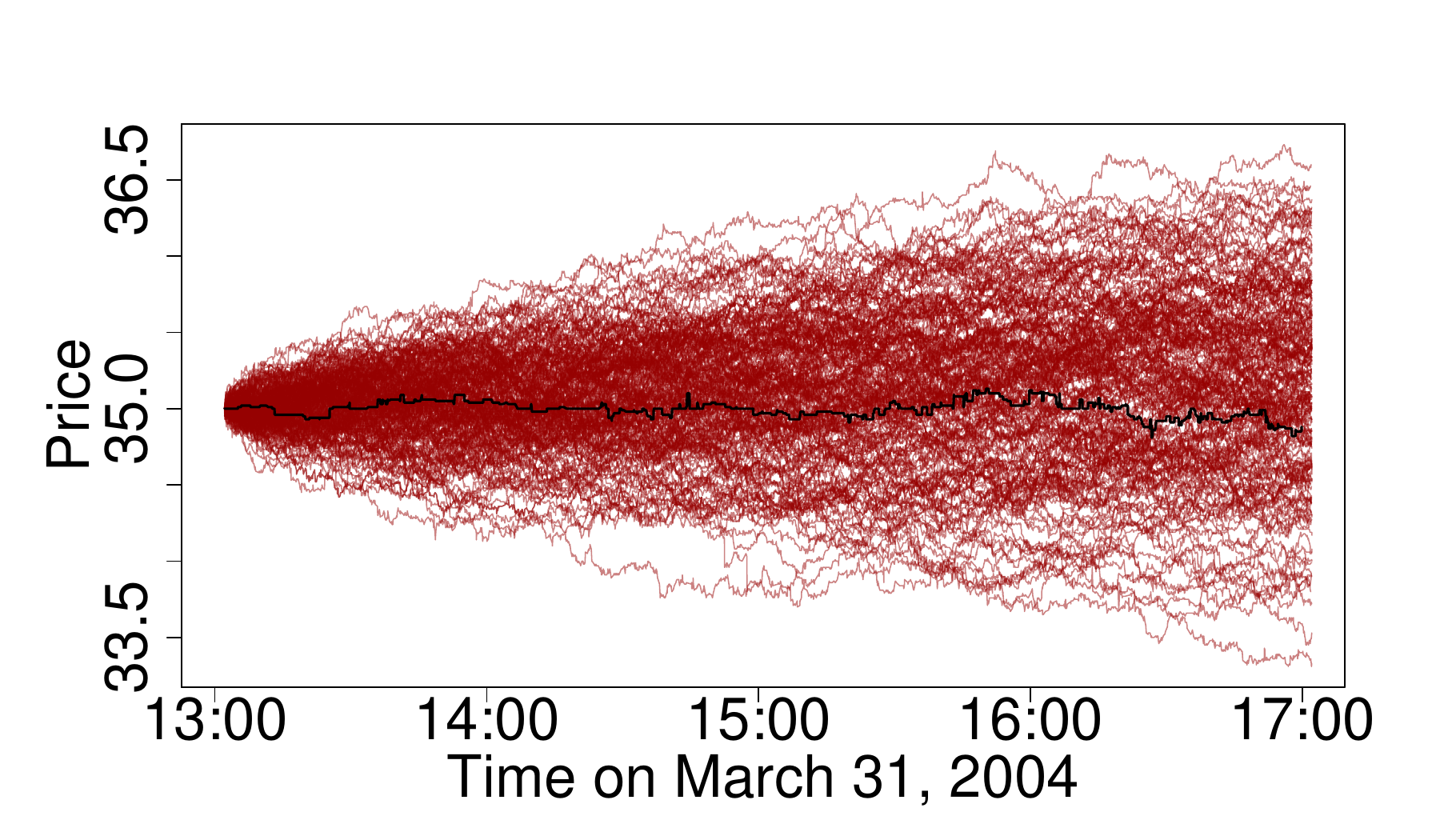}
 \caption{Power law}
 \label{fig:simul_emp_pow}
\end{subfigure}
\end{figure}

\begin{figure}
 \caption{Estimated DGX parameters and average spread}
 \label{fig:Param_DGX_vs_spread}
 \centering
 \begin{minipage}{0.9\linewidth}
  The scatter plot depicts the average spread of DAX components in the sampling interval (first quarter of 2004) during continuous trading of the XETRA order book against the estimated parameters of a DGX-distribution fitted to the unconditional frequencies of order arrivals (and cancellations) across price levels.
  The parameters were estimated using the log-likelihood method described in \cite{BiFCK01}.
 \end{minipage}
 \begin{subfigure}{0.45\linewidth}
  \includegraphics[width=\linewidth]{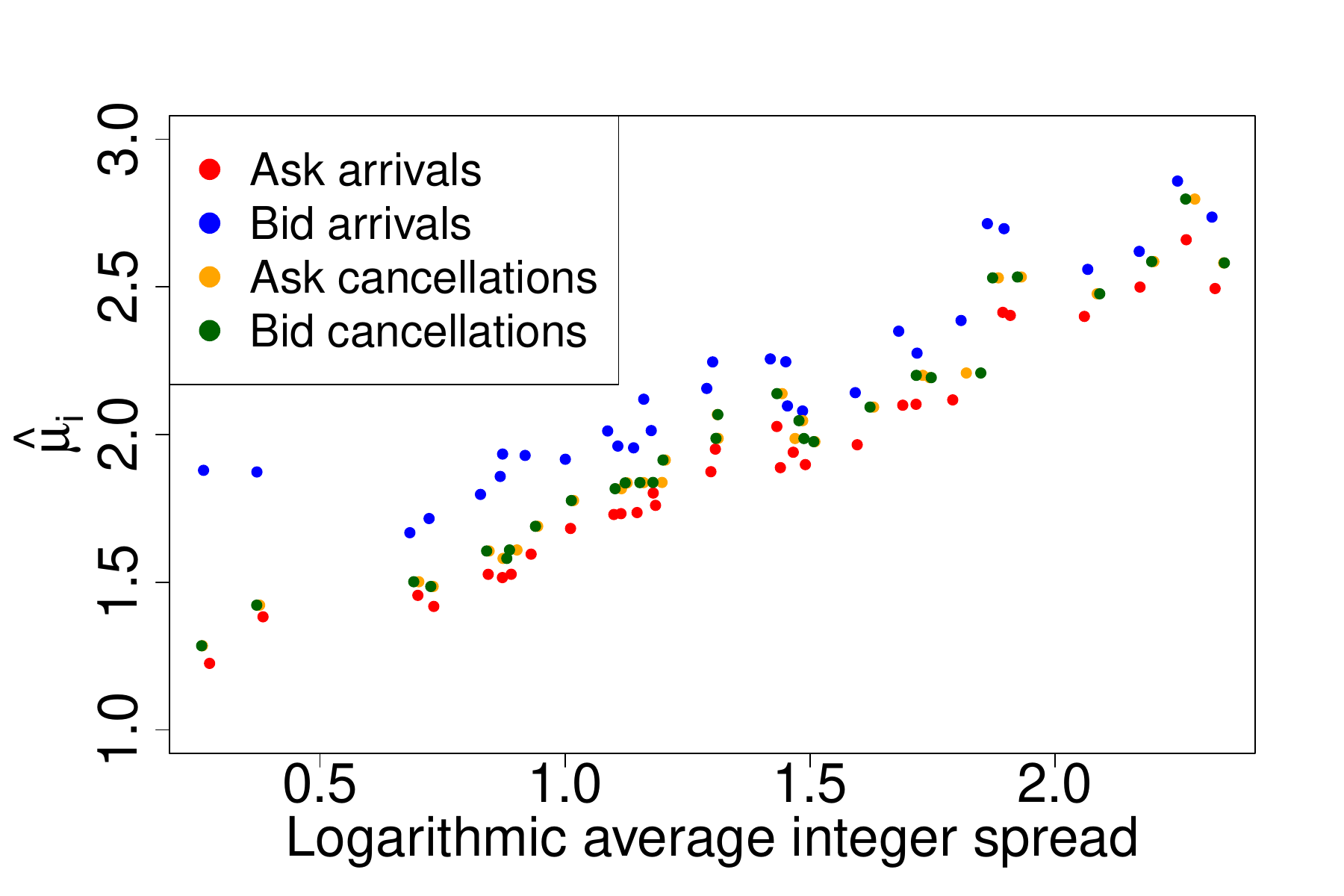}
  \subcaption{$\hat{\mu}_i$ against average spread}
 \end{subfigure}
  \begin{subfigure}{0.45\linewidth}
  \includegraphics[width=\linewidth]{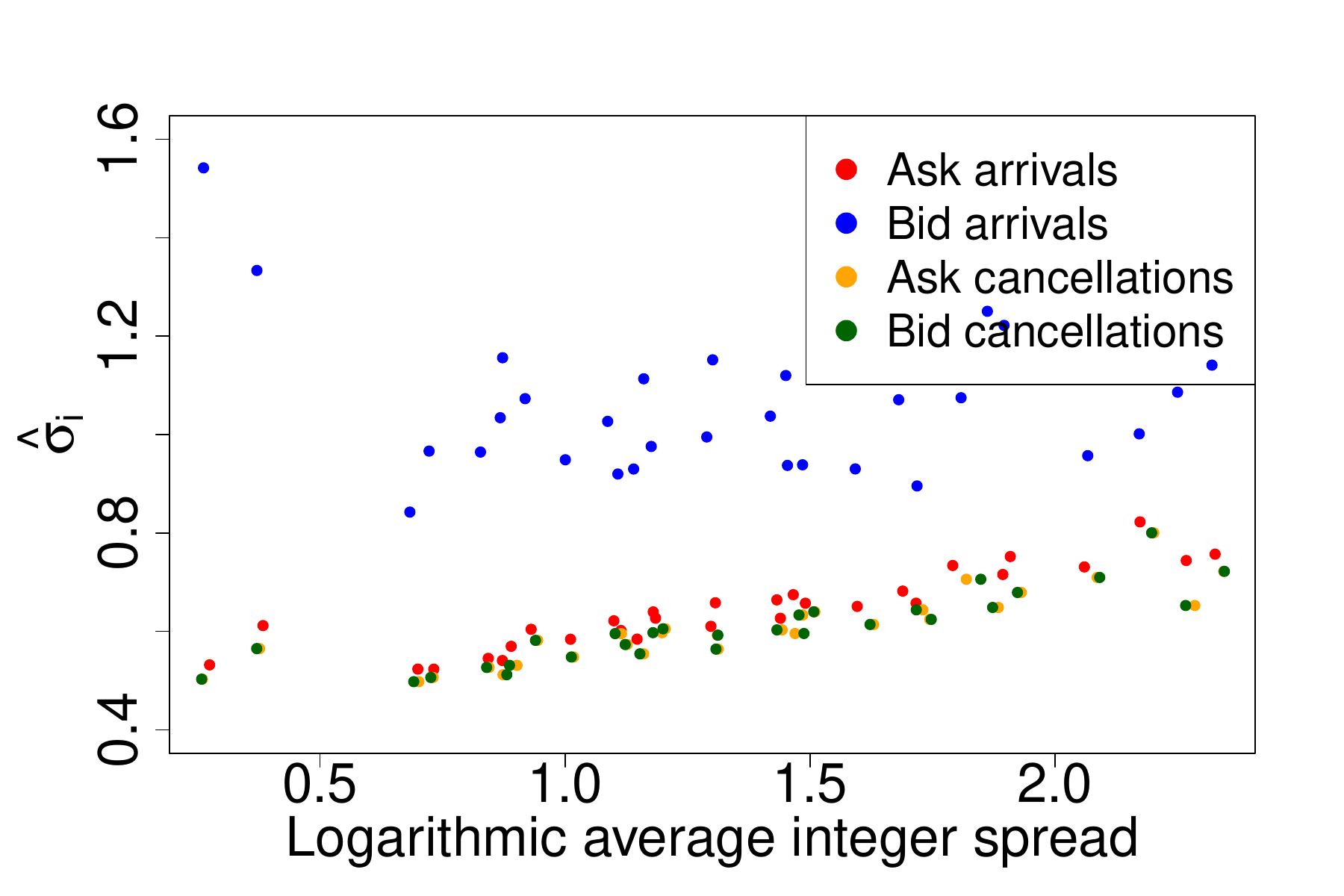}
  \subcaption{$\hat{\sigma}_i$ against average spread}
 \end{subfigure}

\end{figure}

Fourth, using the unconditional empirical frequency distributions as well as the empirical rates $\bar{r}_{0,M,i,j,\cdot}$, the results presented in \Cref{fig:simul_emp} deviate from empirical stylized facts in that the resulting paths are much more volatile.
The imbalance between ask and bid order arrivals and cancellations exhibited by some stocks, together with the observation that on average ask orders arrive 
closer to the best bid than bid orders to the best ask, drags transaction prices on average slightly down (see \Cref{tab:MeanStdSimulated_Tprices}).
The smaller distance to the best quote, on average, of ask orders can be seen in \Cref{fig:bin_freq} as well as in \Cref{fig:Param_DGX_vs_spread}.
It seems that in our sample, sellers tend to seek quicker order execution by placing their orders close to the buy side.
Buyers on the other hand, test their fortunes and patiently wait for a good deal to occur deeper in the book.
This can be clearly seen in \Cref{fig:Param_DGX_vs_spread}: For the same average spread, arriving bid orders are placed on average further away from the opposite market side than arriving ask orders.

\section{Empirical Analysis}
\label{sec:EmpiricalAnalysis}
In our empirical analysis, we focus on three variables that we deem most important to market participants.
The first is the logarithmic return which could be achieved based on a buy-and-sell strategy in subsequent intervals.
The second is the return of a sell-and-buy strategy.
The third one is the exchange liquidity measure (XLM) as introduced in \Cref{sec:Observables}, calculated with a round-trip of EUR 100{.}000.
To implement them, we sample the data in intervals of fixed length $\Delta t$ (1, 2, 5, 10, 15, 30, 45, 50, 60, 120, 240 minutes).
Let $t$ denote the last point in time of some arbitrary interval and $t-1$ the last point in the previous interval.
Then the logarithmic returns of a buy-and-sell ($\Delta p_{t,b}$) and a sell-and-buy strategy ($\Delta p_{t,s}$) are given as
\begin{align*}
 \Delta p_{t,b} = \log(\beta_{A,t}) - \log(\beta_{B,t-1}) ,\\
 \Delta p_{t,s} = \log(\beta_{B,t}) - \log(\beta_{A,t-1}) .
\end{align*}
For the $XLM$ the last observation in the respective interval is taken.

We have shown in \Cref{eq:MomentProjection} that the moments of some observable $O$ of the LOB system can be expressed as
\begin{align*}
 {E}_{t_0}[O^\nu]  = \sum_{\ket{z}} \bra{z} O^\nu e^{H(t -t_0)} \ket{\psi_0} .
\end{align*}
We can perceive the right hand side of this equation as an intricate function of the arrival and cancellation rates.
These rates depend on the event rates $\bar{r}_{0,M,i,j,e}$ of arrivals and cancellations of the various order types, the relative logarithmic integer price level $d_l$, the order size $q$, the
spread $\Delta$, and possibly other variables.
Thus, we can formulate a linear approximation for the expectation of any observable in the moments of exactly these variables.

By repeated Taylor series approximations of the terms in \Cref{eq:rate_decomp} in the variables $d_l$, $q$, and $\Delta$ around their respective mean, collecting terms, taking expectations and
neglecting terms with an order higher than four (see \Cref{appendix:Linearization}), we get for the expected value of some observable a linear approximation of the form
\begin{align}
 {\E}_{t_0}[O_{i,t}] &= \gamma_{0,i} +  \bigg(\delta_{i,0}+\delta_{i,1} {\E}_{t_0}[\Delta_t]\,\,+ \sum_{v=2}^{4}\delta_{i,v}
 {\E}_{t_0}[(\Delta_t-\mu_{\Delta,t})^v]\bigg) \,\,\times \notag\\
 &\quad\sum_{M,j,e} \rho_{0,M,i,j,e}  {\E}_{t_0}[\bar{r}_{0,M,i,j,e,t}]\,\,\times \notag\\
 &\hspace{1cm}\bigg(\kappa_{M,i,j,e,0}+ \kappa_{M,i,j,e,1} {\E}_{t_0,M,i,j,e}[d_{l,t}]\,\,+   \notag\\
 &\hspace{2.5cm}\sum_{v=2}^{4} \kappa_{M,i,j,e,v} {\E}_{t_0,M,i,j,e}[(d_{l,t}-\mu_{d_l,t})^v]\bigg) \,\,\times \notag\\
  &\hspace{1cm}\bigg(\xi_{M,i,j,e,0}+ \xi_{M,i,j,e,1} {\E}_{t_0,M,i,j,e}[q_{t,M,j,e}]\,\,+  \notag\\
  &\hspace{2.5cm}\sum_{v=2}^{4}\xi_{M,i,j,e,v} {\E}_{t_0,M,i,j,e}[(q_{t,M,j,e}-\mu_{q,t})^v]\bigg)+ \varepsilon_i,
 \label{eq:LinApproximation1}
\end{align}
where $\mu_{x,t} ={\E}_{t_0,M,i,j,e}[x_t]$.
Note that the indices in the subscript of the expected values indicate their conditioning set as outlined in \Cref{sec:Observables}.
Hence, the tuple of subscripts $(t_0,M,i,j,e)$ indicates that the conditional expectation is formed with information available at time $t_0$ for stock $i$ given the market side $M$, order type $j$,
and event type $e$.
As we are not interested in $\gamma_{0,i}$, $\rho_{0,M,i,j,e}$, $\kappa_{M,i,j,e,v}$, $\xi_{M,i,j,e,v}$, and $\delta_{i,v}$, we can collect all possible products in \Cref{eq:LinApproximation1} in the
parameters $\gamma_{0,i},\gamma_{1,i}, \ldots \gamma_{1,R}$.
In total, this yields 1{.}971 parameters.
Hence, the estimation of the parameters is sensibly feasible using ordinary least-squares up to non-overlapping intervals with a length of 15 minutes.
For non-overlapping 15 minute intervals, we can get 2{.}164 observations from the 64 trading days in our sample.
Increasing the interval length beyond 15 minutes makes the use of overlapping intervals and rolling variable calculation necessary.
While this is in principle feasible, we restrict the analysis for the specification in \Cref{eq:LinApproximation1} to non-overlapping intervals and sampling frequencies below 15 minutes.

To increase the length of the intervals, we use three alternative specifications which entail less parameters.
In the first alternative, the moments of the spread are only included additively:
\begin{align}
 {\E}_{t_0}[O_{i,t}] &= \gamma_{0,i} +  \delta_{i,0}+\delta_{i,1} {\E}_{t_0}[\Delta_t] \,\,+ \sum_{v=2}^{4}\delta_{i,v} {\E}_{t_0}[(\Delta_t-\mu_{\Delta,t})^v] +\notag \\
 &\quad \sum_{M,j,e} \rho_{0,M,i,j,e}  {\E}_{t_0}[\bar{r}_{0,M,i,j,e,t}] \,\, \times \notag\\
 &\hspace{1cm} \bigg(\kappa_{M,i,j,e,0}+ \kappa_{M,i,j,e,1} {\E}_{t_0,M,i,j,e,t}[d_{l,t}] \,\,+ \notag\\
 &\hspace{2.5cm}  \sum_{v=2}^{4} \kappa_{M,i,j,e,v} {\E}_{t_0,M,i,j,e,t}[(d_{l,t}-\mu_{d_l,t})^v]\bigg)\,\,\times \notag\\
 &\hspace{1cm}\bigg(\xi_{M,i,j,e,0}+ \xi_{M,i,j,e,1} {\E}_{t_0,M,i,j,e,t}[q_{t,M,j,e}]\,\,+ \notag\\
  &\hspace{2.5cm}\sum_{v=2}^{4}\xi_{M,i,j,e,v} {\E}_{t_0,M,i,j,e,t}[(q_{t,M,j,e}-\mu_{q,t})^v]\bigg) + \varepsilon_i .
 \label{eq:LinApproximation2}
\end{align}
This specification entails the estimation of 399 parameters.
This enables us to estimate the model on interval lengths of up to one hour.

In the second alternative, the moments of the spread and the moments of the order size are included additively:
\begin{align}
 {\E}_{t_0}[O_{i,t}] &= \gamma_{0,i} +  \delta_{i,0}+\delta_{i,1} {\E}_{t_0}[\Delta_t] \,\, + \sum_{v=2}^{4}\delta_{i,v} {\E}_{t_0}[(\Delta_t-\mu_{\Delta,t})^v] +\notag \\
 &\sum_{M,j,e} \rho_{0,M,i,j,e}  {\E}_{t_0}[\bar{r}_{0,M,i,j,e,t}]  \,\, \times \notag\\
 &\hspace{1cm}\bigg(\kappa_{M,i,j,e,0}+ \kappa_{M,i,j,e,1} {\E}_{t_0,M,i,j,e,t}[d_{l,t}] \,\,+  \notag\\
 &\hspace{2.5cm} \sum_{v=2}^{4} \kappa_{M,i,j,e,v} {\E}_{t_0,M,i,j,e,t}[(d_{l,t}-\mu_{d_l,t})^v]\bigg) \,\, + \notag\\
  &\quad\xi_{M,i,j,e,0}+ \xi_{M,i,j,e,1} {\E}_{t_0,M,i,j,e,t}[q_{t,M,j,e}] \,\, + \notag\\
  &\hspace{2.5cm}\sum_{v=2}^{4}\xi_{M,i,j,e,v} {\E}_{t_0,M,i,j,e,t}[(q_{t,M,j,e}-\mu_{q,t})^v] + \varepsilon_i \ .
 \label{eq:LinApproximation3}
\end{align}
This reduces the number of parameters to 95 and makes interval lengths of up to 4 hours possible.

In the third alternative, we employ a completely additive structure:
\begin{align}
 {\E}_{t_0}[O_{i,t}] &= \gamma_{0,i} +  \delta_{i,0}+\delta_{i,1} {\E}_{t_0}[\Delta_t] \,\, + \sum_{v=2}^{4}\delta_{i,v} {\E}_{t_0}[(\Delta_t-\mu_{\Delta,t})^v] +\notag \\
 &\sum_{M,j,e} \rho_{0,M,i,j,e}  {\E}_{t_0}[\bar{r}_{0,M,i,j,e,t}]  \,\, + \notag\\
 &\quad\kappa_{M,i,j,e,0}+ \kappa_{M,i,j,e,1} {\E}_{t_0,M,i,j,e,t}[d_{l,t}] \,\,+  \notag\\
 &\hspace{2.5cm} \sum_{v=2}^{4} \kappa_{M,i,j,e,v} {\E}_{t_0,M,i,j,e,t}[(d_{l,t}-\mu_{d_l,t})^v] \,\, + \notag\\
  &\quad\xi_{M,i,j,e,0}+ \xi_{M,i,j,e,1} {\E}_{t_0,M,i,j,e,t}[q_{t,M,j,e}] \,\, + \notag\\
  &\hspace{2.5cm}\sum_{v=2}^{4}\xi_{M,i,j,e,v} {\E}_{t_0,M,i,j,e,t}[(q_{t,M,j,e}-\mu_{q,t})^v] + \varepsilon_i \ .
 \label{eq:LinApproximation4}
\end{align}
This reduces the number of parameters which have to be estimated further to 43.

For each of the four specifications in Equations~\eqref{eq:LinApproximation1}~to~\eqref{eq:LinApproximation4}, we use a formulation in which the moments on both sides of the equations are estimated contemporaneously, i.e., at the same time $t$.
Naturally, this is only possible in-sample.
Additionally, we also investigate specifications of Equations~\eqref{eq:LinApproximation1}~to~\eqref{eq:LinApproximation4} in which the moments on the right hand side are estimated at $t-1$ to
describe the expectation of the observable on the left hand side.
This formulation allows for out-of-sample evaluation of the model in terms of predictive power assuming that the moments in one interval anchor the moments of the subsequent interval.

We evaluate our model across several measures in- and out-of-sample.
\subsection{In-sample Analysis}
For the in-sample evaluation, we consider the adjusted and unadjusted $R^2$ as well as the root mean squared error ($RMSE$).
Furthermore, following  \cite{ZhouPHTZ18}, we also use the direction prediction accuracy ($DPA$) defined as
\begin{align}
 DPA = \frac{100}{T} \sum_{t=1}^T \frac{\max\left(0,\Delta p_{\cdot, t} \cdot \Delta \hat{p}_{\cdot,t} \right)}{\Delta p_{\cdot,t} \cdot \Delta \hat{p}_{\cdot,t}}  .
 \label{eq:DPA}
\end{align}

\begin{figure}
\caption{DPA}
\label{fig:DPA_inS}
\centering
\begin{minipage}{0.9\linewidth}
  The figures  report the in-sample direction prediction accuracy (DPA) (as defined in \Cref{eq:DPA}).
  The in-sample DPA is reported for the estimated model equations specified in \Cref{eq:LinApproximation1} (blue), \Cref{eq:LinApproximation2} (red),  \Cref{eq:LinApproximation3} (green)  and \Cref{eq:LinApproximation4} (orange) for the
sampling frequencies 1, 2, 5, 10, 15, 20, 30, 45, 60, 120 and 240.
\end{minipage}
 \begin{subfigure}{0.49\linewidth}
  \includegraphics[width=\linewidth]{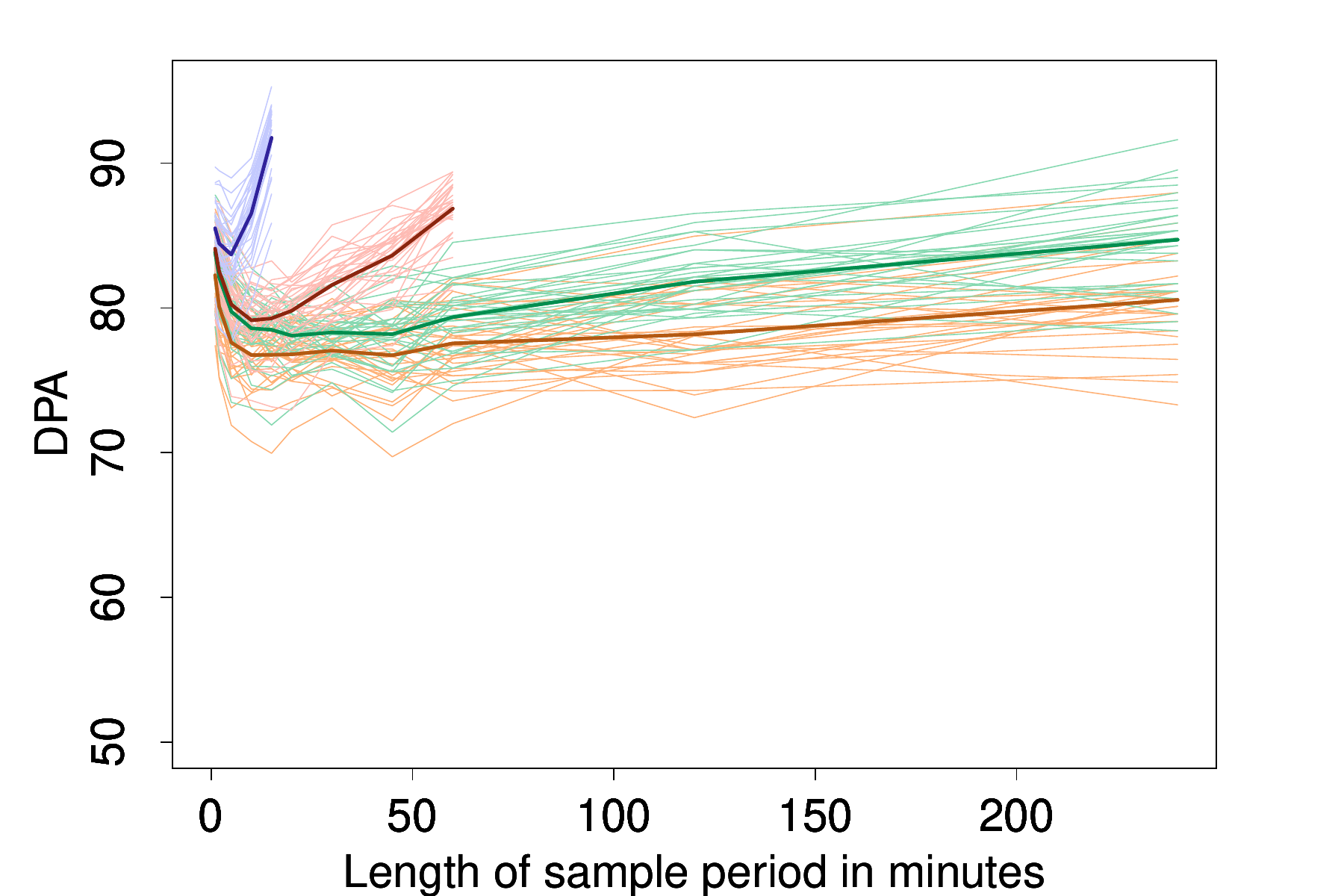}
  \subcaption{Contemporaneous Buy}
 \end{subfigure}
 \begin{subfigure}{0.49\linewidth}
  \includegraphics[width=\linewidth]{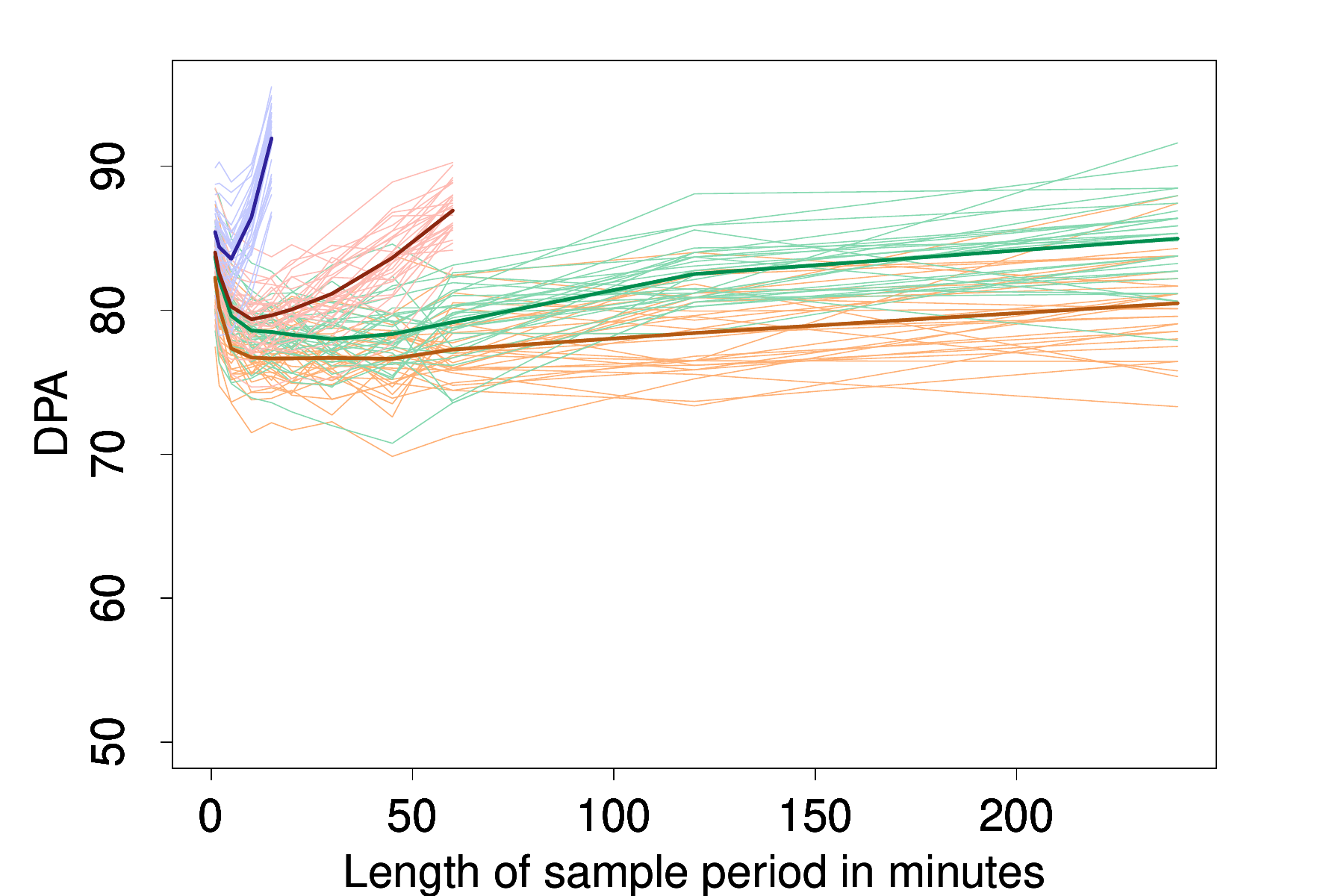}
   \subcaption{Contemporaneous Sell}
 \end{subfigure}
  \begin{subfigure}{0.49\linewidth}
  \includegraphics[width=\linewidth]{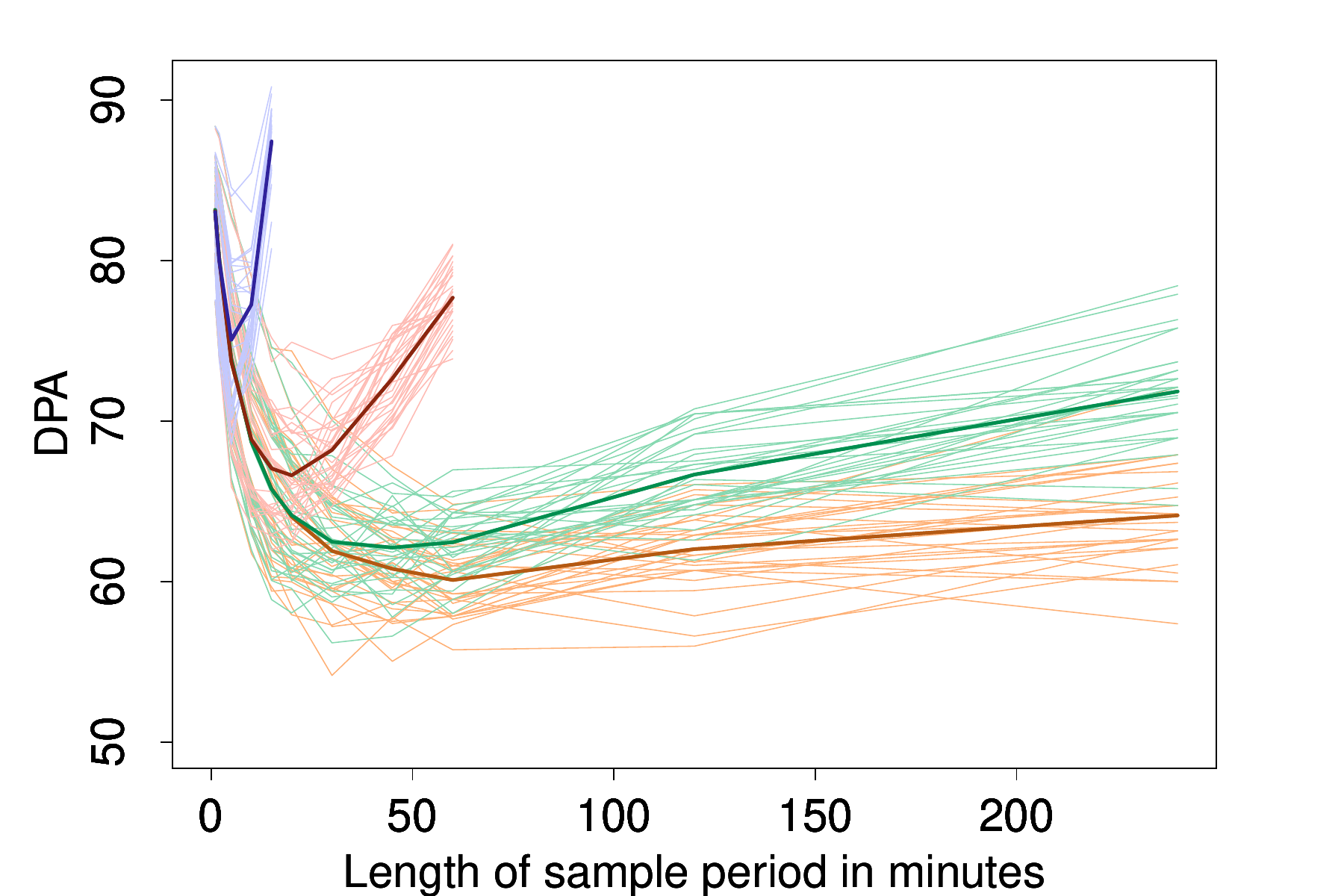}
    \subcaption{Lagged Buy}
 \end{subfigure}
 \begin{subfigure}{0.49\linewidth}
  \includegraphics[width=\linewidth]{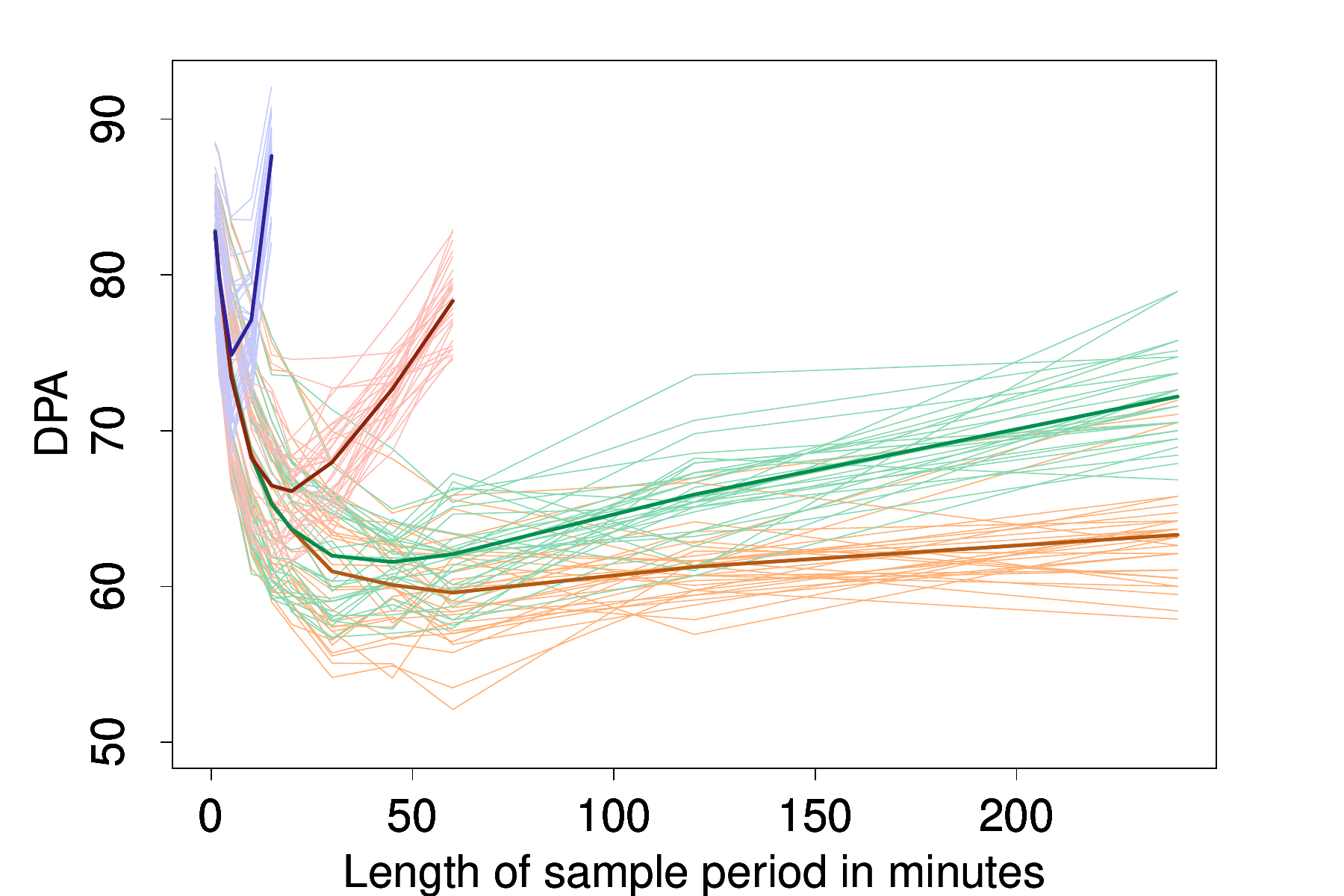}
  \subcaption{Lagged Sell}
 \end{subfigure}
   \begin{subfigure}{0.49\linewidth}
  \includegraphics[width=\linewidth]{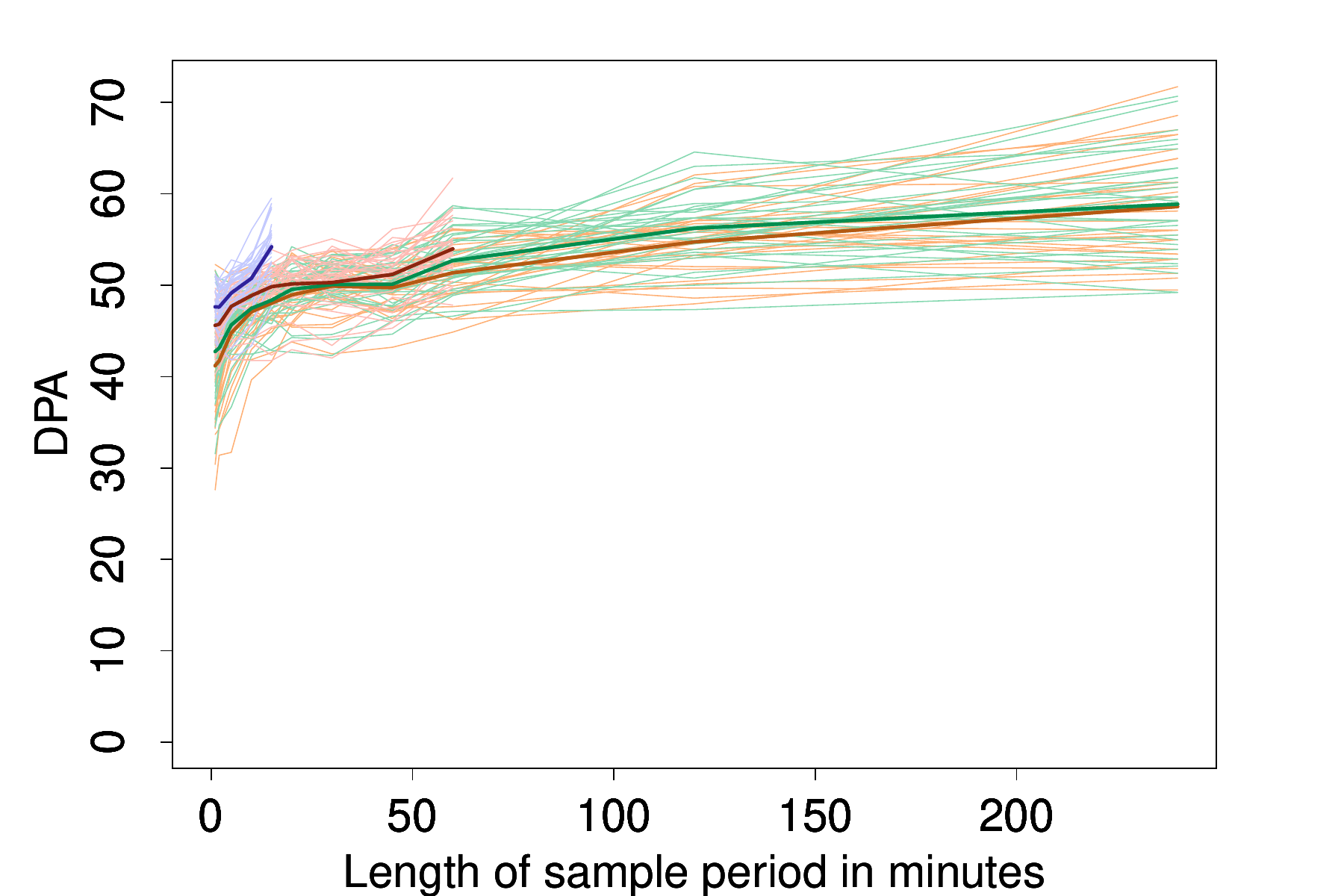}
    \subcaption{Contemporaneous $XLM$}
 \end{subfigure}
 \begin{subfigure}{0.49\linewidth}
  \includegraphics[width=\linewidth]{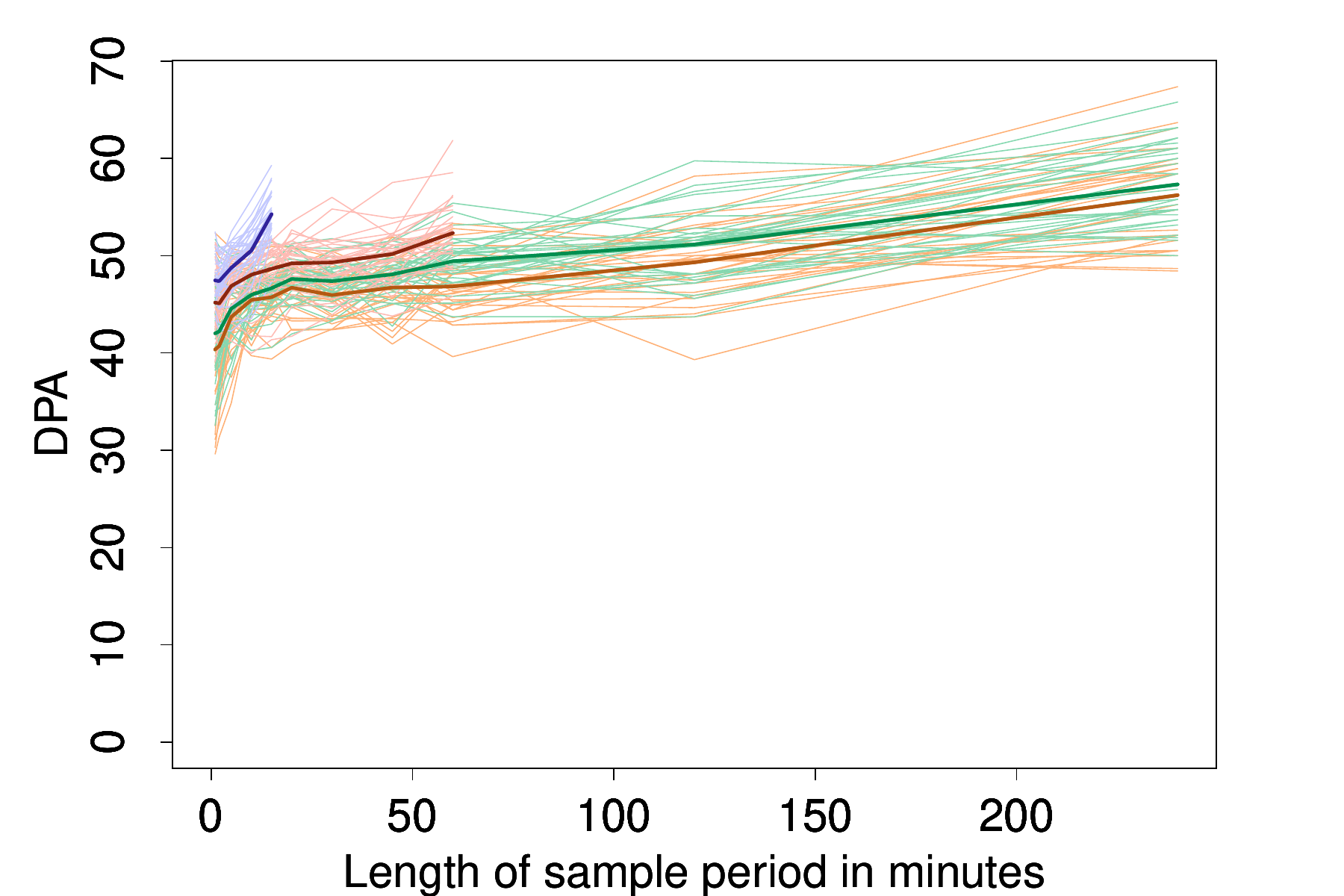}
  \subcaption{Lagged $XLM$}
 \end{subfigure}
\end{figure}

\begin{figure}
\caption{Adjusted $R^2$}
\label{fig:R2adj_inS}
\centering
\begin{minipage}{0.9\linewidth}
  The figures below report the in-sample, adjusted $R^2$ for the estimated model equations specified in \Cref{eq:LinApproximation1} (blue), \Cref{eq:LinApproximation2} (red),  \Cref{eq:LinApproximation3} (green)  and \Cref{eq:LinApproximation4} (orange) for the sampling frequencies 1, 2, 5, 10, 15, 20, 30, 45, 60, 120 and 240.
\end{minipage}

 \begin{subfigure}{0.49\linewidth}
  \includegraphics[width=\linewidth]{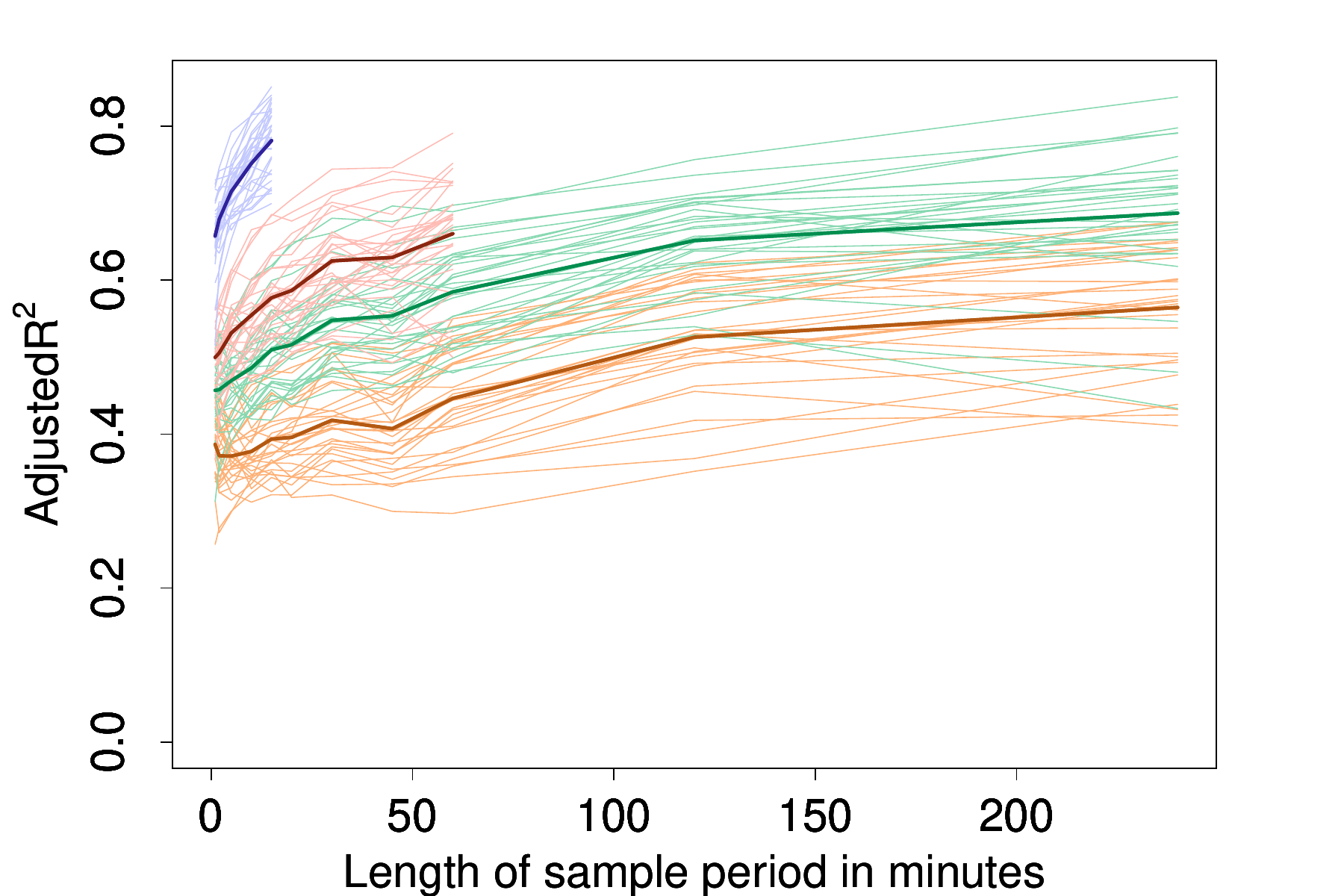}
  \subcaption{Contemporaneous Buy}
 \end{subfigure}
 \begin{subfigure}{0.49\linewidth}
  \includegraphics[width=\linewidth]{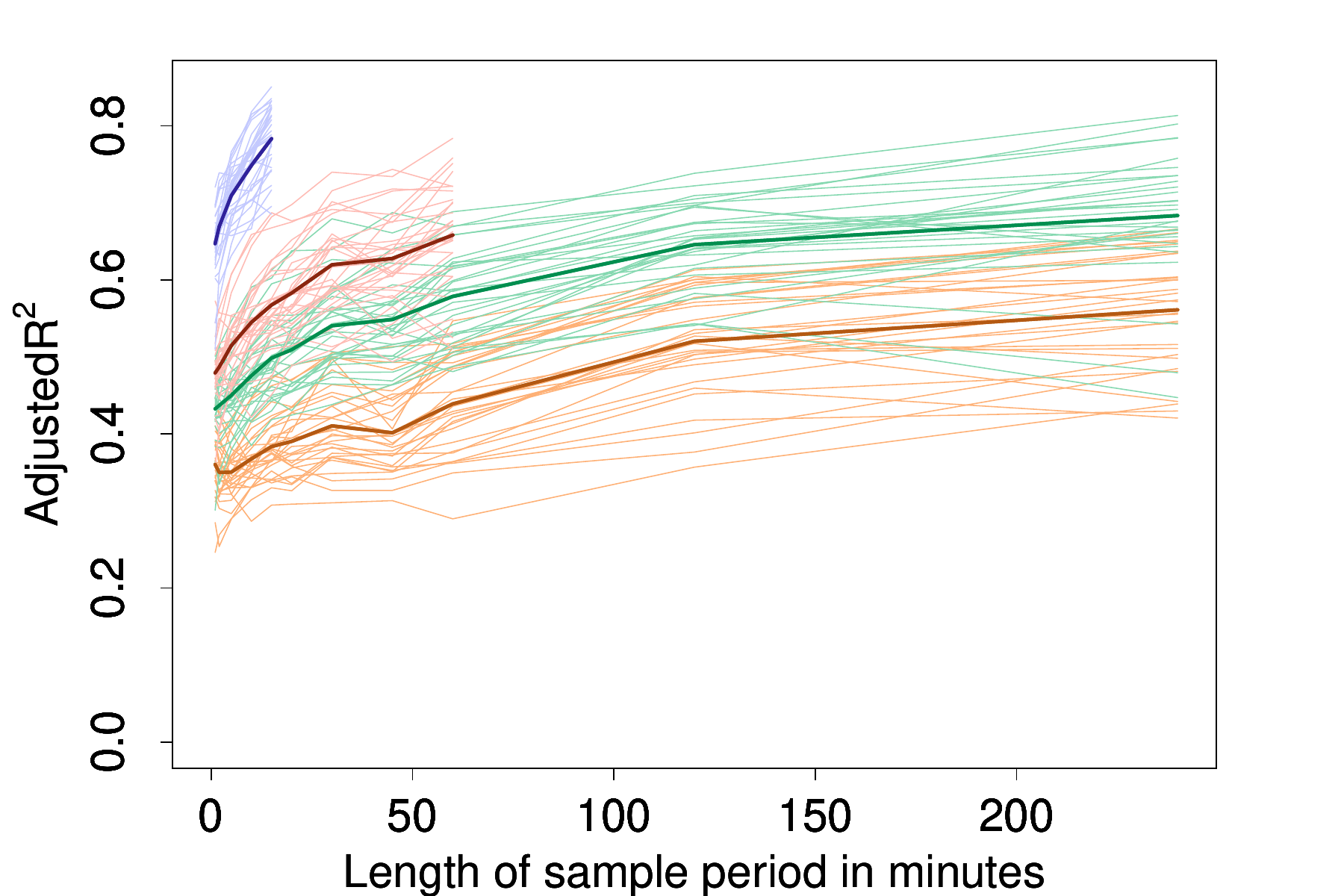}
   \subcaption{Contemporaneous Sell}
 \end{subfigure}
  \begin{subfigure}{0.49\linewidth}
  \includegraphics[width=\linewidth]{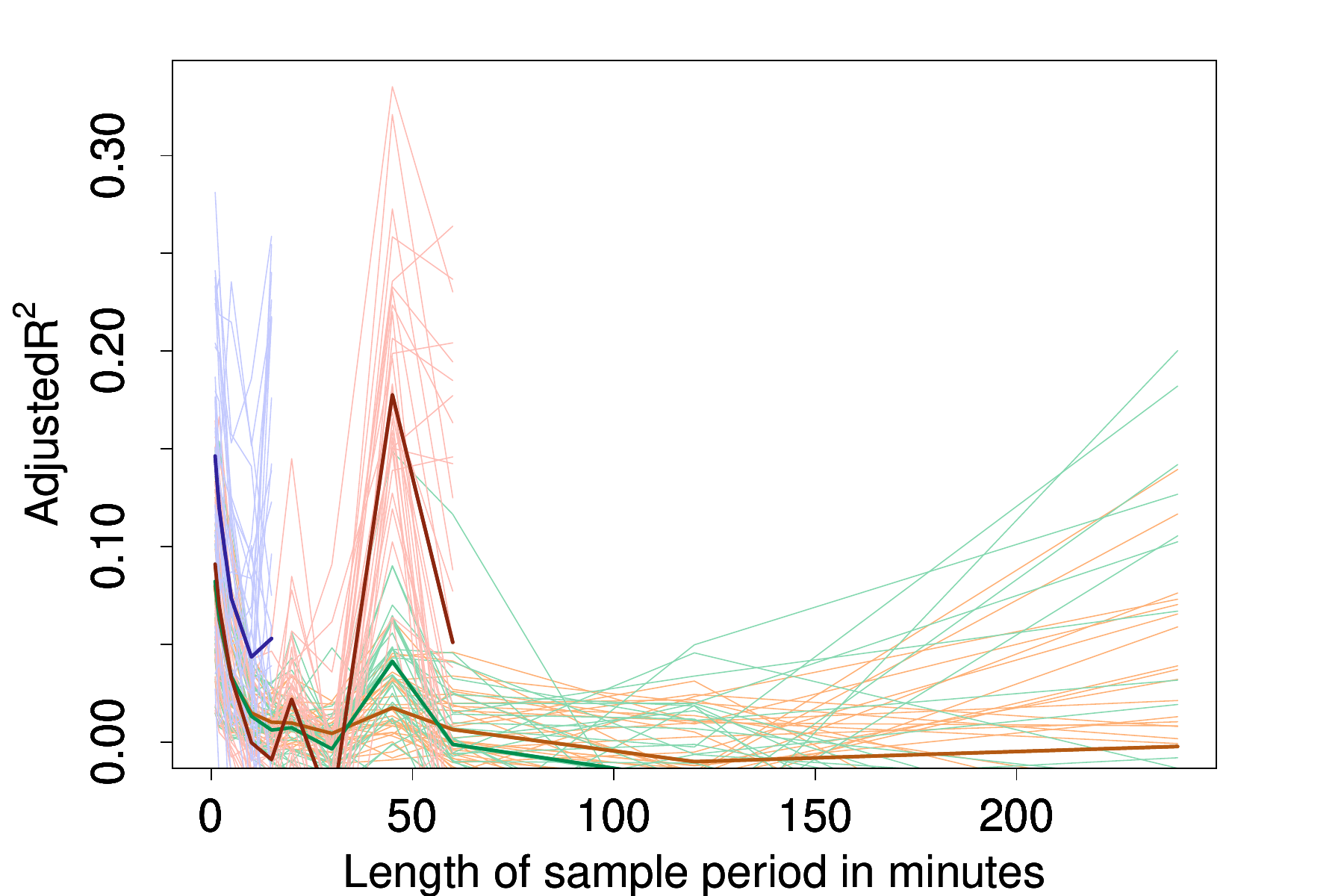}
   \subcaption{Lagged Buy}
 \end{subfigure}
 \begin{subfigure}{0.49\linewidth}
  \includegraphics[width=\linewidth]{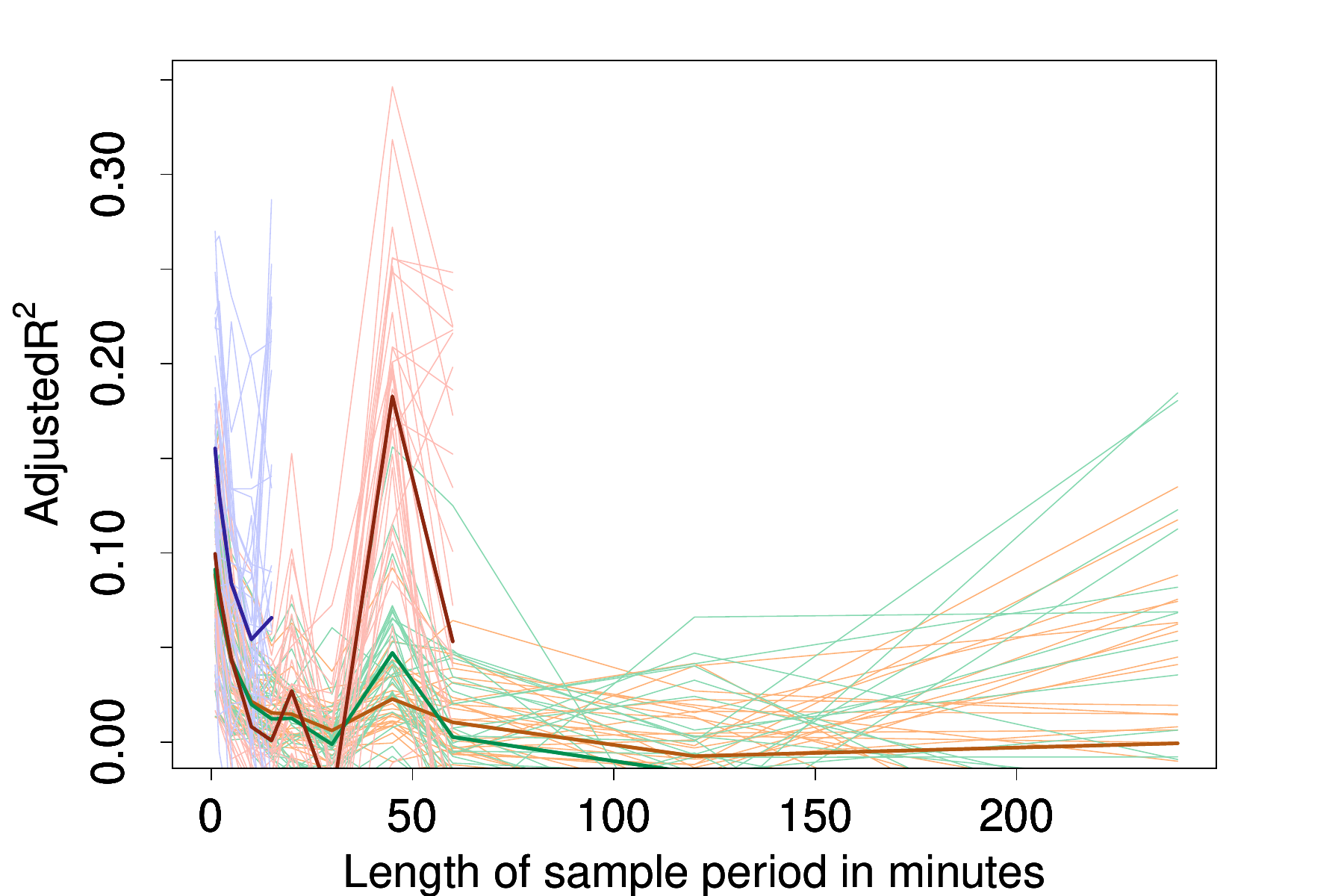}
  \subcaption{Lagged Sell}
 \end{subfigure}
    \begin{subfigure}{0.49\linewidth}
  \includegraphics[width=\linewidth]{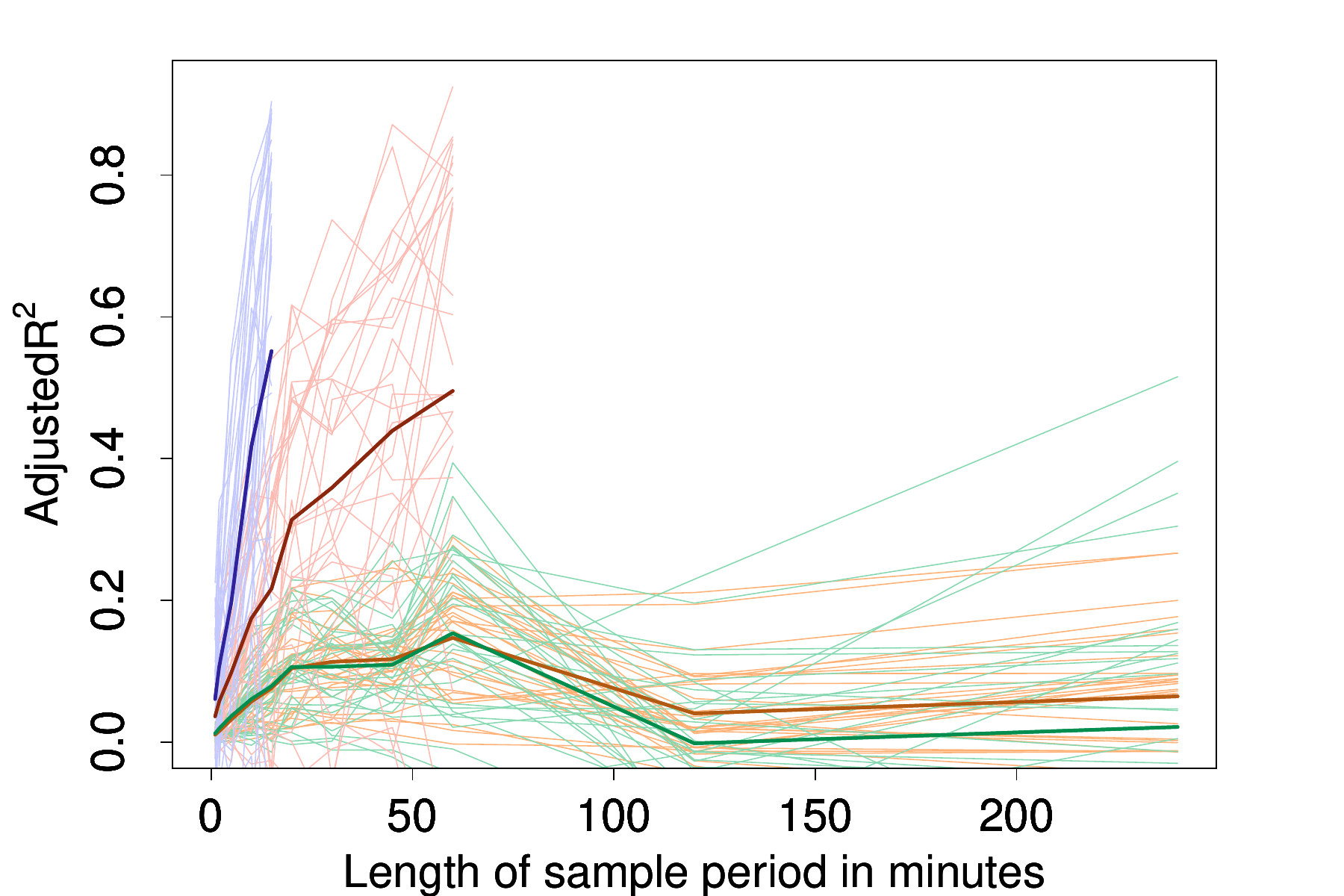}
    \subcaption{Contemporaneous $XLM$}
 \end{subfigure}
 \begin{subfigure}{0.49\linewidth}
  \includegraphics[width=\linewidth]{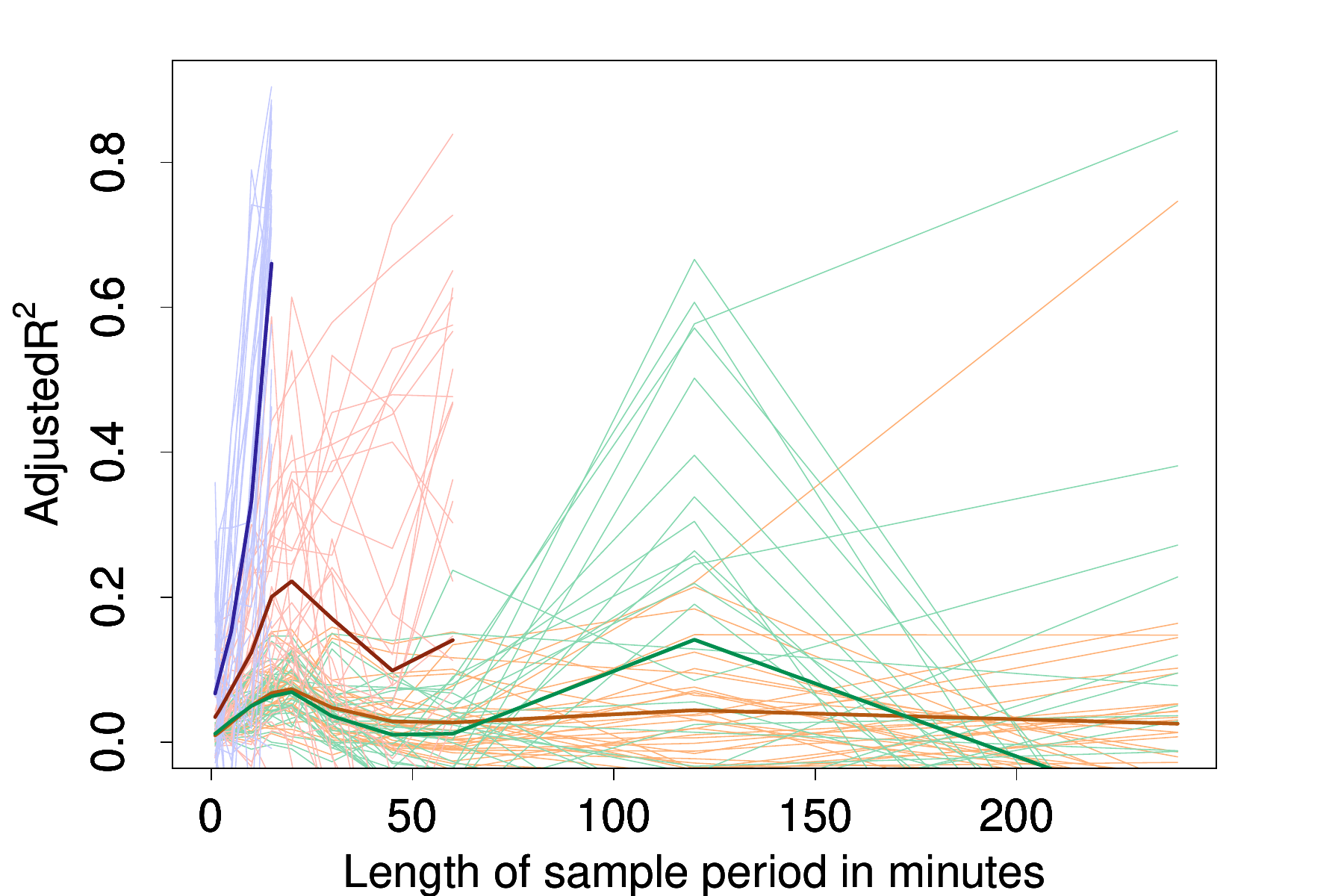}
  \subcaption{Lagged $XLM$}
  \end{subfigure}
\end{figure}

\begin{figure}
\caption{$R^2$}
\label{fig:R2_inS}
\centering
\begin{minipage}{0.9\linewidth}
 The figures  report the in-sample $R^2$ for the estimated model equations specified in \Cref{eq:LinApproximation1} (blue), \Cref{eq:LinApproximation2} (red),  \Cref{eq:LinApproximation3} (green)  and \Cref{eq:LinApproximation4} (orange) for the sampling frequencies 1, 2, 5, 10, 15, 20, 30, 45, 60, 120 and 240.
\end{minipage}
 \begin{subfigure}{0.49\linewidth}
  \includegraphics[width=\linewidth]{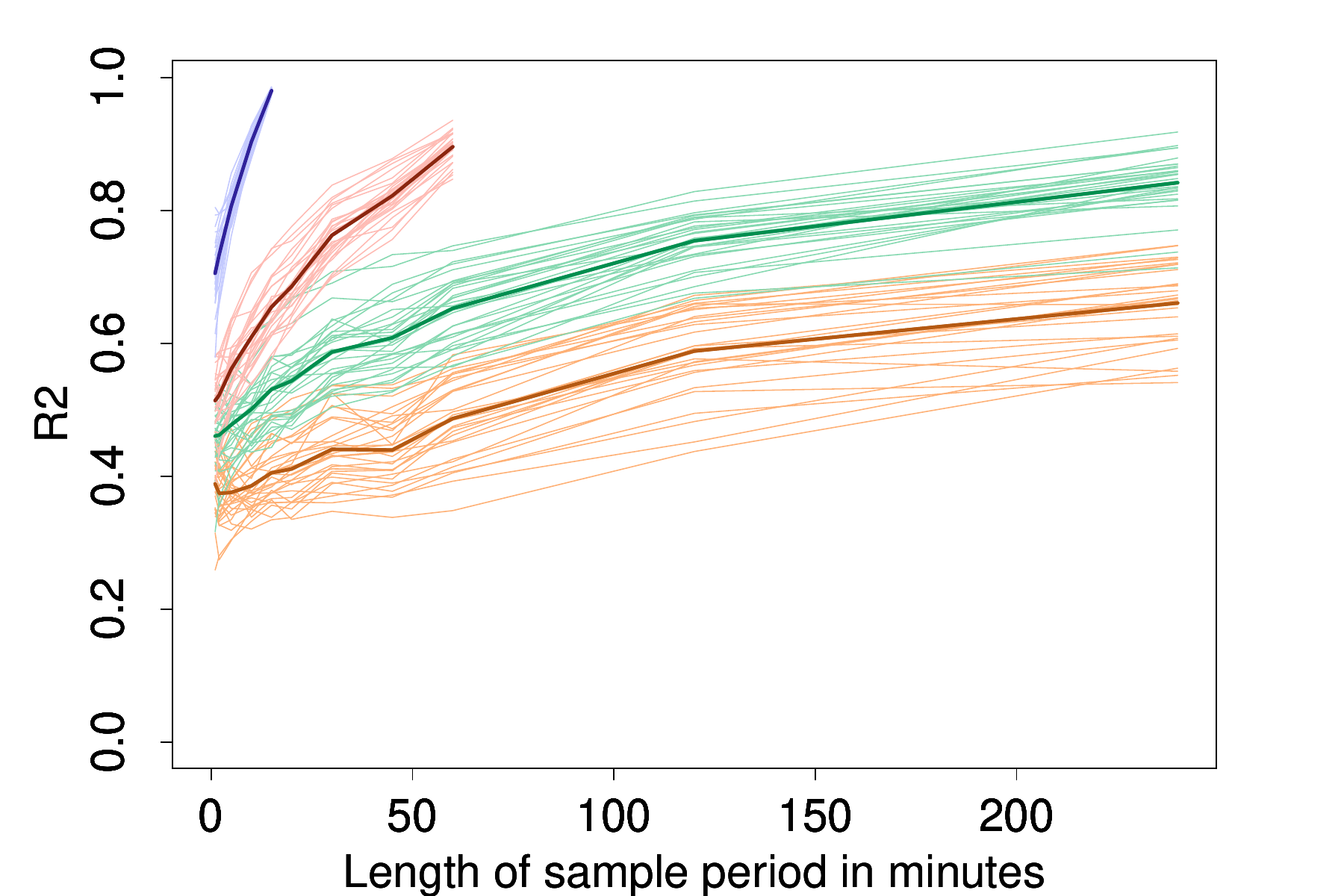}
  \subcaption{Contemporaneous Buy}
 \end{subfigure}
 \begin{subfigure}{0.49\linewidth}
  \includegraphics[width=\linewidth]{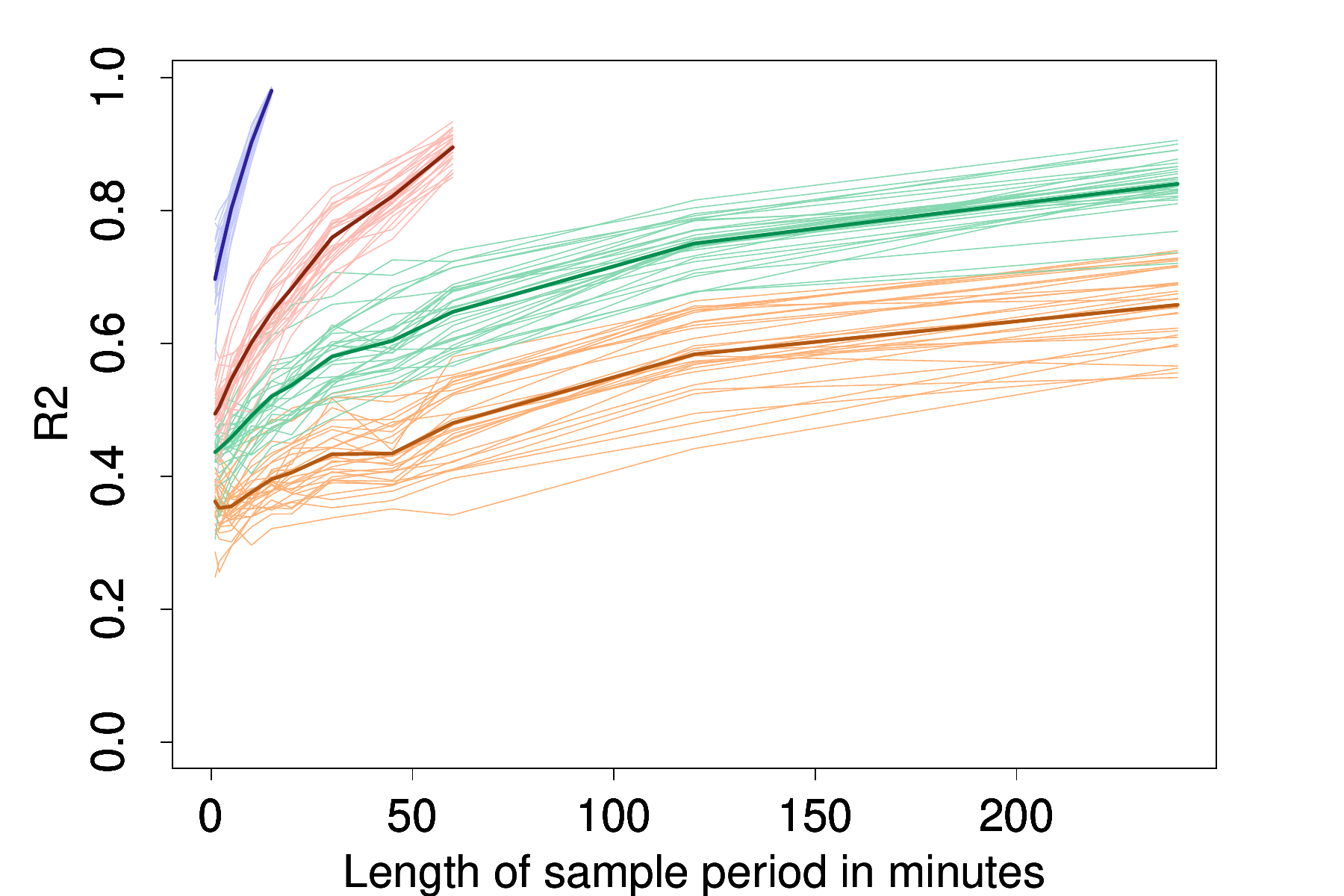}
   \subcaption{Contemporaneous Sell}
 \end{subfigure}
  \begin{subfigure}{0.49\linewidth}
  \includegraphics[width=\linewidth]{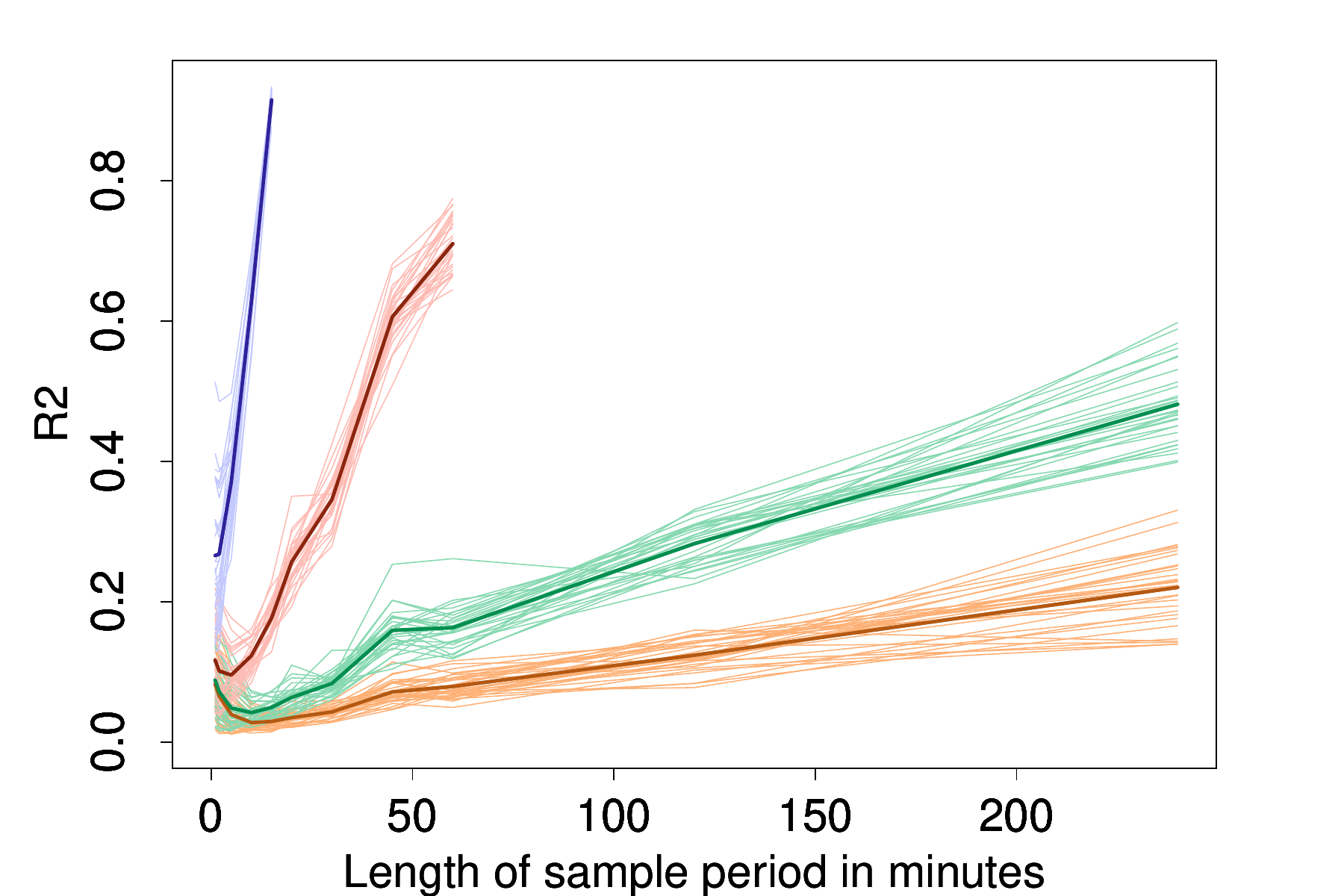}
   \subcaption{Lagged Buy}
 \end{subfigure}
 \begin{subfigure}{0.49\linewidth}
  \includegraphics[width=\linewidth]{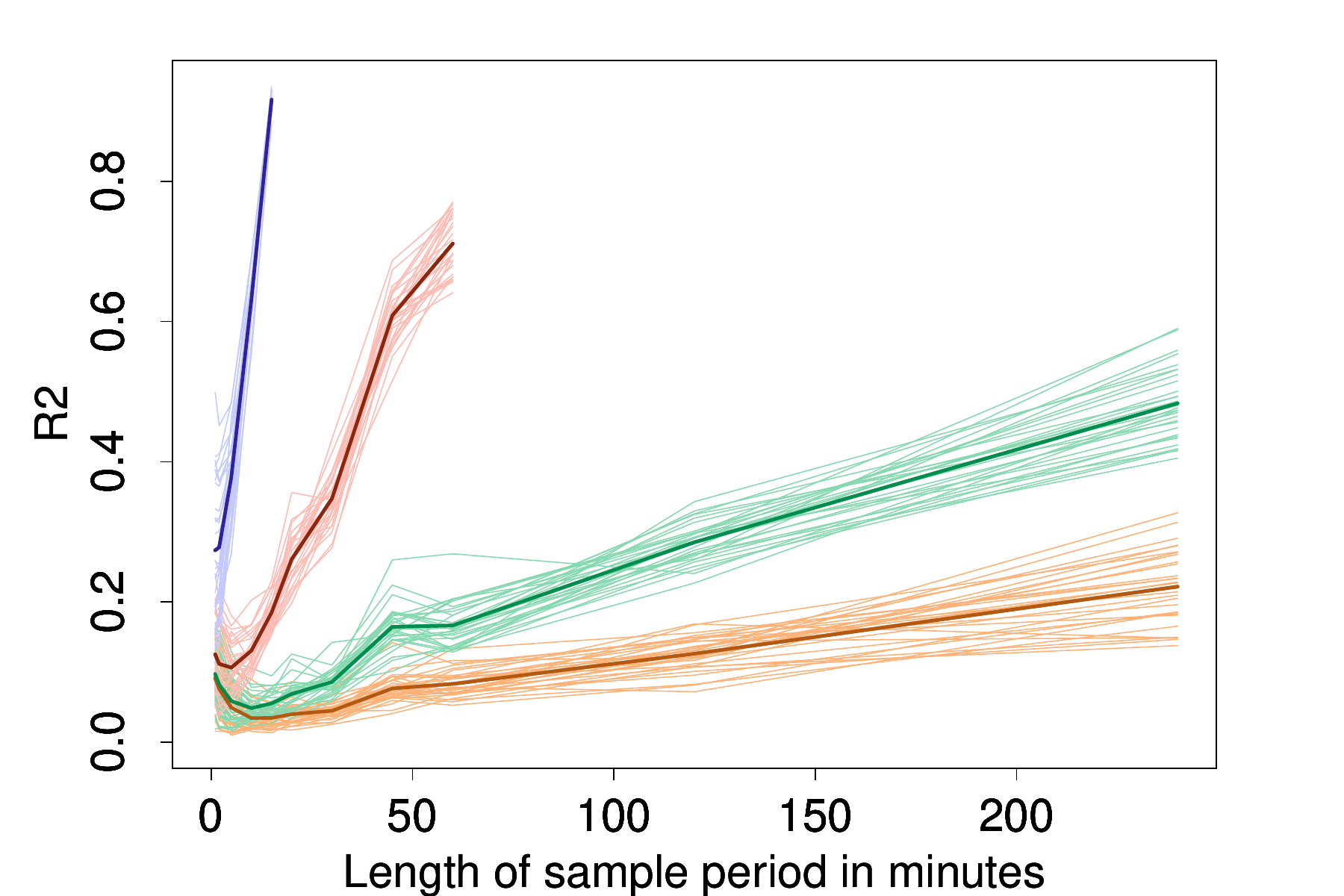}
  \subcaption{Lagged Sell}
 \end{subfigure}
     \begin{subfigure}{0.49\linewidth}
  \includegraphics[width=\linewidth]{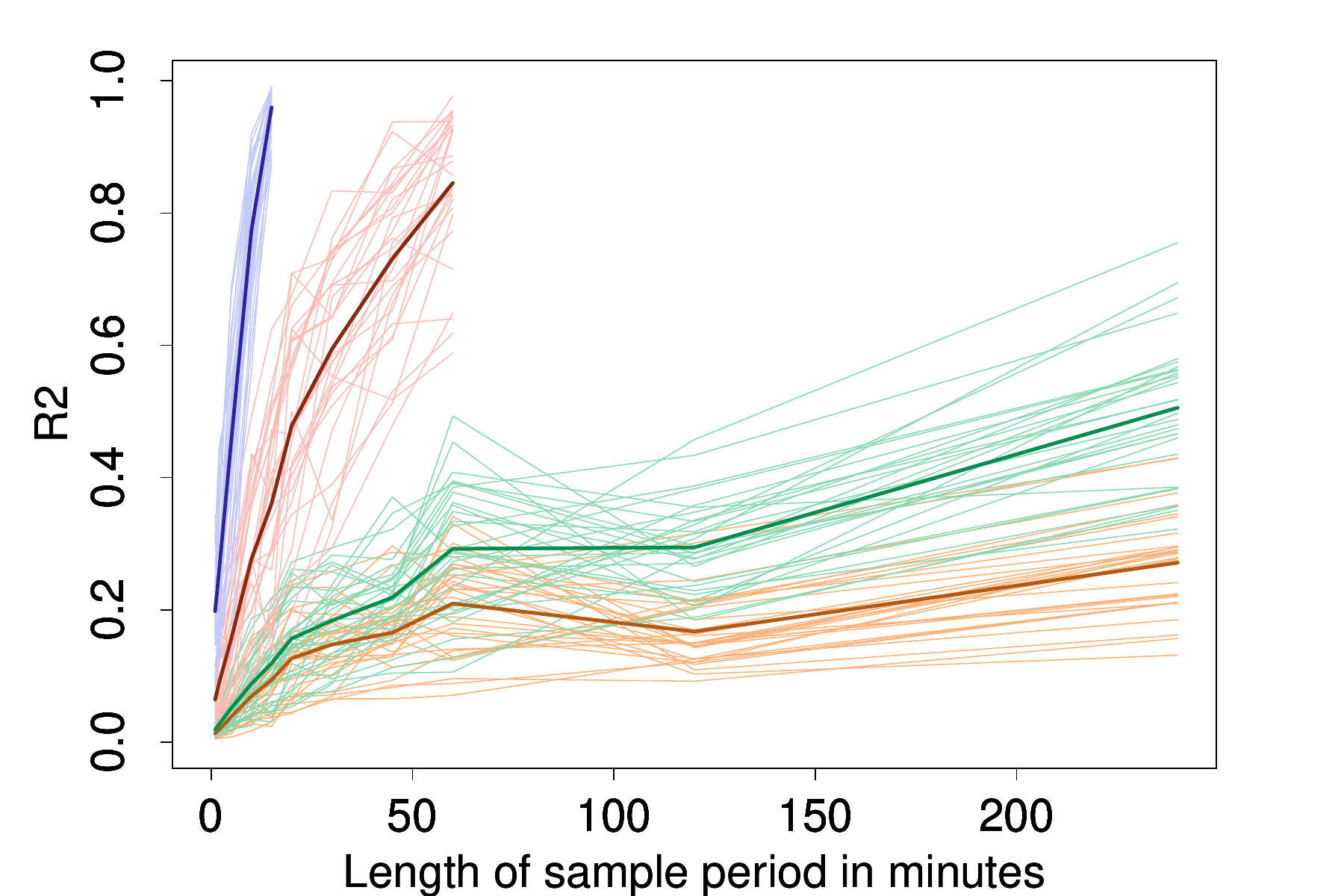}
    \subcaption{Contemporaneous $XLM$}
 \end{subfigure}
 \begin{subfigure}{0.49\linewidth}
  \includegraphics[width=\linewidth]{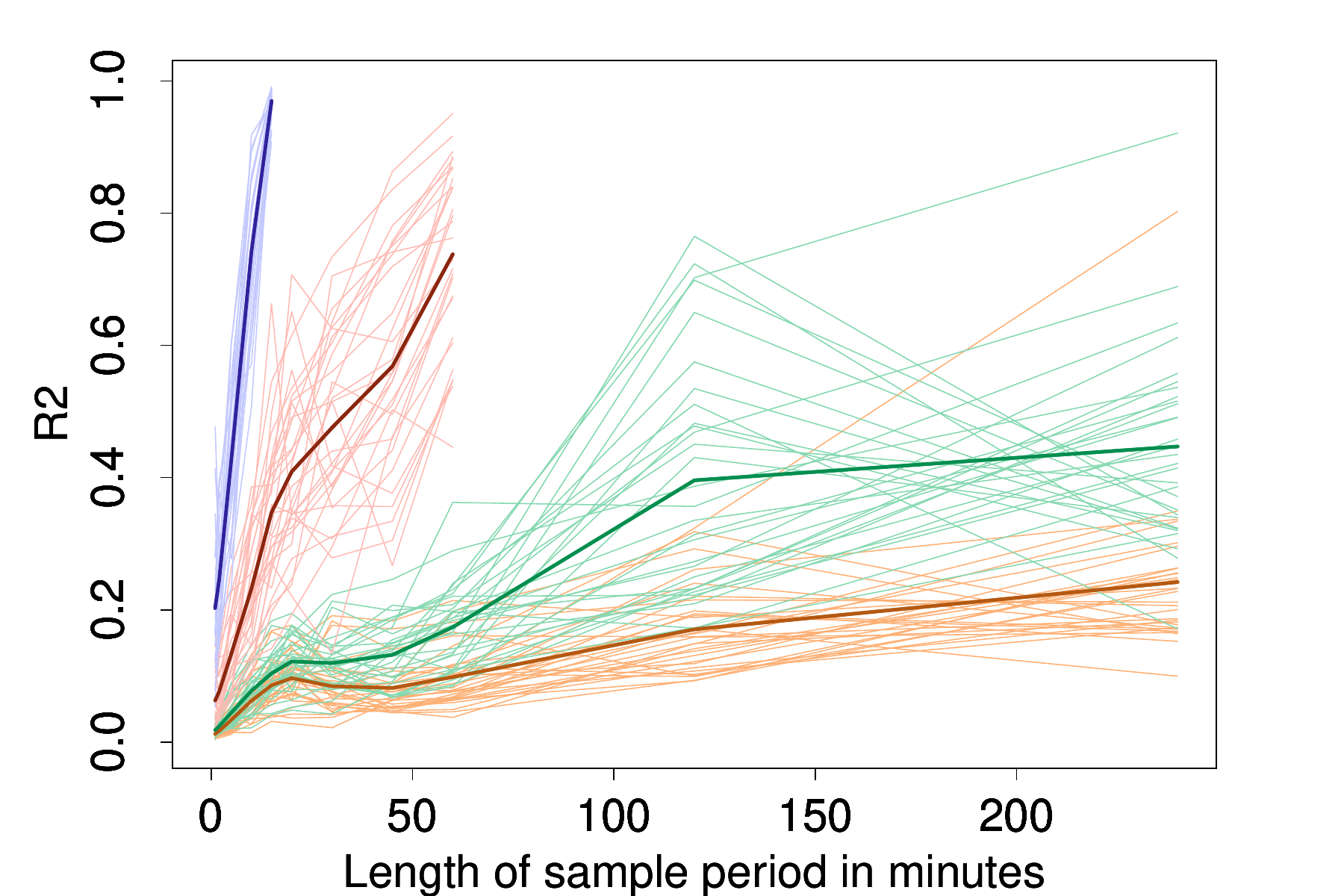}
  \subcaption{Lagged $XLM$}
  \end{subfigure}
\end{figure}

\begin{figure}
\caption{$RMSE$}
\label{fig:rmse_inS}
\centering
\begin{minipage}{0.9\linewidth}
 The figures  report the in-sample $RMSE$ for the estimated model equations specified in \Cref{eq:LinApproximation1} (blue), \Cref{eq:LinApproximation2} (red),  \Cref{eq:LinApproximation3} (green)  and \Cref{eq:LinApproximation4} (orange) for the sampling frequencies 1, 2, 5, 10, 15, 20, 30, 45, 60, 120 and 240.
\end{minipage}
 \begin{subfigure}{0.49\linewidth}
  \includegraphics[width=\linewidth]{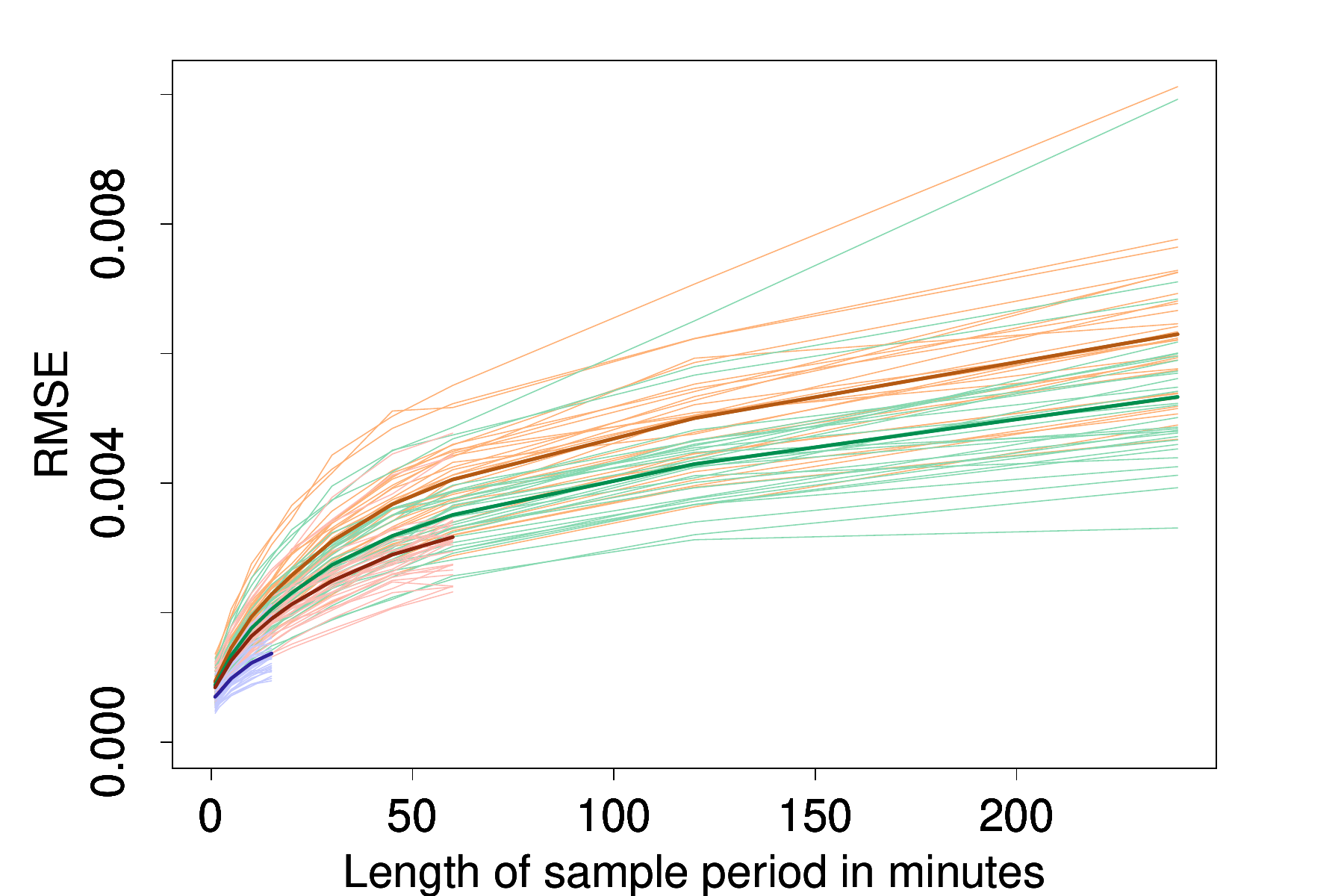}
  \subcaption{Contemporaneous Buy}
 \end{subfigure}
 \begin{subfigure}{0.49\linewidth}
  \includegraphics[width=\linewidth]{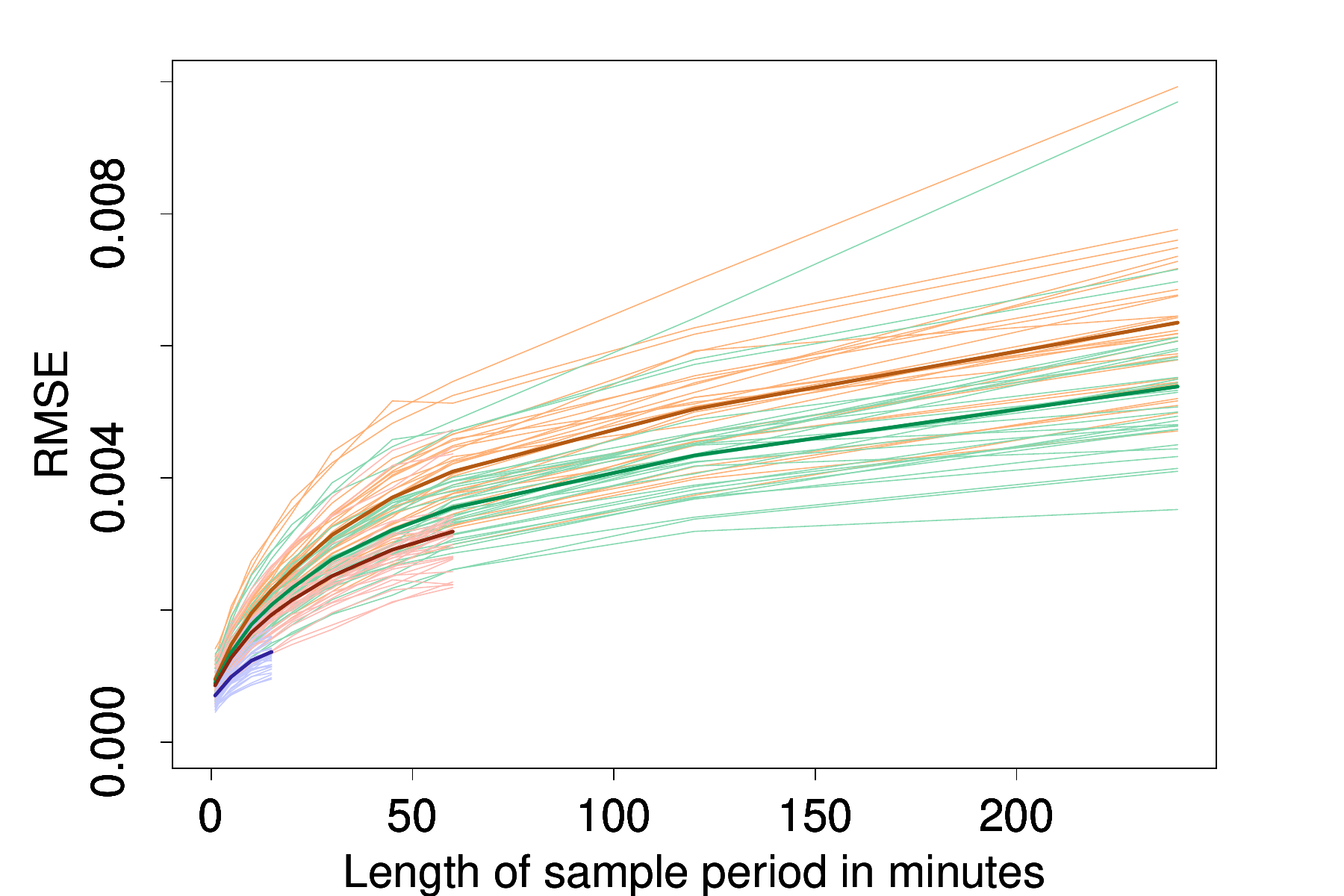}
   \subcaption{Contemporaneous Sell}
 \end{subfigure}
  \begin{subfigure}{0.49\linewidth}
  \includegraphics[width=\linewidth]{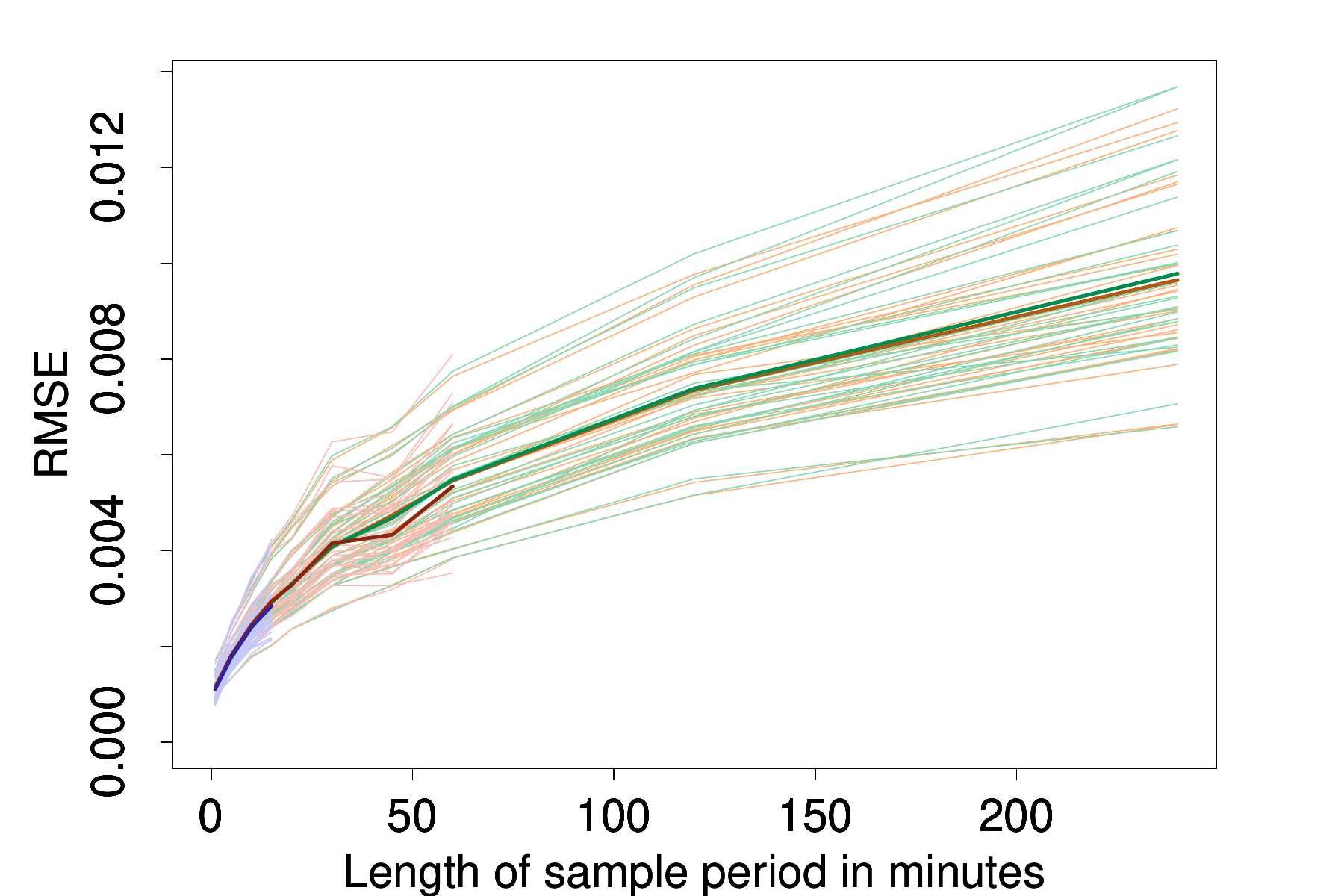}
   \subcaption{Lagged Buy}
 \end{subfigure}
 \begin{subfigure}{0.49\linewidth}
  \includegraphics[width=\linewidth]{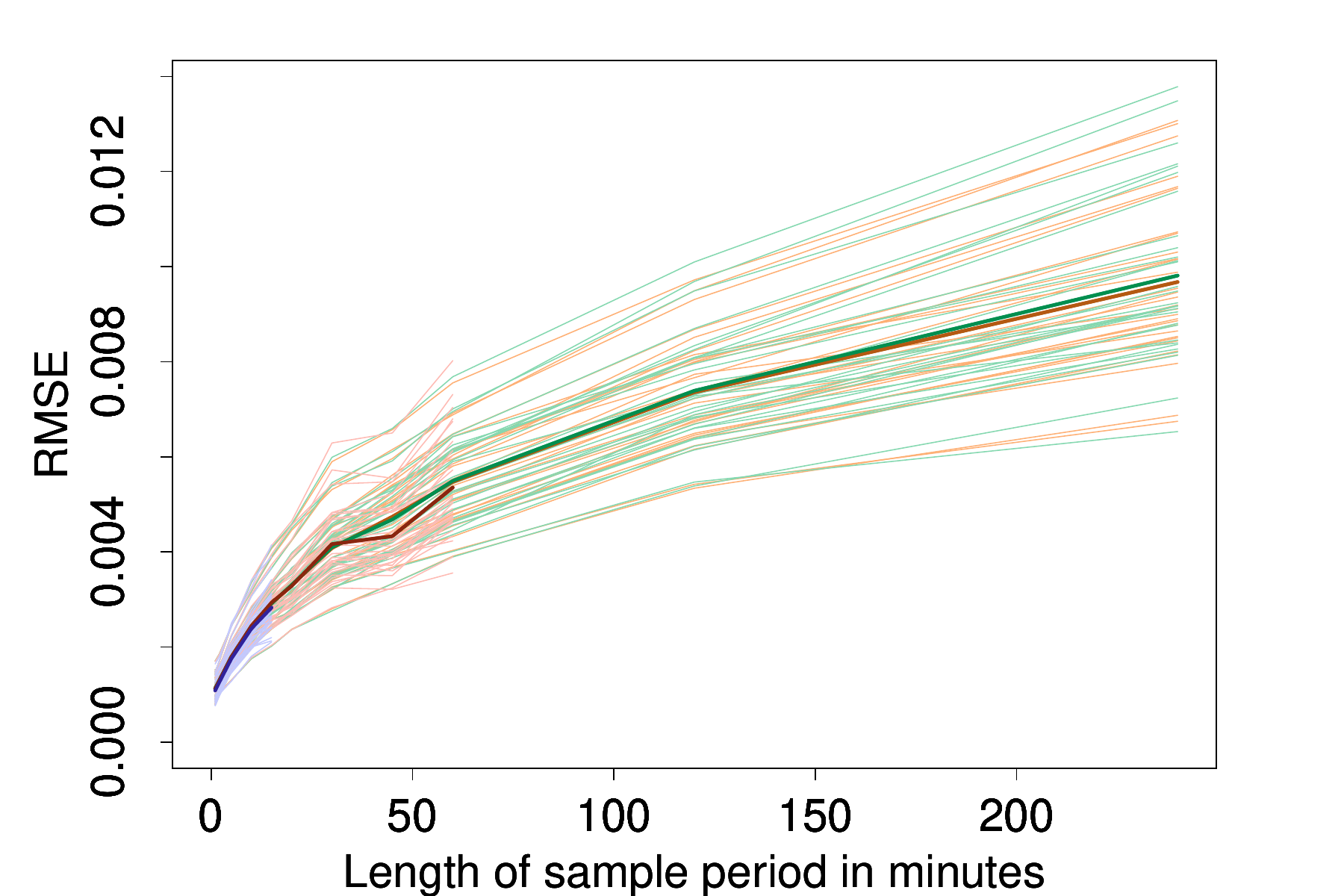}
  \subcaption{Lagged Sell}
 \end{subfigure}
     \begin{subfigure}{0.49\linewidth}
  \includegraphics[width=\linewidth]{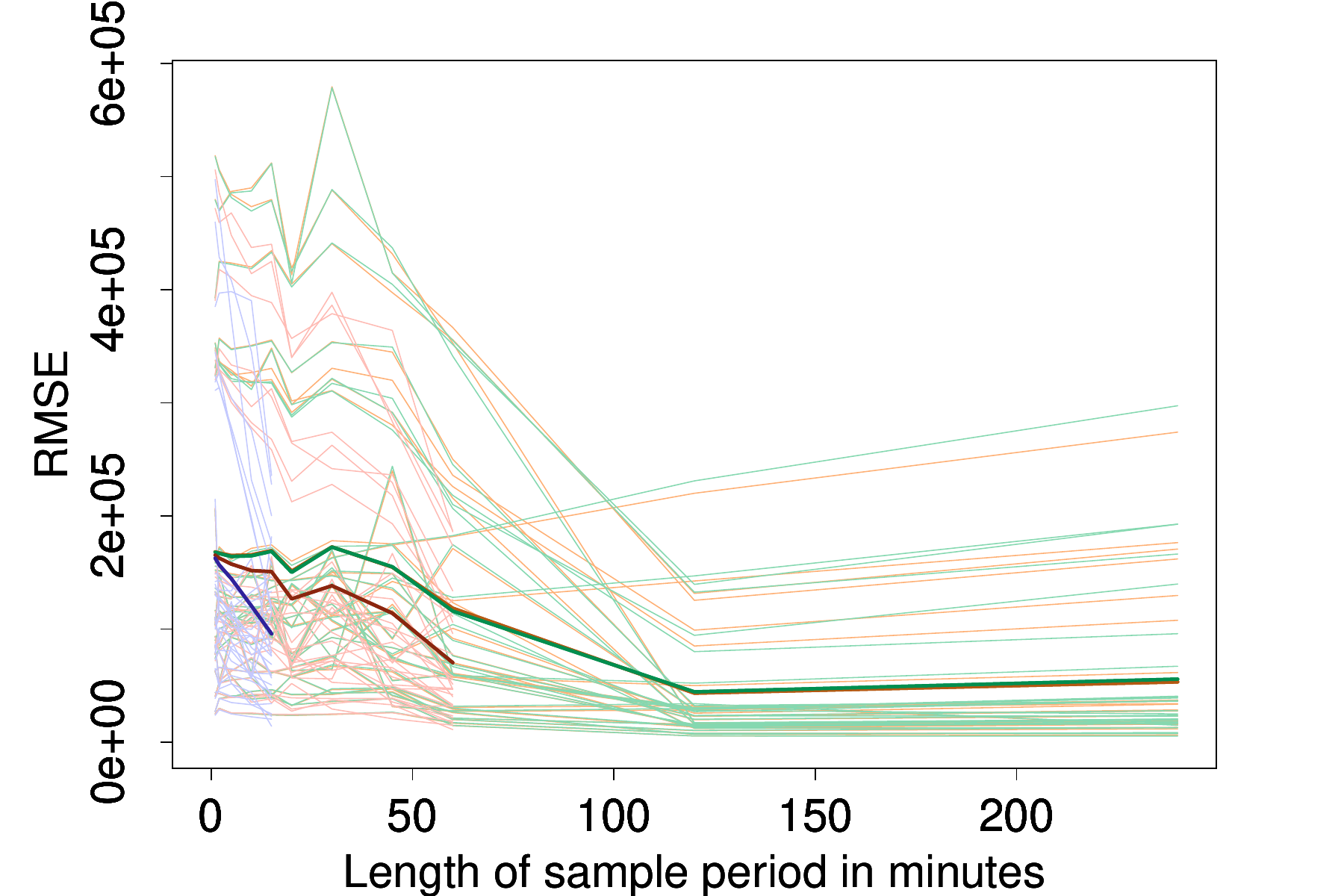}
    \subcaption{Contemporaneous $XLM$}
 \end{subfigure}
 \begin{subfigure}{0.49\linewidth}
  \includegraphics[width=\linewidth]{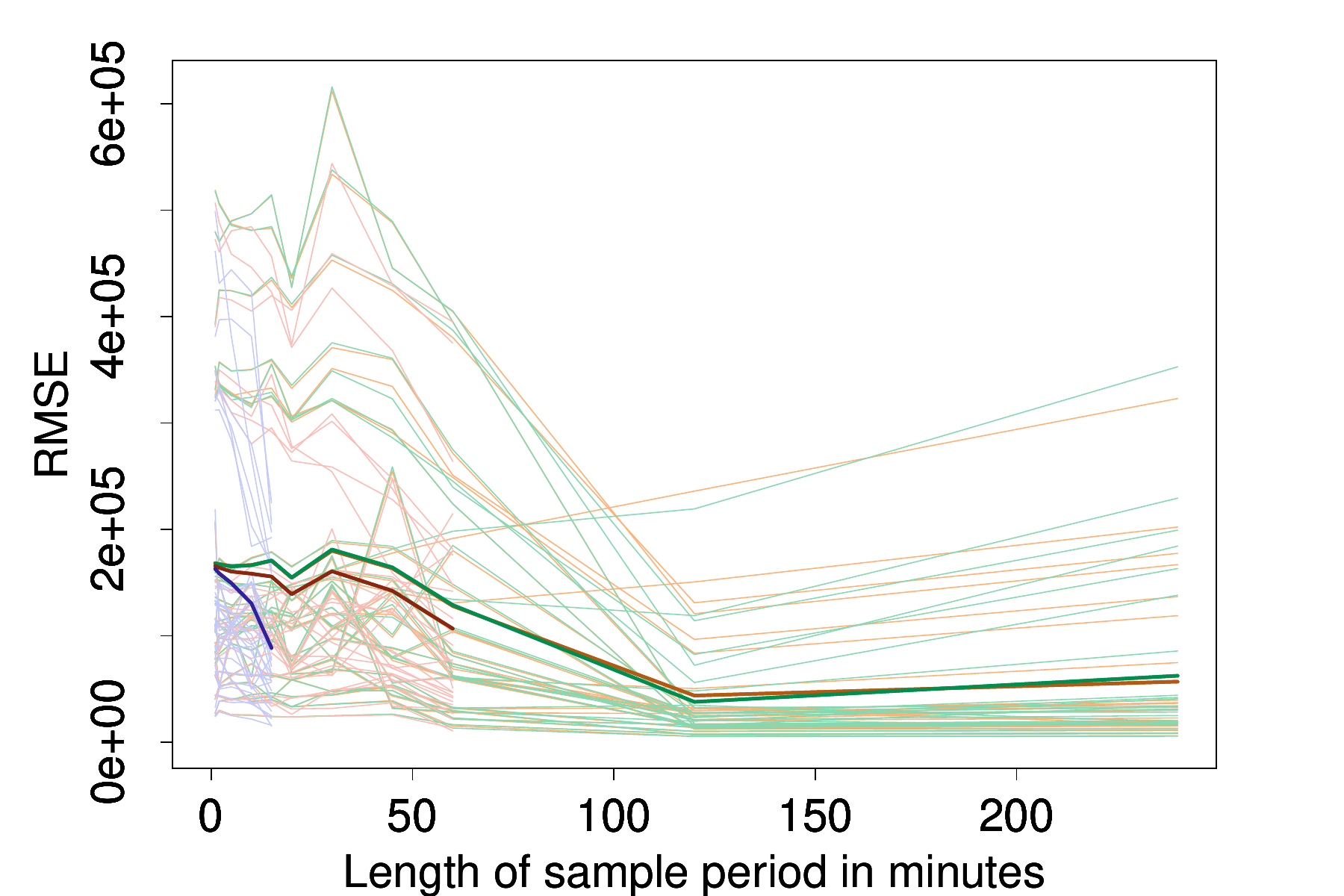}
  \subcaption{Lagged $XLM$}
  \end{subfigure}
\end{figure}

The in-sample results are depicted in Figures \ref{fig:DPA_inS} to~\ref{fig:rmse_inS}.
In each figure, the contemporaneous models are depicted in subfigures (a) and (b) for returns and (e) for the $XLM$ measure while the results for the specifications which use only past information to
model the current state are presented in subfigures (c) and (d) for returns and (f) for the $XLM$ measure.
Every line in the graph represents the results for one stock.
The highlighted thicker line is the average of the respective measure across all stocks.
There are a few noteworthy results.
First, the DPA as well as the $R^2$ measures (adjusted and unadjusted) allow us to reject the hypothesis that the contemporaneous as well as the lagged models have no significance in explaining the
data.
Our results suggest that the extensive model in \Cref{eq:LinApproximation1} overfits the data with growing sampling frequency and, hence, less observations.
This can be seen by the drastically increasing $R^2$ and $DPA$ values when the sampling frequency is increased.
The adjusted $R^2$ should account for this effect, and indeed remains rather stable.
However, when the degrees of freedom of the model become sparse, the adjusted $R^2$ is not able to correct the full extent of the overfitting.
Nonetheless, the high values at the highest frequencies indicate that the contemporaneous model describes high-frequency returns very well.
In all measures, the contemporaneous specification in \Cref{eq:LinApproximation1} (blue) turns out to be superior to the specifications in \Cref{eq:LinApproximation2} (red),
\Cref{eq:LinApproximation3} (green) and \Cref{eq:LinApproximation4} (orange) , i.e., it has a higher direction prediction accuracy, better fit in terms of higher $R^2$ values, and results in a lower root mean squared error.

Subfigures (c), (d), and (f) in Figures~\ref{fig:DPA_inS} to~\ref{fig:rmse_inS} present the results using the lagged specifications of
Equations~\eqref{eq:LinApproximation1}~to~\eqref{eq:LinApproximation4}, i.e., they compare their out-of-sample forecast performance.
Recall that these specifications rely heavily on the assumption that the arrival rates stay constant for some (very) short time horizons.
Therefore, the results regarding the performance of the different models turns out to be different compared to the in-sample evaluation above.
Now, the specifications in Equations~\eqref{eq:LinApproximation3} and \eqref{eq:LinApproximation4}, which use far less interaction terms and have a rather small number of parameters, capture the
dynamics of returns almost as well as the other two specifications in Equations~\eqref{eq:LinApproximation1} and \eqref{eq:LinApproximation2} which becomes apparent, for example, when looking at
the $DPA$ in Figure~\ref{fig:DPA_inS}.
Only considering the adjusted $R^2$ (Figure~\ref{fig:R2adj_inS}), the specification in Euqation~\eqref{eq:LinApproximation1} has a better performance.
It also slightly decreases the $RMSE$.

For the $XLM$ measure, the performance of the highly parameterized models is better.
In our opinion, this is due to the following two reasons.
First, the $XLM$ measure changes with each event, so that on a high frequency the volatility of the $XLM$ measure is higher than the volatility of returns.
The second reason is rooted in the definition of the $XLM$ measure as a fraction of observables which in general requires a higher polynomial degree to arrive at a sensible approximation.
The more complex models in Equations~\eqref{eq:LinApproximation1}~and~\eqref{eq:LinApproximation2} are, thus, better suited to provide such an approximation.
Note also that we take the last observation for the $XLM$ within an interval which is highly volatile.
From a trading perspective, the average $XLM$ over the interval may be better suited to describe liquidity of the market during the interval and might be a less volatile measure.
Nevertheless, we chose the more volatile measure for our analysis.
Our results should therefore pose a lower bound with respect to the modeling accuracy.

Nevertheless, for intraday returns, the relatively parsimonious models in Equations~\eqref{eq:LinApproximation3} and \eqref{eq:LinApproximation4} perform remarkably well.
Compared to other studies like \cite{ZhouPHTZ18} who use high-dimensional neural networks (without contemporaneous information), we find a decisively smaller $RMSE$.
For example, the out-of-sample forecast error reported by \cite{ZhouPHTZ18} for their best model (GAN for minimizing forecast error loss and direction prediction loss with training sets with a length
of 20 days and test sets with a length of 5 days) is 0.0079 with a DPA of 69\% on a 1 minute interval.
Even though the in-sample results in our case are not exactly comparable, we can note that for the models with lagged information on 1 minute intervals an average $RMSE$ of 0.0010 can be achieved
together with a DPA slightly above 85\% on average.
We also see that, on high frequencies, the models with lagged information all result in similar $RMSE$ and $DPA$ values.
Again the results for $R^2$ show that overfitting is a problem for lower sampling frequencies.
These only differ in $R^2$ (adjusted and unadjusted).
Nevertheless, the size of the adjusted $R^2$ of around 15\% on 1 minute intervals is remarkable.

All models perform better at intervals of 45 minutes than on other frequencies.
This is rooted in an above-average precision of the prediction of the return in the last interval of the day.
When using 45 minutes intervals, the last interval of the day is only 30 minutes long.
Hence,  estimated moments based on the observations within the previous 45min interval are used to predict the last, shorter, 30min-long interval.
If these shorter end-of-day intervals are not excluded from the data, the model performs better compared to other frequencies.
This suggests another avenue for further research: Can the model performance be improved by using longer intervals to estimate sample moments to then make predictions on smaller time horizons.
Put differently, is it in general beneficial to calculate moments on the last hour of data to forecast the next ten minutes?
The result for the 45min interval suggest there may be some benefits.
However, answering the question would entail a rolling calculation of means on overlapping intervals which is beyond the scope of the current paper.

\subsection{Out-of-sample Analysis}

We also evaluate our models based on out-of-sample predictions using rolling windows.
Depending on the size of the non-overlapping intervals, we vary the number of intervals included in the rolling window.\footnote{
Recall that when we talk about using \textit{non-overlapping intervals}, we mean the procedure to use the observations associated to the last interval of the rolling window to predict the observation of the next interval.
Alternatively, one could use \textit{overlapping-intervals}, i.e., update the observation at each new event to predict the next interval.
However, this procedure is computationally very demanding and, therefore, out of the scope of the current paper.
}
\Cref{tab:RollingWindow} presents the lengths of these windows for the frequencies considered.
\begin{table}
\caption{Rolling windows}
\label{tab:RollingWindow}
\begin{minipage}{0.8\linewidth}
 The table lists the number of intervals (and thus observations) within the rolling windows used to fit the model.
 The third column lists the approximate number of trading days over which the rolling window is spanned.
 For each window, we conduct an out-of-sample prediction.
 In the last column, the potential total number of non-overlapping intervals, i.e., the number of available observations is reported.
 The actual number of observations for which an out-of-sample forecast is produced depends on the availability of the necessary moments for the estimation of Equations~\eqref{eq:LinApproximation1}~to~\eqref{eq:LinApproximation4}.
 \vspace{0.2cm}
\end{minipage}
\centering
 \begin{tabular}{rrrr}
 \toprule
 \bfseries Frequency & \bfseries Intervals & \bfseries Days& \bfseries Total\\
 \midrule
 1 min & 10{.}000   & 21& 32{.}503\\
 2 min & 5{.}000    &21& 16{.}252\\
 5 min & 4{.}000    &42& 6{.}501\\
 10 min & 2{.}500   &52& 3{.}251\\
 15 min & 1{.}500   &47& 2{.}167\\
 20 min & 750       &31& 1{.}626\\
30 min & 500        &31& 1{.}084\\
 45 min & 500       &47& 723\\
 60 min & 300       &38& 542\\
 120 min & 150      &38&271\\
  240 min & 100     &50& 136\\
   \bottomrule
\end{tabular}
\end{table}
In addition to the root mean squared prediction error ($RMSPE$) and the out-of-sample $DPA$, we also use the $R^2$ of a Mincer-Zarnowitz regression \cite[]{MincerZ69} to evaluate the model.

Results are reported in Table~\ref{tab:1and5minint} for selected 1 and 5 minute intervals.
As can be seen, for the return series, the precision of the out-of-sample prediction is remarkably high.
On a 1 minute frequency, we are able to predict the direction of the next price change with an average accuracy at or over 80\%, irrespective of the model.
But even on the 5 minute frequency, the accuracy only falls slightly below 75\%.
For both buy and sell returns, we deem the $R^2$ rather high and the $RMSPE$ rather low given that we intend to predict financial returns on a ultra-high frequency.%
\footnote{
For the stock ADS, we also observe a 100\% accuracy when predicting the direction of the next price change.
Also the $R^2$ of the Mincer-Zarnowitz regression is around 50\%.
However, it needs to be mentioned that for this stock, only 19 out-of-sample predictions are made in total.
Since, based on 1 minute intervals, some moments that enter the right hand side of Equations~\eqref{eq:LinApproximation1}~to~\eqref{eq:LinApproximation3} cannot be calculated, we drop the observations for these intervals from our sample.
In effect ADS has 10{.}019 valid observations on a 1 minute frequency.
Due to such missing values also out-of-sample 1 minute results for DB1, FME and HEN3 are not reported since less than 10{.}000 observations are valid for these stocks.
This is why we do not include the 1-min out-of-sample results for ADS in the figures.
}

For the $XLM$, results are somewhat different.
The precision of the forecast is poor which is in line with the in-sample results.
Again, this is due to the high variability of the measure and its structure.
The $XLM$ changes with each event and is a highly nonlinear function in the arrival rates (see Equations~\eqref{eq:XLM_all}~to~\eqref{eq:XLMV}).
Therefore, the linear approximation may be poor and the approximation for longer time horizons may be especially poor.
In this line of argument, it is worth mentioning that the model with the highest complexity in our considerations (specified in \Cref{eq:LinApproximation1}) performs best in all evaluation measures.

\begin{table}
\caption{Out-of-sample results: 1 and 5 minute intervals}
\label{tab:1and5minint}
\centering
\begin{minipage}{0.9\linewidth}
 The table presents the  out-of-sample results for the 1 and 5 minute intervals for the stocks ADS and FME.
 The model alternatives in \Cref{eq:LinApproximation1}, \Cref{eq:LinApproximation2}, \Cref{eq:LinApproximation3} and \Cref{eq:LinApproximation4} are referred to in the rows A1 - A4 respectively.
 For each model, the $RMSPE$, the fit of the Mincer-Zarnowitz regression ($R^2_{MZ}$) and the direction prediction accuracy ($DPA$) are reported, both for the returns of the buy strategy $\Delta p_{t,b}$ and the ones of the sell strategy $\Delta p_{t,s}$.
 ALV is the most liquid stock with the highest number of events in the sample period while FME is one of the less liquid stocks.
\vspace*{0.2cm}
\end{minipage}
\setlength{\extrarowheight}{6pt}
 \begin{tabular}{c c c ccc c ccc}
  \toprule
                &&&\multicolumn{3}{c}{\bfseries 1min (ALV)} &&\multicolumn{3}{c}{ \bfseries 5min (FME)}\\
                &\bfseries Model&&$\Delta p_{t,s}$ &$\Delta p_{t,b}$ &$XLM$ &&$\Delta p_{t,s}$ &$\Delta p_{t,b}$&$XLM$\\
     \cline{2-2} \cline{4-6} \cline{8-10}
  \multirow{4}{*}{$RMSPE$} &A1 &&0.00103&0.00104&146989    &&0.00202 &0.00199&216247\\
   &A2 &&0.00095&0.00096&161087 &&0.00183 &0.00174&220908\\
   &A3 &&0.00094&0.00094&164625 &&0.00180 &0.00170&224286\\
   &A4 &&0.00094 &0.00094 &165198  &&0.00179  &0.00170 &224677 \\
   \midrule
   \multirow{4}{*}{$R^2_{MZ}$} &A1 &&0.0101&0.0071&0.2546 &&0.0357 & 0.0150&0.1076\\
   &A2 &&0.0131&0.0099&0.0802 &&0.0307 &0.0316&0.0372\\
   &A3 &&0.0146&0.0128&0.0396 &&0.0321 &0.0378&0.0079\\
   &A4 &&0.0145&0.0137 &0.0213 &&0.0323 &0.0393 &0.0050\\
  \midrule
   \multirow{4}{*}{$DPA$} &A1 &&77.02&77.75&47.11 &&74.66 & 72.96&51.58\\
   &A2 &&80.22&80.45&46.56 &&78.28 &79.19&41.18\\
   &A3 &&80.67&80.80&45.94 &&80.32 &79.75&34.16\\
   &A4 &&80.72 &80.92    &46.32 &&80.54 &79.86 &32.13 \\
   \bottomrule
 \end{tabular}
\end{table}

\begin{figure}
\caption{DPA}
\label{fig:DPA_OOS}
\centering
\begin{minipage}{0.9\linewidth}
  The figures report the out-of-sample direction prediction accuracy (DPA) (as defined in \Cref{eq:DPA}).
  The out-of-sample DPA is reported for the estimated model equations specified in \Cref{eq:LinApproximation1} (blue), \Cref{eq:LinApproximation2} (red),  \Cref{eq:LinApproximation3} (green)  and \Cref{eq:LinApproximation4} (orange)  for the
sampling frequencies 1, 2, 5, 10, 15, 20, 30, 45, 60, 120 and 240 minutes, estimated with a rolling window one-step ahead forecast.
  The respective window lengths are listed in \Cref{tab:RollingWindow}.
\end{minipage}
  \begin{subfigure}{0.49\linewidth}
  \includegraphics[width=\linewidth]{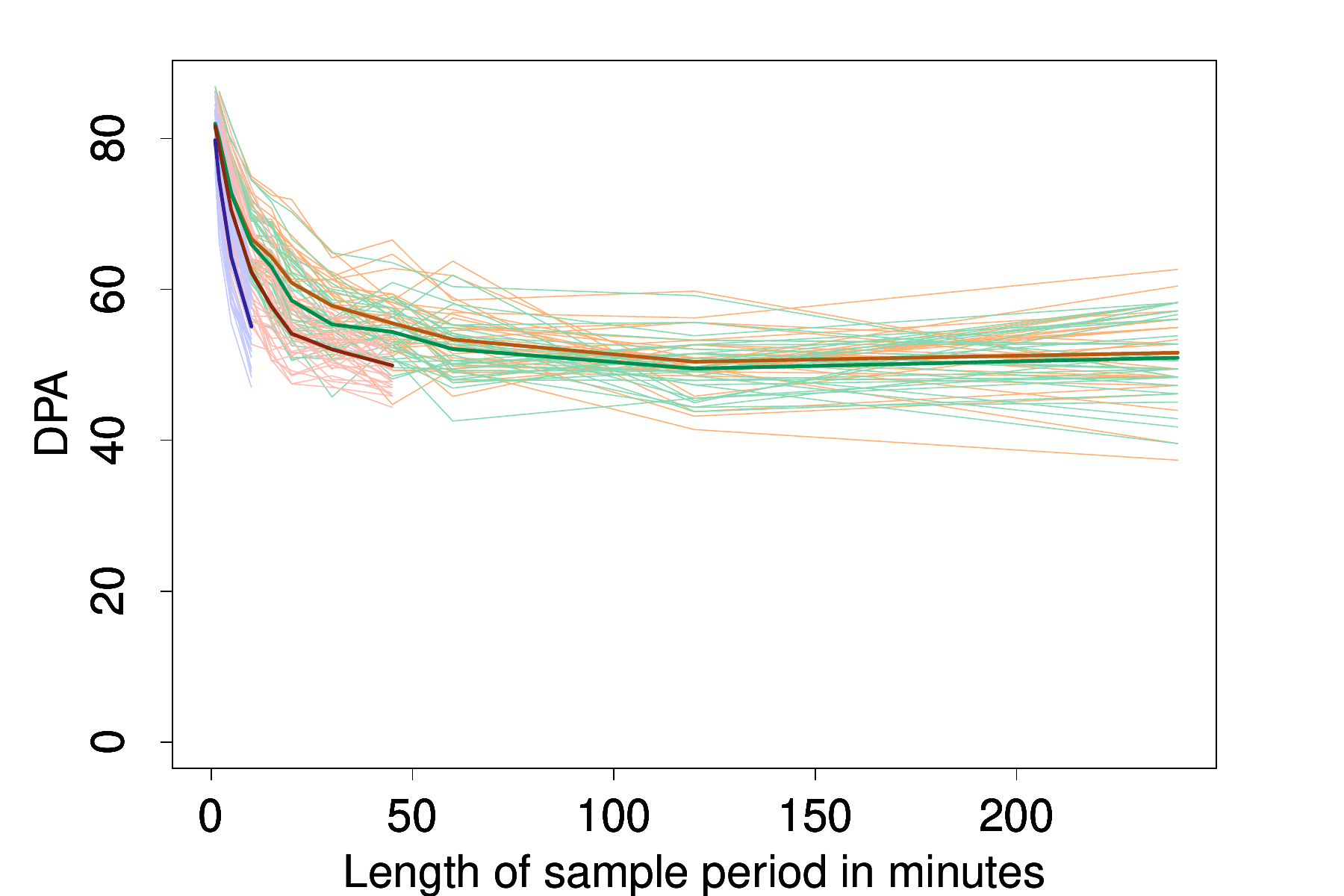}
    \subcaption{Buy}
 \end{subfigure}
 \begin{subfigure}{0.49\linewidth}
  \includegraphics[width=\linewidth]{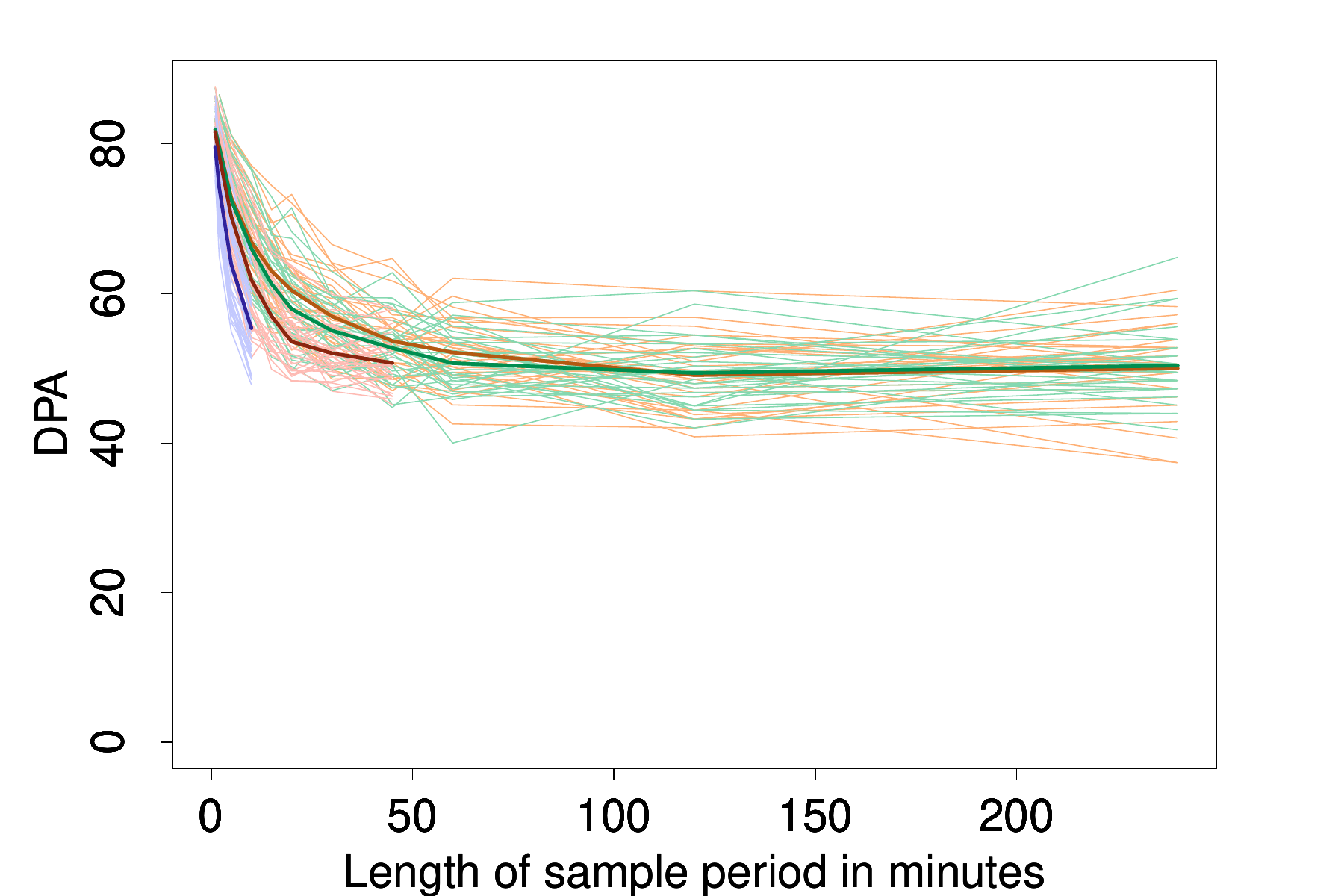}
  \subcaption{Sell}
 \end{subfigure}
  \begin{subfigure}{0.49\linewidth}
  \includegraphics[width=\linewidth]{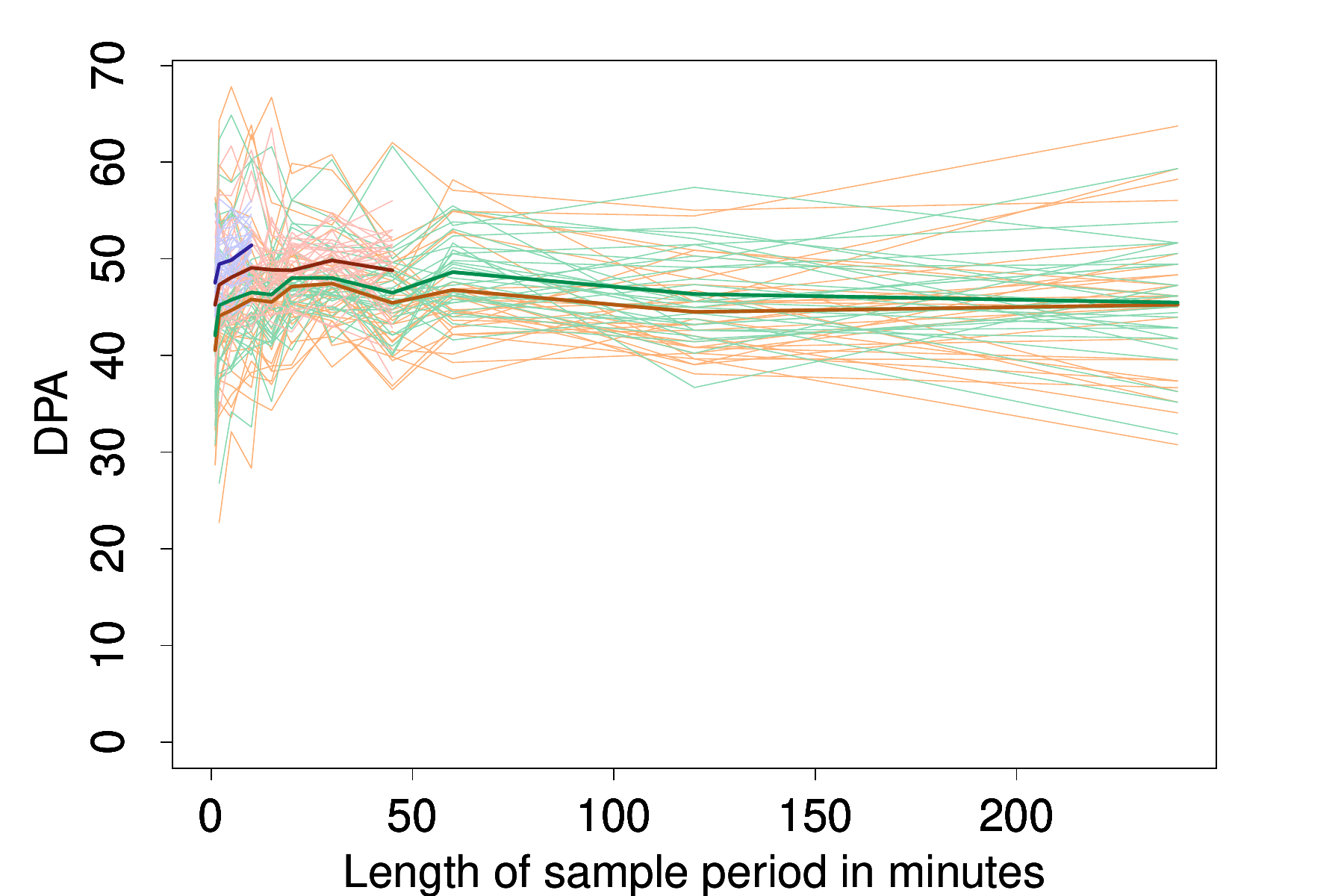}
  \subcaption{XLM}
 \end{subfigure}
\end{figure}

\begin{figure}
\caption{Adjusted $R^2$}
\label{fig:R2MZ_OOS}
\centering
\begin{minipage}{0.9\linewidth}
  The figures report the $R^2_{MZ}$ of the Mincer-Zarnowitz regression \cite{MincerZ69} based on the out-of-sample one-step ahead rolling window forecast.
  The $R^2_{MZ}$ is reported for the estimated model equations specified in \Cref{eq:LinApproximation1} (blue), \Cref{eq:LinApproximation2} (red) ,  \Cref{eq:LinApproximation3} (green)  and \Cref{eq:LinApproximation4} (orange) for the
sampling frequencies 1, 2, 5, 10, 15, 20, 30, 45, 60, 120 and 240 minutes.
  The respective window lengths are listed in \Cref{tab:RollingWindow}.
\end{minipage}
  \begin{subfigure}{0.49\linewidth}
  \includegraphics[width=\linewidth]{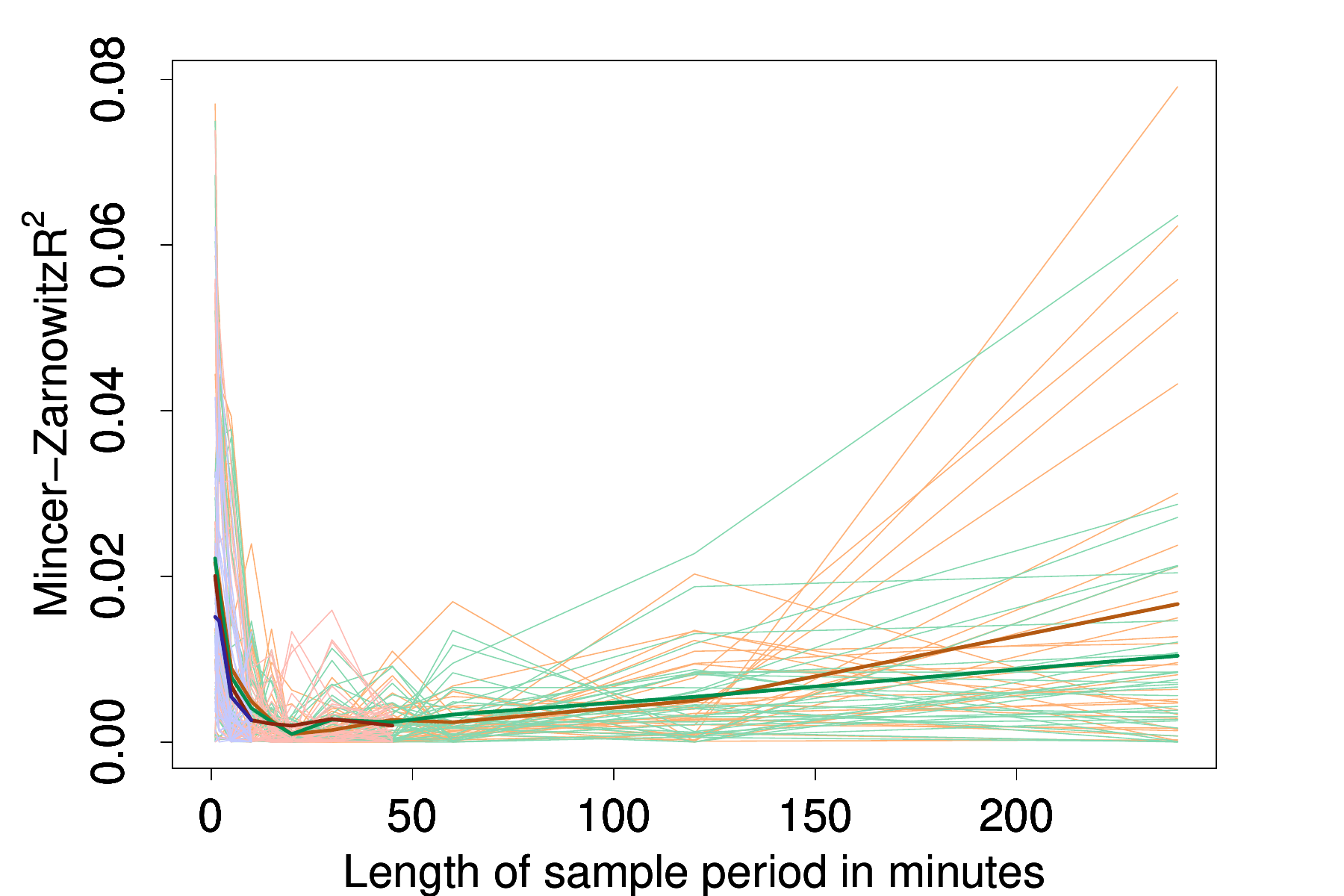}
   \subcaption{Buy}
 \end{subfigure}
 \begin{subfigure}{0.49\linewidth}
  \includegraphics[width=\linewidth]{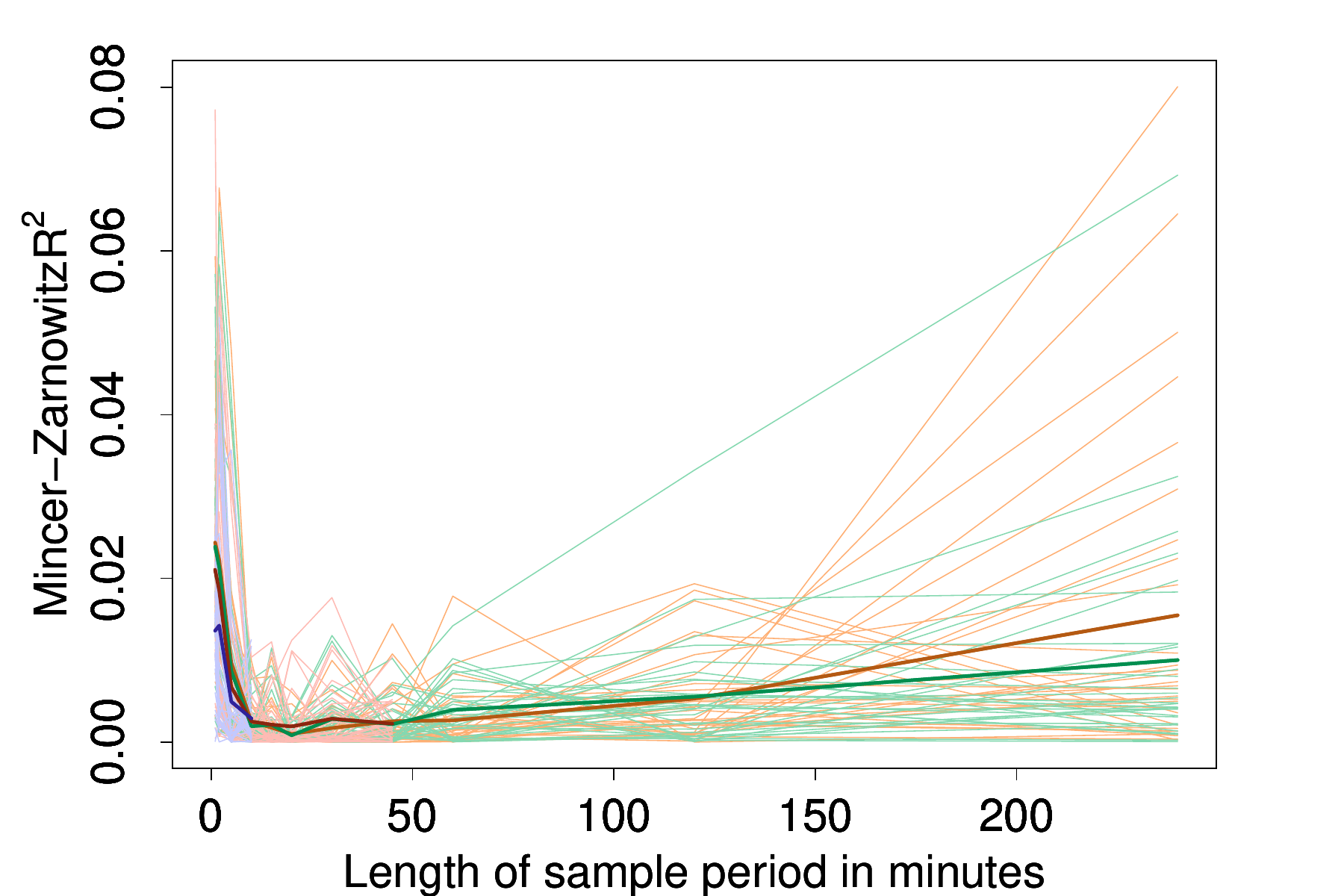}
  \subcaption{Sell}
 \end{subfigure}
 \begin{subfigure}{0.49\linewidth}
  \includegraphics[width=\linewidth]{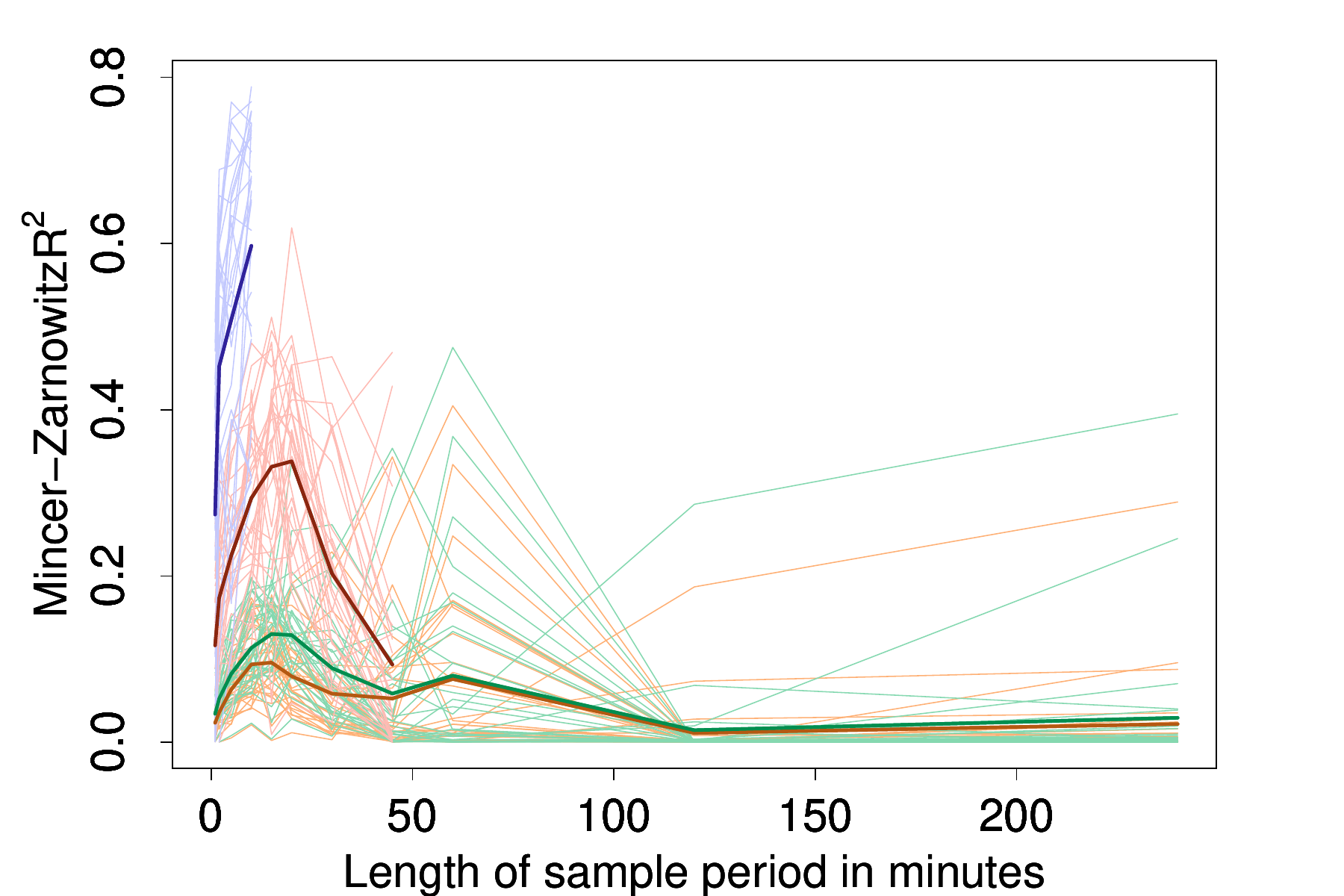}
  \subcaption{XLM}
 \end{subfigure}
\end{figure}

\begin{figure}
\caption{$RMSPE$}
\label{fig:rmse_OOS}
\centering
\begin{minipage}{0.9\linewidth}
 The figures below report the out-of-sample $RMSPE$ for the estimated model equations specified in \Cref{eq:LinApproximation1} (blue), \Cref{eq:LinApproximation2} (red),  \Cref{eq:LinApproximation3} (green)  and \Cref{eq:LinApproximation4} (orange) for the sampling frequencies 1, 2, 5, 10, 15, 20, 30, 45, 60, 120 and 240.
 The predictions are based on the out-of-sample one-step ahead rolling window forecast.
  The respective window lengths are listed in \Cref{tab:RollingWindow}.
\end{minipage}
 \begin{subfigure}{0.49\linewidth}
  \includegraphics[width=\linewidth]{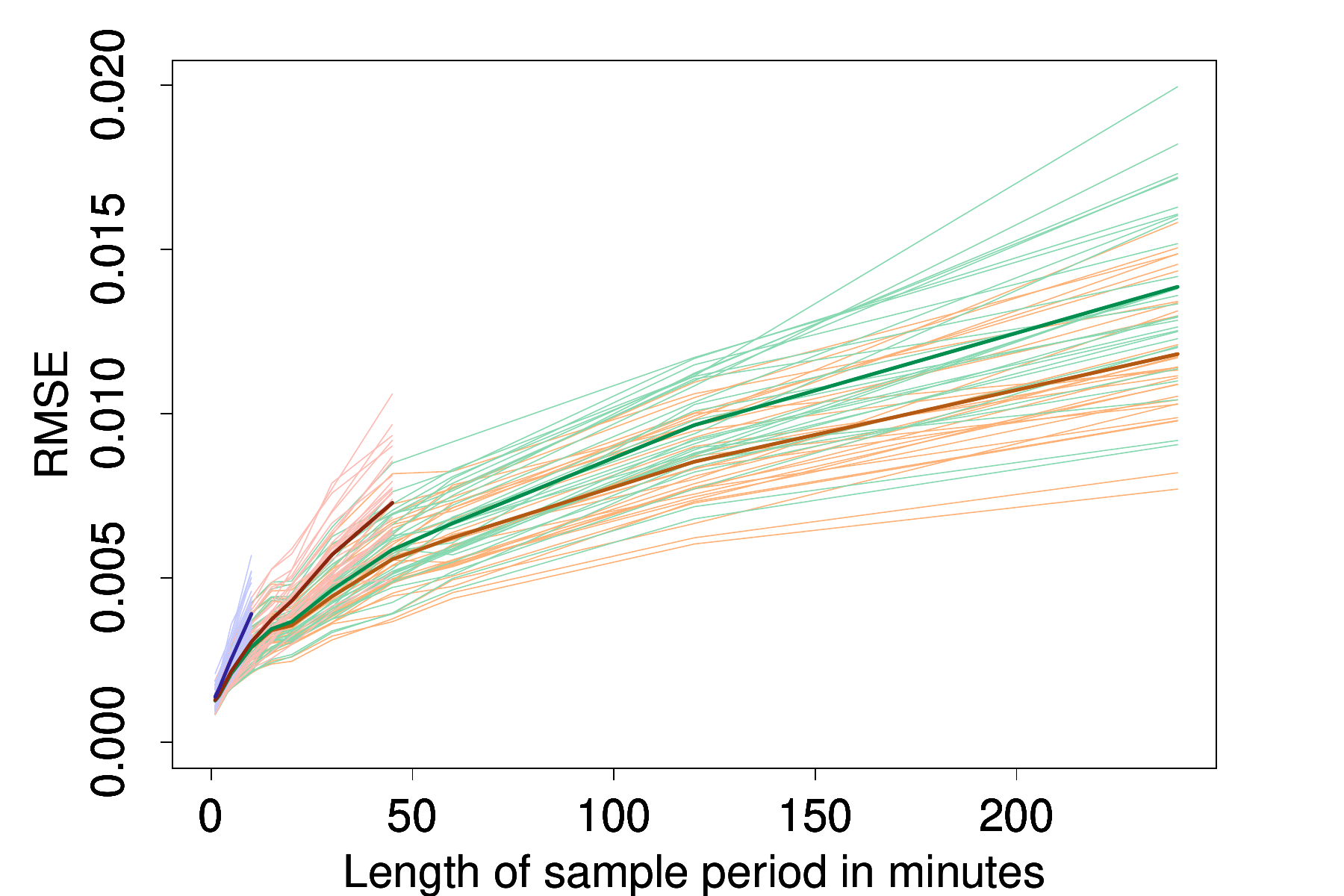}
  \subcaption{Buy}
 \end{subfigure}
 \begin{subfigure}{0.49\linewidth}
  \includegraphics[width=\linewidth]{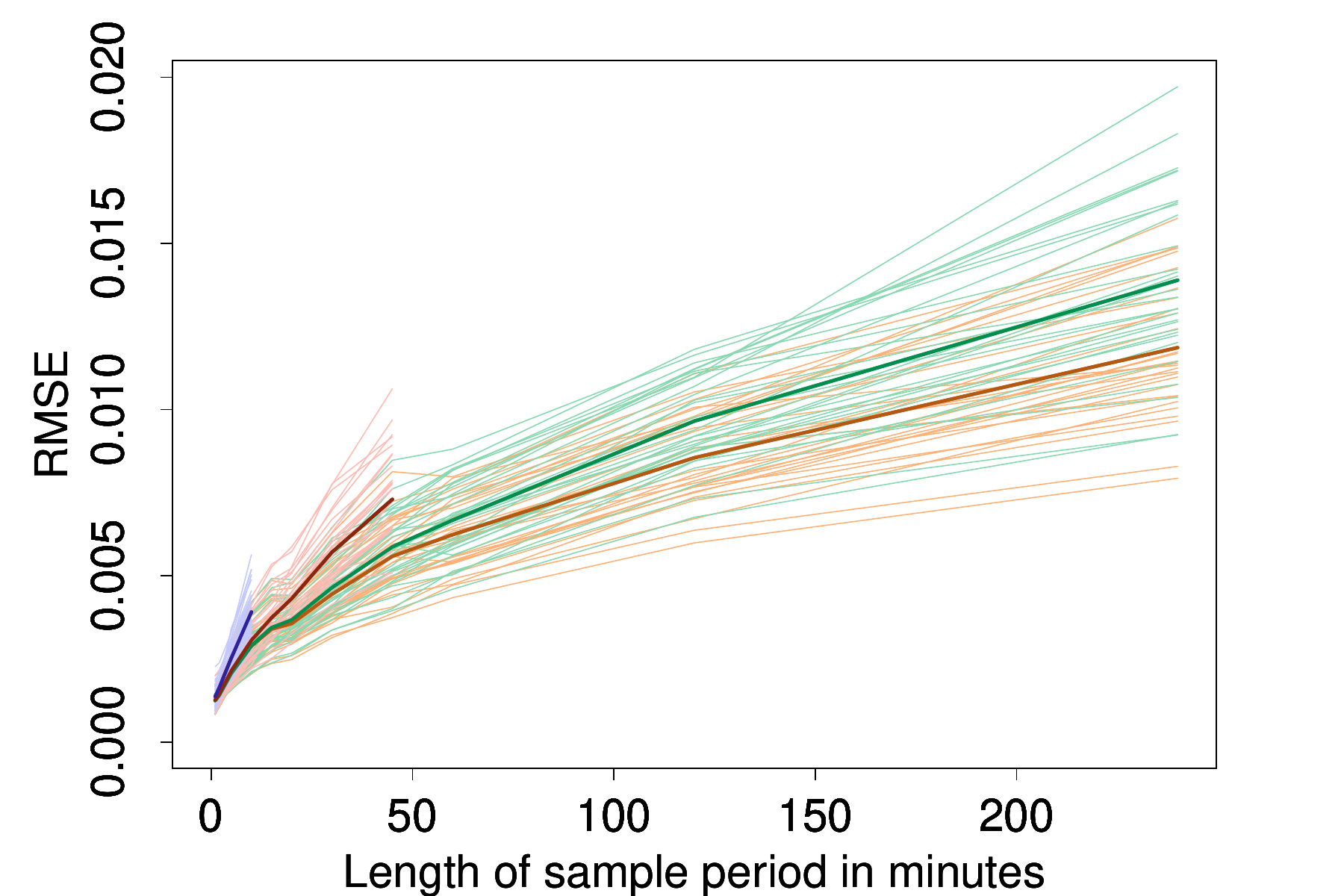}
   \subcaption{Sell}
 \end{subfigure}
  \begin{subfigure}{0.49\linewidth}
  \includegraphics[width=\linewidth]{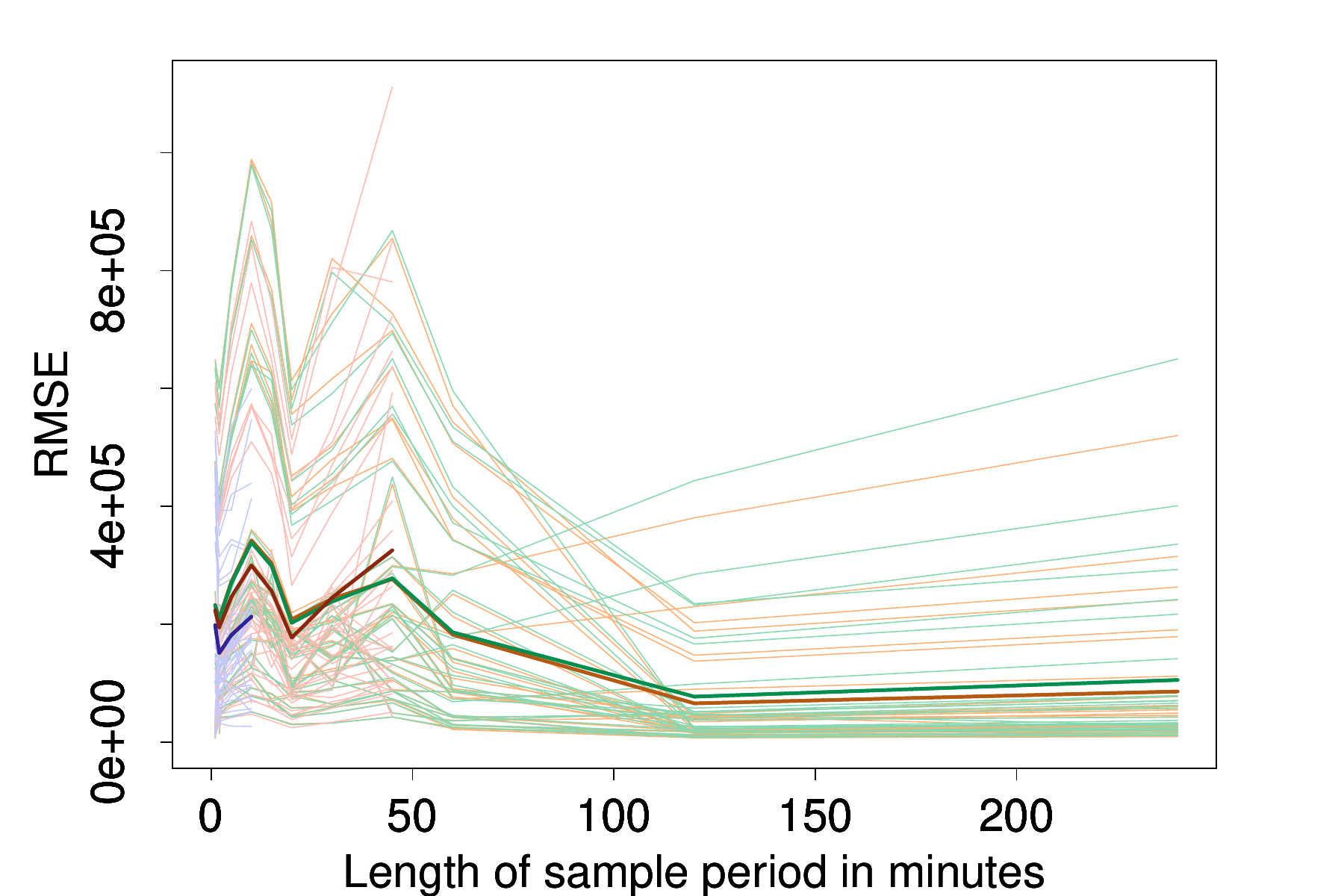}
   \subcaption{XLM}
 \end{subfigure}
\end{figure}

The results for all stocks and all frequencies are presented in Figures~\ref{fig:DPA_OOS} to \ref{fig:rmse_OOS}.
As we can see the smaller the interval, the better the forecasting ability of all our linear models.
The extensive linear approximation in \Cref{eq:LinApproximation1} predicts  the direction of the returns very well on small intervals.
The $R^2_{MZ}$ of the Mincer-Zarnowitz regression of above 2\% for the sell strategy is above what we had expected for returns on ultra-high frequencies.
On intervals longer than 10 minutes, the predictive ability of all three linear approximations is, however, poor.
It can also be noted, that the sparse model formulations in Equations~\eqref{eq:LinApproximation3} and \eqref{eq:LinApproximation4} perform just as well, or even better in some situations, than the
heavily parameterized formulation in \Cref{eq:LinApproximation1} in the case of the return series.
This is not true for the $XLM$.
For the $XLM$, the more complex formulations in Equation~\eqref{eq:LinApproximation1} and \eqref{eq:LinApproximation2} perform better in all measures.
Especially, the constant $RMSE$ and the increasing $DPA$ and $R^2_{MZ}$ up to 5-min intervals are remarkable.
The variance of the $DPA$ in \Cref{fig:DPA_OOS}c shows how noisy the $XLM$ and the associated forecasts are and by how much the more complex model is able to reduce this variability.

\section{Conclusion}
\label{sec:Conclusion}
In the paper, we propose a model for the limit order book which describes the dynamics in the book by a continuous Markov process and allows to forecast returns.
The mathematical formulation is based on the operator algebra which we borrow from physics.
Our model closely describes the reality of the order book and identifies the arrival and cancellation rates as the key ingredients of the book's dynamics.
By means of a simulation study, we show that the distribution of order arrival rates across price levels determines the shape of the book and, as a consequence, the transaction price evolution.
Varying the type and shape of arrival and cancellation rates across prices and volume, we find that the moments of price levels and quantity levels of incoming and canceled orders are
important determinants for the evolution of the book.

In an empirical study which is based on a linearized version of our model, we estimate three different specifications on non-overlapping intervals of various frequencies.
As we have the entire record of the XETRA order book for 3 months at our disposition, we can include a large number of parameters in the estimation such that an evaluation of the model becomes
feasible.
In-sample, all considered models exhibit a good fit in terms of $R^2$, $RMSE$ and direction prediction accuracy (DPA).
Our fully parameterized model seems to overfit the data on lower frequencies.
Nevertheless, when using only past information, the values for the adjusted $R^2$ range for the minute-by-minute intervals around or over 10\% whereas the direction is correctly predicted in around
70\% of all cases.

To evaluate the robustness of our results we also conduct an out-of-sample test of the model.
We use one-step-ahead forecasts on various frequencies and evaluate the accuracy with the $R^2_{\text{MZ}}$ of a Mincer-Zarnowitz regression, the $DPA$ as well as the $RMSPE$.
We find an extraordinarily good fit for ultra-high frequency returns as the $R^2_{\text{MZ}}$ is generally above $2\%$.
In addition, on low frequencies, the $DPA$ is around 80\% which suggests that we can predict the directional change of the next return very precisely.
The time varying estimates of the parameters as well as the short forecasting horizon make the return prediction astonishingly accurate.
We also try to predict liquidity at the end of each interval with the $XLM$ measure.
The measure cannot be forecasted well for longer time intervals with adequate accuracy.
Even on very short time horizons, the best fitting model is barely able to predict the direction of the next change of the $XLM$ in more than 50\% of the cases.
This result is rooted in the definition and the very volatile nature of the $XLM$ measure.

On the basis of the event log of XETRA for the first quarter of 2004, we have, nevertheless, shown that our model describes the LOB data well, both in- and out-of-sample.
The data requirements are rather high as knowledge about price and quantity levels of incoming and canceled orders are required.
This sort of data is usually not available.
Even though returns may be predicted, market impact of actual trading strategies as well as order costs may hamper profitability of a trading strategy based on our model.
Nevertheless, we are convinced that our empirical analysis provides new lower limits of forecast accuracy, as we have made several approximating decisions in the course of this paper.
In addition, for time horizons beyond 1 minute other variables may possibly help to predict returns or any other measure in the order book.

\section*{Acknowledgments}
The authors acknowledge support by the state of Baden-W\"urttemberg through bwHPC.

\section*{Data availability statement}
The data that support the findings of this study are available from the corresponding author upon reasonable request.

\newpage

     \onehalfspace
    \addcontentsline{toc}{section}{References}
    \bibliographystyle{kluwer}
    \bibliography{libPhD3}

\newpage
\appendix

\section{Distribution of events}
\label{appendix:dist_of_events}
{\singlespacing
\begin{longtable}{crrrrcrrrrrrrrrrr}
\caption{Number of events related to order type and market side}
\label{tab:OrderNumbers}\\
\multicolumn{11}{p{0.95\linewidth}}{
\footnotesize
 The table lists the total number of events related to limit (\#L), market (\#M), iceberg (\#I), and market-to-limit (\#T) orders for each stock in our sample.
 In the columns \%L, \%M, \%I, and \%T the percentage of events occurring on the sell (S) and buy (B) market side is tabled for each order type.
 The last column (\%Total) reports the share of all events for each market side.
}&
\\
\\
\toprule
\bfseries Ticker&
\multicolumn{1}{c}{\bfseries \#L}
&\multicolumn{1}{c}{\bfseries \#M}
&\multicolumn{1}{c}{\bfseries \#I}
&\multicolumn{1}{c}{\bfseries \#T}
&
&\multicolumn{1}{c}{\bfseries \%L}
&\multicolumn{1}{c}{\bfseries \%M}
&\multicolumn{1}{c}{\bfseries \%I}
&\multicolumn{1}{c}{\bfseries \%T}
&\multicolumn{1}{c}{\bfseries \%Total}\\
\midrule
\endfirsthead

\toprule
\bfseries Ticker&
\multicolumn{1}{c}{\bfseries \#L}
&\multicolumn{1}{c}{\bfseries \#M}
&\multicolumn{1}{c}{\bfseries \#I}
&\multicolumn{1}{c}{\bfseries \#T}
&
&\multicolumn{1}{c}{\bfseries \%L}
&\multicolumn{1}{c}{\bfseries \%M}
&\multicolumn{1}{c}{\bfseries \%I}
&\multicolumn{1}{c}{\bfseries \%T}
&\multicolumn{1}{c}{\bfseries \%Total}\\
\midrule
\endhead
 \multirow{2}{*}{ADS}&\multirow{2}{*}{ 1,129,682}&\multirow{2}{*}{ 16,976}&\multirow{2}{*}{  4,749}&\multirow{2}{*}{ 1,969}&B&48.4&55.1&49.7&41.4&48.5\\* 

&&&&&S&51.6&44.9&50.3&58.6&51.5\\ 
\midrule
\multirow{2}{*}{ALT}&\multirow{2}{*}{ 1,094,414}&\multirow{2}{*}{ 16,785}&\multirow{2}{*}{  9,552}&\multirow{2}{*}{ 1,412}&B&46.4&50.1&60.6&39.9&46.6\\* 

&&&&&S&53.6&49.9&39.4&60.1&53.4\\ 
\midrule
\multirow{2}{*}{ALV}&\multirow{2}{*}{ 4,237,243}&\multirow{2}{*}{ 68,446}&\multirow{2}{*}{ 39,416}&\multirow{2}{*}{ 2,105}&B&48.3&55.3&44.4&51.2&48.4\\* 

&&&&&S&51.7&44.7&55.6&48.8&51.6\\ 
\midrule
\multirow{2}{*}{BAS}&\multirow{2}{*}{ 2,585,776}&\multirow{2}{*}{ 31,450}&\multirow{2}{*}{ 36,082}&\multirow{2}{*}{ 1,885}&B&49.5&50.7&58.7&45.3&49.7\\* 

&&&&&S&50.5&49.3&41.3&54.7&50.3\\ 
\midrule
\multirow{2}{*}{BAY}&\multirow{2}{*}{ 2,199,894}&\multirow{2}{*}{ 34,187}&\multirow{2}{*}{ 29,943}&\multirow{2}{*}{ 1,721}&B&49.6&50.9&42.9&46.5&49.5\\* 

&&&&&S&50.4&49.1&57.1&53.5&50.5\\ 
\midrule
\multirow{2}{*}{BMW}&\multirow{2}{*}{ 2,087,167}&\multirow{2}{*}{ 28,557}&\multirow{2}{*}{ 34,625}&\multirow{2}{*}{ 1,707}&B&48.6&60.5&47.0&45.0&48.7\\* 

&&&&&S&51.4&39.5&53.0&55.0&51.3\\ 
\midrule
\multirow{2}{*}{CBK}&\multirow{2}{*}{ 1,676,325}&\multirow{2}{*}{ 23,601}&\multirow{2}{*}{ 24,753}&\multirow{2}{*}{ 1,413}&B&49.8&48.7&44.2&41.7&49.7\\* 

&&&&&S&50.2&51.3&55.8&58.3&50.3\\ 
\midrule
\multirow{2}{*}{CONT}&\multirow{2}{*}{ 1,130,309}&\multirow{2}{*}{ 15,866}&\multirow{2}{*}{  9,641}&\multirow{2}{*}{ 1,419}&B&48.5&49.2&51.0&43.6&48.6\\* 

&&&&&S&51.5&50.8&49.0&56.4&51.4\\ 
\midrule
\multirow{2}{*}{DB1}&\multirow{2}{*}{   936,959}&\multirow{2}{*}{ 16,205}&\multirow{2}{*}{ 16,381}&\multirow{2}{*}{ 1,315}&B&48.5&50.8&49.2&39.4&48.5\\* 

&&&&&S&51.5&49.2&50.8&60.6&51.5\\ 
\midrule
\multirow{2}{*}{DBK}&\multirow{2}{*}{ 3,339,752}&\multirow{2}{*}{ 48,314}&\multirow{2}{*}{ 47,215}&\multirow{2}{*}{ 2,278}&B&49.3&46.2&50.9&45.8&49.3\\* 

&&&&&S&50.7&53.8&49.1&54.2&50.7\\ 
\midrule
\multirow{2}{*}{DCX}&\multirow{2}{*}{ 2,711,327}&\multirow{2}{*}{ 44,301}&\multirow{2}{*}{ 57,847}&\multirow{2}{*}{ 2,003}&B&50.1&41.8&56.0&49.5&50.1\\* 

&&&&&S&49.9&58.2&44.0&50.5&49.9\\ 
\midrule
\multirow{2}{*}{DPW}&\multirow{2}{*}{ 1,001,394}&\multirow{2}{*}{ 26,360}&\multirow{2}{*}{ 28,571}&\multirow{2}{*}{ 1,468}&B&47.6&50.1&50.5&40.2&47.8\\* 

&&&&&S&52.4&49.9&49.5&59.8&52.2\\ 
\midrule
\multirow{2}{*}{DTE}&\multirow{2}{*}{ 2,349,138}&\multirow{2}{*}{ 87,942}&\multirow{2}{*}{ 49,129}&\multirow{2}{*}{ 3,581}&B&49.7&51.8&48.1&39.0&49.7\\* 

&&&&&S&50.3&48.2&51.9&61.0&50.3\\ 
\midrule
\multirow{2}{*}{EOA}&\multirow{2}{*}{ 2,701,672}&\multirow{2}{*}{ 35,484}&\multirow{2}{*}{ 34,695}&\multirow{2}{*}{ 2,106}&B&51.2&54.3&53.2&47.9&51.3\\* 

&&&&&S&48.8&45.7&46.8&52.1&48.7\\ 
\midrule
\multirow{2}{*}{FME}&\multirow{2}{*}{   801,834}&\multirow{2}{*}{ 11,156}&\multirow{2}{*}{  4,159}&\multirow{2}{*}{ 1,223}&B&48.9&50.4&56.2&39.8&49.0\\* 

&&&&&S&51.1&49.6&43.8&60.2&51.0\\ 
\midrule
\multirow{2}{*}{HEN3}&\multirow{2}{*}{ 1,101,152}&\multirow{2}{*}{ 11,405}&\multirow{2}{*}{  3,565}&\multirow{2}{*}{ 1,492}&B&47.1&51.7&39.3&39.5&47.1\\* 

&&&&&S&52.9&48.3&60.7&60.5&52.9\\ 
\midrule
\multirow{2}{*}{HVM}&\multirow{2}{*}{ 1,482,520}&\multirow{2}{*}{ 29,616}&\multirow{2}{*}{ 41,955}&\multirow{2}{*}{ 1,392}&B&50.4&54.1&50.9&41.6&50.5\\* 

&&&&&S&49.6&45.9&49.1&58.4&49.5\\ 
\midrule
\multirow{2}{*}{IFX}&\multirow{2}{*}{ 1,594,470}&\multirow{2}{*}{ 50,125}&\multirow{2}{*}{ 63,497}&\multirow{2}{*}{ 1,584}&B&48.8&54.1&42.4&40.7&48.7\\* 

&&&&&S&51.2&45.9&57.6&59.3&51.3\\ 
\midrule
\multirow{2}{*}{LHA}&\multirow{2}{*}{ 1,169,415}&\multirow{2}{*}{ 23,026}&\multirow{2}{*}{ 28,737}&\multirow{2}{*}{ 1,570}&B&48.8&48.6&50.3&42.2&48.8\\* 

&&&&&S&51.2&51.4&49.7&57.8&51.2\\ 
\midrule
\multirow{2}{*}{LIN}&\multirow{2}{*}{ 1,157,591}&\multirow{2}{*}{ 13,496}&\multirow{2}{*}{  7,384}&\multirow{2}{*}{ 1,807}&B&48.3&48.6&63.2&46.0&48.4\\* 

&&&&&S&51.7&51.4&36.8&54.0&51.6\\ 
\midrule
\multirow{2}{*}{MAN}&\multirow{2}{*}{ 1,023,998}&\multirow{2}{*}{ 14,952}&\multirow{2}{*}{ 15,252}&\multirow{2}{*}{ 1,697}&B&47.3&49.0&60.2&39.1&47.5\\* 

&&&&&S&52.7&51.0&39.8&60.9&52.5\\ 
\midrule
\multirow{2}{*}{MEO}&\multirow{2}{*}{ 1,144,291}&\multirow{2}{*}{ 16,064}&\multirow{2}{*}{ 15,028}&\multirow{2}{*}{ 1,460}&B&48.8&52.0&52.8&39.3&48.9\\* 

&&&&&S&51.2&48.0&47.2&60.7&51.1\\ 
\midrule
\multirow{2}{*}{MUV2}&\multirow{2}{*}{ 2,896,094}&\multirow{2}{*}{ 46,036}&\multirow{2}{*}{ 37,068}&\multirow{2}{*}{ 1,908}&B&49.2&57.0&43.7&45.4&49.3\\* 

&&&&&S&50.8&43.0&56.3&54.6&50.7\\ 
\midrule
\multirow{2}{*}{RWE}&\multirow{2}{*}{ 2,061,625}&\multirow{2}{*}{ 31,746}&\multirow{2}{*}{ 35,398}&\multirow{2}{*}{ 2,014}&B&51.6&44.6&53.3&47.5&51.5\\* 

&&&&&S&48.4&55.4&46.7&52.5&48.5\\ 
\midrule
\multirow{2}{*}{SAP}&\multirow{2}{*}{ 2,800,569}&\multirow{2}{*}{ 36,907}&\multirow{2}{*}{ 20,332}&\multirow{2}{*}{ 1,530}&B&49.6&49.4&57.1&42.9&49.7\\* 

&&&&&S&50.4&50.6&42.9&57.1&50.3\\ 
\midrule
\multirow{2}{*}{SCH}&\multirow{2}{*}{ 1,312,153}&\multirow{2}{*}{ 24,385}&\multirow{2}{*}{ 17,908}&\multirow{2}{*}{ 1,439}&B&48.3&50.3&51.6&41.6&48.4\\* 

&&&&&S&51.7&49.7&48.4&58.4&51.6\\ 
\midrule
\multirow{2}{*}{SIE}&\multirow{2}{*}{ 3,444,640}&\multirow{2}{*}{ 58,410}&\multirow{2}{*}{ 54,186}&\multirow{2}{*}{ 2,172}&B&48.3&49.5&53.8&48.0&48.4\\* 

&&&&&S&51.7&50.5&46.2&52.0&51.6\\ 
\midrule
\multirow{2}{*}{TKA}&\multirow{2}{*}{ 1,130,506}&\multirow{2}{*}{ 23,019}&\multirow{2}{*}{ 19,060}&\multirow{2}{*}{ 1,797}&B&48.5&52.8&45.2&41.1&48.5\\* 

&&&&&S&51.5&47.2&54.8&58.9&51.5\\ 
\midrule
\multirow{2}{*}{TUI}&\multirow{2}{*}{   970,118}&\multirow{2}{*}{ 21,965}&\multirow{2}{*}{ 15,737}&\multirow{2}{*}{ 1,269}&B&48.7&53.8&40.5&40.4&48.7\\* 

&&&&&S&51.3&46.2&59.5&59.6&51.3\\ 
\midrule
\multirow{2}{*}{VOW}&\multirow{2}{*}{ 1,966,460}&\multirow{2}{*}{ 26,478}&\multirow{2}{*}{ 42,894}&\multirow{2}{*}{ 1,715}&B&48.6&49.3&53.0&44.4&48.7\\* 

&&&&&S&51.4&50.7&47.0&55.6&51.3\\ 
\bottomrule
\multirow{2}{*}{TOTAL}&\multirow{2}{*}{55,238,488}&\multirow{2}{*}{933,260}&\multirow{2}{*}{844,759}&\multirow{2}{*}{52,451}&B&49.1&51.2&50.2&43.5&49.1\\* 

&&&&&S&50.9&48.8&49.8&56.5&50.9\\ 
\midrule

\end{longtable}
}

\section{Simulation Specification}
\label{appendix:Simulation}
This appendix presents the theoretical and algorithmical details of the stochastic simulation algorithm (SSA) used to simulate an artificial history of the LOB.

The starting point of the SSA is based on the probability that within the next interval $\delta\tau$ no event occurs which we can denote in our notation as
\begin{align}
P_0(\delta\tau) &= \sum_{z \in \mathcal{H}} \bra{z} \exp(\diag(H) \delta\tau) \ket{\Psi(t_0)} \label{eq:NothingHappens},
\end{align}
where the diagonal elements in $H$ are obtained by
\begin{align*}
\diag(H) \delta \tau = \bra{z}H\delta\tau\ket{z} = - \sum_{k,q,M} \alpha_M(k,q;z)\delta\tau  - \sum_{k,q,M} \omega_M(k,q;z)\delta\tau  .
\end{align*}
This is the negative sum of the rates of all possible events conditional on the book being in state $\ket{z}$.

\cite{Gillespie1977} shows how to formulate this probability for some event $\mu$ to happen during the interval $\tau$ without the operator algebra.
First, the probability that an order arrives during the interval $d\tau$ is $r_\mu d\tau$, where $r_\mu$ is the rate corresponding to the event.
In our case, $r_\mu$ may be some rate from the set of arrival or cancellation rates, $\alpha_M(k,q)\delta\tau$ or $\omega_M(k,q)\delta\tau$.
In fact, we may label all possible events with integer numbers and let $\mu$ be a specific integer denoting a specific event.
Setting $\tau = \delta \tau + d\tau$, the probability that given the state $\ket{\Psi(t_0)}$ at time $t_0$ the next reaction $\mu$ will happen during the next interval of $\tau$, denoted
$P(\tau,\mu)$, can be written as the product of the probability that nothing will happen during $\delta \tau$ and the probability that $\mu$ will happen during $d\tau$:
\begin{align}
 P(\tau,\mu) = P_0(\delta\tau)  r_\mu  d\tau   . \label{eq:NothingHappens0}
\end{align}

From \Cref{eq:NothingHappens0}, \cite{Gillespie1977} deduces that the probability that nothing happens during $\tau$, can be formulated as
\begin{align}
  P_0(\tau) = P_0(\delta\tau)  \left(1 - \sum_{\nu\neq\mu}r_\nu  d\tau  \right)  . \label{eq:NothingHappens2}
\end{align}
Noting that $\tau = \delta \tau + d\tau$ by definition, bringing all terms involving $P_0$ to the left hand side, dividing both sides by $d\tau$ and taking limits for $\delta \tau \to 0$, yields a differential equation that is solved by setting
\begin{align}
 P_0(\tau) = \exp\left(-\sum_\nu r_\nu \right) \label{eq:NothingHappens3}  .
\end{align}

Substituting \Cref{eq:NothingHappens3} into \Cref{eq:NothingHappens2}, the probability that $\mu$ will happen during the next time interval $\tau$ is given by
\begin{align}
 P(\tau,\mu) =   r_\mu \exp(-r_0\tau) =  r_\mu \sum_{z \in \mathcal{H}} \bra{z} \exp(\diag(H) \tau) \ket{\Psi(t_0)}  , \label{eq:SomethinHappens}
\end{align}
where in our case $r_0 = \sum_{k,q,M} \alpha_M(k,q;z) + \sum_{k,q,M} \omega_M(k,q;z) $.

From \Cref{eq:SomethinHappens}, we may randomly generate the pair $(\tau,\mu)$, i.e., the time when an event occurs $\tau$ and which event will happen $\mu$.
As we have set up the rates as price and size specific, by generating the event $\mu$ we also specify the price location and the size which are affected by the event.
By noting that \Cref{eq:SomethinHappens} determines an exponential distribution with scale parameter $r_0$, we can first sample $\tau$ by drawing $u_1$ from a uniform distribution $\mathcal{U}(0,1)$ and calculating
\begin{align*}
 \tau = \frac{1}{r_0} \log\left(\frac{1}{u_1}\right)  .
\end{align*}
Having determined when an event occurs, we may now ask the question what will happen.
By numerically specifying the rates for all possible events $r_\nu$ and drawing a second realization $u_2$ from a uniform distribution $\mathcal{U}(0,1)$, we may find the integer $\mu$ by solving
\begin{align*}
 \sum_{\nu=1}^{\mu-1} \frac{r_{\nu}}{r_0} <  u_2 \leq  \sum_{\nu=1}^{\mu} \frac{r_{\nu}}{r_0}
\end{align*}
for $\mu$.
In other words, by drawing $u_1$ and $u_2$, we can simulate an answer to the question \emph{when} something will happen with $u_1$ and, with $u_2$, \emph{what} as well as \emph{where} it will take place.
In fact, we also draw a third realization from a uniform distribution $u_3$, to answer the question what size is affected (see \Cref{appendix:Simulation} for details).
Having drawn an event and it's characteristics, the current state of the system can be updated.
This may change the rates $r_\nu$ and their sum $r_0$.
Note that by sampling the events in this fashion, the events are conditionally independent.
They may not be independent as the rates are conditional on the current state (and under the assumption of a higher Markov order also on finitely many previous states) of the LOB.

In order to simulate the LOB dynamics, we have to specify the rates of all possible events and how they depend on the current state.
In our case, all possible events comprise the order arrivals and order cancellations.
Thus, we have to find a functional form for the respective rates $\alpha_M$ and $\omega_M$.
In our specifications presented in \Cref{appendix:Simulation}, we let $\alpha_M$ and $\omega_M$ be functions of the quantity $q$ and the price level $k$ (or more precisely of the integer distance to the opposite best quote $d_l$).
For simplicity, we will assume that all rates $\alpha_M(k,q)$ and $\omega_M(k.q)$ are separable in $k$ and $q$ such that
\begin{align*}
\alpha_M(k,q) = \alpha_{1,M}(k) \alpha_{2,M}(q) \quad \text{  and  } \quad  \omega_M(k.q) = \omega_{1,M}(k) \omega_{2,M}(q)  .
\end{align*}
As the rates are proportional to the probability distribution of arriving (or canceled) orders across price and size, this means that the size of arriving (or canceled orders) is stochastically independent of the price level they concern.
In \Cref{fig:bin_vs_size}, we see that for lower distances to the opposite best quote the size of arriving and canceled order is equally spread out across possible size levels.
A clear relationship between the price level and the size level is not visible.
In the absence of such a clear relationship, we find the approximating assumption that the size and price level are stochastically independent justifiable.

We also decompose the arrival rates further by setting the general intensity of events for each market side $\bar{r}_{0,M,i}$ to the average event rate over the entire sample of stock $i$, where $\bar{r}_{0,M,i}$ is defined as
\begin{align*}
 \bar{r}_{0,M,i} = \sum_{k,q,j} \alpha_{M,i}(k,q)+\omega_{M,i}(k,q)
\end{align*}
which is calculated as the number of events on one market side divided by the total number of events.
Note that since we have several order types, the arrival rates may be split into market orders as well as marketable and non-marketable limit orders.
The empirical frequencies for $\bar{r}_{0,M,i}$ are reported in the last column in \Cref{tab:OrderNumbers}.

Hence, the arrival and cancellation rates for limit orders can be described by the partitioning of the average event rate $\bar{r}_{0,M,i,j,a}$ across price levels $k$ and order sizes $q$:
\begin{align}
 \alpha_M(k,q) &= \bar{r}_{0,M,i,j,a} \,\, p_{K,M}(k;\boldsymbol{\theta}_{M,a}) \,\,  p_{Q,M}(q;\boldsymbol{\phi}_{M,a}) , \notag \\
\omega_M(k.q) &=  \bar{r}_{0,M,i,L,c} \,\, p_{K,M}(k;\boldsymbol{\theta}_{M,c}) \,\, p_{Q,M}(q;\boldsymbol{\phi}_{M,c})  , \label{eq:rate_decomp}
\end{align}
where $\bar{r}_{0,M,i,j,a}$ is the rate for an order of type $j$ (market or limit order) for stock $i$ to arrive and $\bar{r}_{0,M,i,L,c}$ is the rate for a limit order (i.e. $j=L$) to be canceled.
$p_{K,M}(k;\boldsymbol{\theta}_{M,a})$ denotes the discrete probability mass function of order arrivals across the integer price levels $k$ given some parameter set $\boldsymbol{\theta}_{M,a}$ and similarly $p_{Q,M}(q;\boldsymbol{\phi}_{M,a})$ is the discrete probability mass function of order arrivals or cancellations across order sizes.
The index $a$ indicates the parameters for order arrivals, $c$ the parameters for cancellations.
The index $M$ denotes the market side.

In order to  simulate the order book, several probabilities and other conventions have to be specified.
Therefore, we go through the terms in \Cref{eq:rate_decomp} and present how we have chosen to specify $\alpha_{M}(k,q)$ and $\omega_{M}(k,q)$.
For convenience, recall \Cref{eq:rate_decomp} as
\begin{align*}
 \alpha_M(k,q) &= \bar{r}_{0,M,i,j,a} \,\, p_{K,M}(k;\boldsymbol{\theta}_{M,a}) \,\,  p_{Q,M}(q;\boldsymbol{\phi}_{M,a}) \notag \\
\omega_M(k.q) &=  \bar{r}_{0,M,i,L,c} \,\, p_{K,M}(k;\boldsymbol{\theta}_{M,c}) \,\, p_{Q,M}(q;\boldsymbol{\phi}_{M,a}).
\end{align*}
\Cref{fig:EventTree} illustrates the components of \Cref{eq:rate_decomp}.

\begin{figure}
\caption{Simulation event tree}
\label{fig:EventTree}
    \centering
    \begin{minipage}{0.9\linewidth}
     The Figure depicts a decision tree to visualize the components of \Cref{eq:rate_decomp}.
     Taking the subtree marked by the red box, each of the leafs that originate the red box has a different rate $\bar{r}_{0,M,i,j,e}$ where the subscript $e$ refers either to $a$ an order arrival or $c$ an order cancellation.
     Also each of the nodes inside the blue and green box has a different $p_{K,M}(\cdot)$
    \vspace{0.5cm}
    \end{minipage}
\begin{forest}
[
[Event occurs, tikz={
                        \draw[{Latex}-, thick] (.north) --++ (0,0.7);
                        \node [draw,red!70!black,inner sep=0,fit=(!111)(!22)(!u)]{};
                        \node [draw,blue!70!black,inner sep=0,fit=(!11111)(!22111)(!11111)]{};
                        \node [draw,green!40!black,inner sep=0,fit=(!1111)(!2211)(!1111)]{};
                        }
    [Ask side moves
        [Arrival
            [Market [Chose $k$
                 [ Chose $q$ ]
             ]]
            [Limit[Chose $k$
                 [ Chose $q$ ]
             ]]
        ]
        [Cancellation
            [ ,tikz={\draw[-] (.north) --++ (0,-0.5);}
                [Chose $k$
                [ Chose $q$ ]
             ]]]
    ]
    [Bid side moves
        [Arrival
            [Market [Chose $k$
                 [ Chose $q$ ]
             ]]
            [Limit[Chose $k$
                 [ Chose $q$ ]
             ]]
        ]
         [Cancellation, tikz={\node at () [red!60!black,right=1.5cm] {$\bar{r}_{0,M,i,j,e}$};}
             [,tikz={\draw[-] (.north) --++ (0,-0.5);}
             [Chose $k$, tikz={\node at () [green!40!black,right=1.5cm] {$p_{K,M}(\cdot)$};}
                 [ Chose $q$, tikz={\node at () [blue!70!black,right=1.5cm] {$p_{Q,M}(\cdot)$};}
                 ]
             ]]
         ]
    ]
]
]
\end{forest}
\end{figure}
Recall that we chose three theoretical scenarios for the distributions across price levels $p_{K,M}(\cdot)$:
First, the uniform distribution (uni), second, a discrete log-normal distribution with fixed parameters (fix), and third,  a discrete log-normal distribution with dynamic parameters where the
parameters depend on the prevailing spread (dyn).
For the distribution across order sizes, we only consider one theoretical specification: a power law distribution.
Additionally, we also consider the unconditional empirical frequencies of incoming and canceled orders as observed in the first quarter of 2004, both across price and size levels.

\subsection{Rates of Order Types $\bar{r}_{0,M,i,j,e}$}
The first element of \Cref{eq:rate_decomp} is $\bar{r}_{0,M,i,j,e}$, the rate for an arrival ($e=a$) or a cancellation ($e=c$) of order type $j$ on market side $M$ for stock $i$.
We first need to specify the order types that we include in the simulation.
In \Cref{fig:bin_freq}, we have depicted limit orders and market orders across relative integer distances to the best quote to show that there is a somewhat stable distribution across price levels
when the best quote is used as a fix point.
At the zero level, we have plotted the marketable orders split up into different types.

\Cref{tab:MarketableOrders} shows the percentages of the different types of marketable orders in detail.
In general, approximately 10\% of all incoming orders (cancellations excluded) are marketable.
In fact, about half of those marketable orders are arriving on the best quote i.e. with $d=0$.
Around a quarter is due to market orders with no limit price $d<-\infty$ and another quarter are marketable limit orders i.e. with $d<0$.
Marketable iceberg and stop order are tiny in comparison.
The inverse of the last column are the limit orders that are submitted before the best quote.
As depicted in \Cref{fig:EventTree}, in our simulation scenarios, we only treat market and limit orders separately.
We do not distinguish iceberg and stop orders, since they are market and limit orders with some additional features.
Thus, when we use the unconditional empirical frequencies for $\bar{r}_{0,M,i,j,e}$, we calculate
\begin{align*}
 \bar{r}_{0,M,i,j,e} = \frac{n_{M,i,j,e}}{\Delta T} ,
\end{align*}
where $n_{M,i,j,e}$ is the number of arrivals ($e=a$) or cancellations ($e=c$) of order type $j$ on market side $M$ for stock $i$ observed during the entire first quarter of 2004.
$\Delta T$ refers to the total trading time during this period.
In our case, $\Delta T$ is specified to be 64 trading days.
As we restrict the simulation to continuous trading, we only include events during the 8h28m of continuous trading to calculate the frequencies.
In our sample, option settlement is conducted in three dates.
On these three days, further 3 minutes have to be subtracted from the continuous trading phase.
In total, we have $(64\cdot(8+28/60)\cdot60-3\cdot3)\cdot60 = 1{,}950{,}180s$ of continuous trading time in our sample.
Note that we can also decompose the unconditional empirical rates according to
\begin{align}
 \bar{r}_{0,M,i,j,e} =  \frac{n_{\cdot,i,\cdot,\cdot}}{\Delta T}\cdot
   \frac{n_{M,i,\cdot,\cdot}}{n_{\cdot,i,\cdot,\cdot}} \cdot
   \frac{n_{M,i,j,\cdot}}{n_{M,i,\cdot,\cdot}} \cdot
   \frac{n_{M,i,j,e}}{n_{M,i,j,\cdot}}, \label{eq:decomposed_rates}
\end{align}
where $n$ refers to a number of events and the indices specify which characteristic is relevant for counting.
$n_{\cdot,i,\cdot,\cdot}$ means that only the index $i$ (referring to the event concerning stock $i$) is relevant to determine the number of events.
Categories marked with a $\cdot$ in the index are summed over.
In other words, $n_{\cdot,i,\cdot,\cdot}$ denotes the number of events concerning stock $i$.
In the empirical scenarios, all elements of \Cref{eq:decomposed_rates} can be observed.
In theory, we can craft theoretical scenarios to investigate, ceteris paribus, the sensitivity of the LOB dynamics to changes in just one conditional frequency in \Cref{eq:decomposed_rates}.
In this paper, we chose to focus on the sensitivity of the order book dynamics to changes in the distribution across price and quantity levels.

In the scenarios that entail a theoretical distribution, we do not use the empirical values observed in our sample.
We also chose to focus on the distribution of arrival rates across price and size levels.
Thus, we set the values summarized in \Cref{tab:EventRates}.
The rates are specified in the unit [orders/second].
They approximately mirror the observed values in reality but we fix them to parity, so that the two sides of the market are symmetric and balanced.
\begin{table}
\begin{center}
\caption{Event rates for order types}
\label{tab:EventRates}
\begin{minipage}{0.9\linewidth}
 \footnotesize
 The table lists the order arrival and cancellation rates imposed in the scenarios 'dyn', 'fix' and 'uni'.
 The separation between marketable limit orders is only used for the 'uni' scenario.
 In the scenarios 'dyn' and 'fix', we only distinguish between market orders (incl.\ marketable limit orders) and limit orders.
 The rates have are given in the unit [orders/second].
 \vspace{0.1cm}
\end{minipage}

  \begin{tabular}{lccr}
 \toprule
    Order type &&Market Side&\multicolumn{1}{c}{Rate}\\
    \midrule
    \multirow{4}{4cm}{Limit order \newline
                      (non-marketable)}& \multirow{2}{2cm}{arrival}&ask&0.12\\
    &&bid&0.12\\
    \cmidrule{2-4}
    & \multirow{2}{2cm}{cancellation}&ask&0.10\\
    &&bid&0.10\\
      \midrule
    \multirow{2}{4cm}{Limit order\newline
    (marketable)}& \multirow{2}{2cm}{arrival}&ask&0.0025\\
    &&bid&0.0025\\
    \midrule
    \multirow{2}{4cm}{Market order}& \multirow{2}{2cm}{arrival}&ask&0.0025\\
    &&bid&0.0025\\
    \bottomrule
 \end{tabular}
\end{center}
\end{table}

One peculiarity in  the theoretical scenarios 'fix' and 'dyn' is that  we treat marketable limit orders below or above the best quote as market orders.
Marketable limit orders on the best-quote, i.e., with $d=0$, are modeled together with the rest of the limit orders as they approximately seem to fit into the discrete logarithmic distributions
across
price levels (cp.\ \Cref{fig:bin_freq}).
In the scenario 'uni', we separate the market orders and the marketable limit orders (strictly) below or above the best quote up to $d=-10$.

{\singlespacing
\begin{longtable}{lcrrrrrrrr}
\caption{Marketable orders by type}
\label{tab:MarketableOrders}\\
\multicolumn{8}{p{0.9\linewidth}}{
\footnotesize
The table reports the share of marketable orders of all incoming orders in percentages across all stocks in the XETRA data.
The column \%L($d<0$) shows the share of marketable limit orders behind the best quote, whereas the column \%L($d=0$) gives the share of all marketable limit orders directly at the best ask or bid.
The column \%M contains the percentages of market orders.
\%I tables the share of marketable iceberg orders and \%T those of stop orders.
The column \%all is the total share of all marketable orders.
}&
\\
\\
\toprule
\bfseries Ticker&
\bfseries Buy/Sell
&\multicolumn{1}{p{2cm}}{\bfseries \%L($d<0$)}
&\multicolumn{1}{p{2cm}}{\bfseries \%L($d=0$)}
&\multicolumn{1}{p{1cm}}{\bfseries \%M}
&\multicolumn{1}{p{1cm}}{\bfseries \%I}
&\multicolumn{1}{p{1cm}}{\bfseries \%T}
&\multicolumn{1}{p{1cm}}{\bfseries \%all}\\
\midrule
\endfirsthead

\bfseries Ticker&
\bfseries Buy/Sell
&\multicolumn{1}{p{2cm}}{\bfseries \%L($d<0$)}
&\multicolumn{1}{p{2cm}}{\bfseries \%L($d=0$)}
&\multicolumn{1}{p{1cm}}{\bfseries \%M}
&\multicolumn{1}{p{1cm}}{\bfseries \%I}
&\multicolumn{1}{p{1cm}}{\bfseries \%T}
&\multicolumn{1}{p{1cm}}{\bfseries \%all}\\
\midrule
\endhead
    \multirow{2}{*}{ADS}&S& 5.40&1.73&1.50&0.14&0.02& 8.79\\* 

&B& 5.70&1.97&1.93&0.00&0.03& 7.69\\ 
\midrule
\multirow{2}{*}{ALT}&S& 6.61&1.61&1.69&0.03&0.05& 9.99\\* 

&B& 7.54&2.06&1.76&0.00&0.06& 9.67\\ 
\midrule
\multirow{2}{*}{ALV}&S& 5.09&2.59&1.99&0.02&0.05& 9.74\\* 

&B& 5.73&3.07&2.58&0.00&0.04& 8.85\\ 
\midrule
\multirow{2}{*}{BAS}&S& 6.67&2.11&1.47&0.03&0.07&10.35\\* 

&B& 6.63&2.15&1.46&0.00&0.10& 8.88\\ 
\midrule
\multirow{2}{*}{BAY}&S& 7.15&1.93&1.90&0.01&0.10&11.08\\* 

&B& 7.76&2.27&1.92&0.00&0.09&10.13\\ 
\midrule
\multirow{2}{*}{BMW}&S& 6.87&1.87&1.22&0.03&0.13&10.12\\* 

&B& 6.97&2.10&2.00&0.00&0.12& 9.19\\ 
\midrule
\multirow{2}{*}{CBK}&S& 6.10&1.34&1.81&0.01&0.11& 9.37\\* 

&B& 6.28&1.56&1.59&0.00&0.09& 7.93\\ 
\midrule
\multirow{2}{*}{CONT}&S& 6.50&1.37&1.57&0.02&0.07& 9.53\\* 

&B& 6.75&1.53&1.35&0.00&0.06& 8.34\\ 
\midrule
\multirow{2}{*}{DB1}&S& 7.14&1.93&1.98&0.03&0.13&11.21\\* 

&B& 7.66&2.08&1.98&0.00&0.14& 9.88\\ 
\midrule
\multirow{2}{*}{DBK}&S& 7.01&2.90&2.07&0.04&0.06&12.08\\* 

&B& 7.37&2.95&1.65&0.00&0.08&10.40\\ 
\midrule
\multirow{2}{*}{DCX}&S& 7.90&2.49&2.50&0.02&0.12&13.02\\* 

&B& 7.85&2.52&1.56&0.00&0.15&10.53\\ 
\midrule
\multirow{2}{*}{DPW}&S& 8.24&1.80&3.10&0.04&0.21&13.40\\* 

&B& 9.56&1.85&3.22&0.00&0.25&11.67\\ 
\midrule
\multirow{2}{*}{DTE}&S&12.56&3.20&5.02&0.21&0.13&21.12\\* 

&B&13.09&3.34&5.43&0.00&0.11&16.54\\ 
\midrule
\multirow{2}{*}{EOA}&S& 6.97&2.42&1.51&0.04&0.06&11.00\\* 

&B& 6.65&2.37&1.69&0.00&0.06& 9.09\\ 
\midrule
\multirow{2}{*}{FME}&S& 5.34&1.56&1.37&0.01&0.03& 8.31\\* 

&B& 5.58&1.72&1.22&0.00&0.04& 7.34\\ 
\midrule
\multirow{2}{*}{HEN3}&S& 4.20&1.31&0.98&0.05&0.01& 6.56\\* 

&B& 4.44&1.68&1.03&0.00&0.01& 6.13\\ 
\midrule
\multirow{2}{*}{HVM}&S& 8.94&1.76&2.23&0.01&0.17&13.12\\* 

&B& 9.32&1.93&2.59&0.00&0.17&11.41\\ 
\midrule
\multirow{2}{*}{IFX}&S&11.40&2.29&3.91&0.05&0.37&18.01\\* 

&B&12.96&2.94&4.54&0.00&0.30&16.20\\ 
\midrule
\multirow{2}{*}{LHA}&S& 8.36&1.42&2.51&0.05&0.19&12.53\\* 

&B& 8.67&1.50&2.28&0.00&0.20&10.37\\ 
\midrule
\multirow{2}{*}{LIN}&S& 5.59&1.19&1.34&0.08&0.03& 8.23\\* 

&B& 5.83&1.30&1.09&0.00&0.05& 7.18\\ 
\midrule
\multirow{2}{*}{MAN}&S& 7.55&1.41&1.63&0.10&0.11&10.81\\* 

&B& 8.06&1.53&1.50&0.00&0.12& 9.71\\ 
\midrule
\multirow{2}{*}{MEO}&S& 7.75&2.01&1.35&0.04&0.07&11.22\\* 

&B& 7.91&2.22&1.33&0.00&0.09&10.22\\ 
\midrule
\multirow{2}{*}{MUV2}&S& 6.41&2.98&1.72&0.02&0.07&11.20\\* 

&B& 6.90&3.32&2.39&0.00&0.06&10.29\\ 
\midrule
\multirow{2}{*}{RWE}&S& 7.67&2.16&2.26&0.04&0.11&12.24\\* 

&B& 7.17&2.04&1.45&0.00&0.12& 9.33\\ 
\midrule
\multirow{2}{*}{SAP}&S& 5.57&2.53&1.67&0.01&0.04& 9.83\\* 

&B& 5.79&2.58&1.61&0.00&0.05& 8.43\\ 
\midrule
\multirow{2}{*}{SCH}&S& 7.57&1.67&2.31&0.03&0.09&11.67\\* 

&B& 8.32&1.99&2.32&0.00&0.11&10.41\\ 
\midrule
\multirow{2}{*}{SIE}&S& 7.31&2.61&2.22&0.03&0.08&12.25\\* 

&B& 8.08&3.02&2.25&0.00&0.10&11.20\\ 
\midrule
\multirow{2}{*}{TKA}&S& 7.54&1.89&2.42&0.10&0.14&12.09\\* 

&B& 8.12&1.64&2.68&0.00&0.13& 9.89\\ 
\midrule
\multirow{2}{*}{TUI}&S& 6.82&1.50&2.62&0.01&0.16&11.12\\* 

&B& 7.67&2.07&2.96&0.00&0.15& 9.88\\ 
\midrule
\multirow{2}{*}{VOW}&S& 8.55&2.85&1.55&0.01&0.16&13.13\\* 

&B& 9.09&3.16&1.43&0.00&0.19&12.44\\ 
\bottomrule

\end{longtable}
}

\subsection{Order Distribution Across Price Levels $p_{K,M}(\cdot)$}
For the probability distribution of order arrivals across price levels specified in the factor $p_{K,M}(\cdot)$, we distinguish three theoretical scenarios and one scenario using unconditional empirical frequencies.

\subsubsection{Uniform Distribution (uni)}
The easiest approach to define the arrival rates across price levels is a uniform distribution.
In this scenario, we assume that the arrivals of orders are concentrated on the first 90 integer price levels before the best quote of the opposite market.
Additionally, marketable limit orders are also allowed to cross the best quote up to 10 price levels.
In essence, this means that the arrivals of bid and ask orders are concentrated on 100 price levels around the best quote of the opposite market where the arrival rate on each price level is 0{.}0012 orders per second.

For the cancellations, we distribute the probability for an order cancellation uniformly among the occupied price levels.

\subsubsection{Fixed Probability Distribution (fix)}
Empirical frequencies of (non-marketable) limit orders across relative price levels exhibit pronounced probability mass at the tails of the distribution.
For the distribution across price levels, in the scenario 'fix', we use a discrete Gaussian exponential distribution (DGX) as presented by \cite{BiFCK01}.
The distribution is especially useful in cases where the random variable to be modeled is discrete and has pronounced probability mass at the tails.
It is particularly nice that the DGX reduces to the generalized Zipf distribution when $\mu \to -\infty$.
Thus, it is flexible enough to incorporate situation where the probability distribution is a straight line in log-log-plots and cases in which it exhibits some curvature.
A short summary of the DGX distribution is given in \Cref{appendix:DGX}.

In the simulation scenario with a fixed probability distribution, we chose to set the values as outlined in Table~\ref{tab:ParamProbSimK}.
\begin{table}
\begin{center}
\caption{Parameters of probability distribution across $k$}
\label{tab:ParamProbSimK}
\begin{minipage}{0.9\linewidth}
 \footnotesize
 \vspace{0.1cm}
\end{minipage}
  \begin{tabular}{lccrr}
 \toprule
    Order type &&Market Side&\multicolumn{1}{c}{$\mu$}&\multicolumn{1}{c}{$\sigma$}\\
    \midrule
    \multirow{4}{4cm}{Limit order \newline
                      (non-marketable)}& \multirow{2}{2cm}{arrival}&ask&1.726301&0.674654\\
    &&bid&1.765909&0.711773\\
    \cmidrule{2-5}
    & \multirow{2}{2cm}{cancellation}&ask&1.619866&0.620127\\
    &&bid&1.674366&0.650024\\
    \bottomrule
 \end{tabular}
\end{center}
\end{table}
The values are the empirical mean and standard deviation across incoming orders of a random sample over several stocks.
Note that the mean of arrivals is slightly higher on the bid side of the market, i.e., orders are more likely to arrive deeper in the book.
Also the variance of order arrivals is higher.
The same holds for cancellations.
So while there are more arrivals deeper in the book, slightly more orders deep in the book are also canceled.

\subsubsection{Dynamical Probability Distribution (dyn)}
Similar to the case in which we use a DGX distribution with fixed parameters $\mu$ and $\sigma$, in the simulation scenario with a dynamical distribution across price levels, we also use the DGX distribution as the fundamental distribution.
However, in this case we specify the parameters of the distribution to depend on the prevailing spread.
The functional relationship we use is the following:
\begin{align*}
 \mu(\Delta) &= \log(100 \cdot \Delta)+\frac{1}{2},\\
 \sigma(\Delta) &= \sqrt{20\cdot\log(100 \cdot \Delta)}+1.2 .\\
\end{align*}
The functional relation is inspired by a scatter plot of $\hat{\mu}_i$ and $\hat{\sigma}_i$ estimated on the unconditional frequencies of order arrivals (and cancellations) across price levels against
the average spread $\Delta_i$ for each stock $i$.\footnote{We are aware of the fact that the expectation of the functional relationship between DGX parameters and spread is not the same as a
function for the log-likelihood in dependence of the expectation of the spread i.e. ${\E}_{t_0}[\mu(\Delta)] \neq \mu({\E}_{t_0}[\Delta])$. }
This scatter plot is depicted in \Cref{fig:Param_DGX_vs_spread}.
Note also, that we have switched the scale of standard deviation and expectation as observed in the data on purpose.
In that way, we hope to get an impression on how an increase in variance and a decrease of the mean may affect the characteristics of the order book evolution.
The 'dyn' scenario is theoretically also motivated by the quest to study the sensitivity of the LOB system to feedback reactions between the state of the book and traders' order submission behavior.

\subsubsection{Empirical Distribution (emp)}
We also simulate one scenario where we take the empirical frequencies observed across price levels into account.
The empirical log-frequencies are depicted in \Cref{fig:bin_freq}.

\subsection{Order Distribution Across Size Levels $p_{Q,M}(\cdot)$}
For the distribution across volume, we employ two different specifications.
In one specification, we use a power law distribution.
Even though, in our data at hand, we find that a power law does not match the volume distribution.
This can be seen by sheer eyeballing of \Cref{fig:size_dist}.
Nevertheless, the good fit of the power law distribution to describe order sizes has been shown in various articles \cite[][]{BouchaudMP02, GopikrishnanPGS00, MaslovM01}.

%

The probability mass function of a discrete power law distribution where the smallest value of the support is 1, is theoretically defined as \cite[cp.][]{ClausetSCN09} 
\begin{align*}
 p(x;\lambda) &= \zeta(\lambda) {x^{-\lambda}} \forall x \in \mathbb{N},
\end{align*}
where is the Hurwitz zeta function $\zeta(\lambda) = \sum_{n=0}^\infty (n+1)^{-\lambda}$.
We fix the parameter in all simulations at $\lambda = 1.6$ which is close to empirically observed values.

According to \citeauthor{ClausetSCN09} (\citeyear{ClausetSCN09}, Appendix D), given a random number $u \in [0,1]$, we can generate an integer realization $\tilde{x}$ from the power law
distribution by calculating
\begin{align*}
\tilde{x}  = \lfloor (1-0.5) (1-u)^{-1/\lambda}+1/2\rfloor -1,
\end{align*}
where $\lfloor\cdot\rfloor$ signify the floor operator which cuts off the decimal places of the argument.

Note that since we assume independence of $p_{Q,M}(\cdot)$ and $p_{K,M}(\cdot)$, and each arriving order surely has to be assigned a size, we simply generate a third realization from a uniform distribution $\mathcal{U}(0,1)$ to determine the volume.
In other words, before randomly generating what size is affected with $u_3$, we answer the question \emph{when} something will happen with $u_1$ and, with $u_2$, \emph{what} as well as \emph{where}
(i.e.\ at which limit price level) it will happen, as described in \Cref{sec:TheSimulation}.

\subsubsection{Empirical Distribution (emp)}
We also simulate one scenario where we use the empirical frequencies observed across quantity levels to simulate the LOB evolution.
The empirical log-frequencies are depicted in \Cref{fig:size_dist}.

\subsection{The Fully Empirical Scenario (emp,emp)}

For the case where both the distribution across price levels and the distribution across price levels are sampled from the empirically observed frequencies, we take the joint frequencies (not the
product of the marginal frequencies) to sample both size and price of an incoming or canceled order.

\section{Discrete Gaussian Exponential Distribution (DGX)}
\label{appendix:DGX}
In the simulation, as described in \Cref{appendix:Simulation}, we use the DGX for the simulation of order arrivals and cancellations across price levels.

The probability mass function of the distribution can be defined according to \cite{BiFCK01} as
\begin{align*}
 p(x=k;\mu,\sigma) &= \frac{A(\mu,\sigma)}{k} \exp\left( - \frac{(\log(k) -\mu)^2}{2\sigma^2}\right), \quad\quad \forall k \in \mathbb{N},
\end{align*}
where the normalizing constant $A(\mu,\sigma)$ is defined as
\begin{align*}
 A(\mu,\sigma)  = \left\{ \sum_{k=1}^\infty \frac{1}{k}\exp\left( - \frac{(\log(k) -\mu)^2}{2\sigma^2}\right)\right\}^{-1}.
\end{align*}

In \Cref{fig:bin_freq}, we use a slightly modified version of the DGX by truncating the distribution at 1 to show how the DGX can be fit to the data.
The truncated DGX can be derived from the truncated (log-)normal distribution for continuous values and has the following probability distribution function:
\begin{align*}
 p(x=k;\mu,\sigma) &= \frac{1}{1-\phi(-\frac{\mu}{\sigma})}\frac{A_T(\mu,\sigma)}{k\sigma} \exp\left( - \frac{(\log(k) -\mu)^2}{2\sigma^2}\right), \quad\quad \forall k \in \mathbb{N},
\end{align*}
The normalization factor $A_T(\mu,\sigma)$ is similarly defined by
\begin{align*}
 A_T(\mu,\sigma)  = \frac{1}{\sigma\left(1-\phi(-\frac{\mu}{\sigma})\right)}\left\{ \sum_{k=1}^\infty \frac{1}{k}\exp\left( - \frac{(\log(k) -\mu)^2}{2\sigma^2}\right)\right\}^{-1}.
\end{align*}

The parameters $\mu$ and $\sigma$ can be estimated using a maximum likelihood specification as described in \cite{BiFCK01}.

\section{Taylor Series Expansion of Linear Models} \label{appendix:Linearization}
In this appendix the Taylor Series expansion that justifies the model specifications in Equations~\eqref{eq:LinApproximation1}~to~\eqref{eq:LinApproximation4} is derived.

Starting point of the derivation is the decomposition of the event rate in \Cref{eq:rate_decomp}.
First, for the sake of brevity, we introduce the intensity of events at the price level $k$ with affected order size $q$ given a prevailing spread $\Delta$.
\begin{align*}
r_{M,i,j,e}(k,q \mid \Delta) = \alpha_{M,i,j}(k,q \mid \Delta) + \omega_{M,i,j}(k,q\mid \Delta)
\end{align*}
where the index $M$ denotes the market side (either bid or ask), $i$ indicates the instrument, $j$ denotes the order type (limit or market) and $e$ denotes the event type (arrival or cancellation).
The right hand side follows when writing arrival $\alpha$ and cancellation rates $\omega$ separately.

As we have seen in \Cref{sec:time_evolution}, the Hamiltonian $H$ is directly constructed from the event rates.
As a direct consequence, the conditional probability to find the system in state $\ket{z}$ given that it has been in $\ket{z_0}$ at time $t_0$ can be expressed as described in \Cref{eq:cond_prob} or in more detail as
\begin{align}
	 p(z,t|z_0,t_0) &= \braket{z \mid \psi(t)} = \bra{z} U (t,t_0) \ket{\psi(t_0)} = \bra{z} \exp\big(\;\int_{t_0}^t H (\tau) d\tau \big) \ket{\psi(t_0)}  \notag \\
	 &=\bra{z} \sum_{w=1}^{\infty} \frac{\bigg(\;\int_{t_0}^t H (\tau) d\tau\bigg)^w}{w!} \ket{\psi(t_0)}
\end{align}

As one can see, by construction the conditional distribution is polynomial in the (time integral over) arrival and cancellation rates, and, depending on the succession of orders in $\ket{z}$ (see e.g. \Cref{eq:mixed_state_split}), further combinatorical factors have to be introduced (which include the factorial $w!$).
Only regarding the terms up to order one the conditional probability can be written as
\begin{align*}
  p(z,t|z_0,t_0) &= \braket{z \mid \psi(t_0) } + \bra{z} \int_{t_0}^t H (\tau) d\tau \ket{\psi(t_0)}
\end{align*}
This first order approximation may fit the conditional probability for short time horizons well, however, for longer time horizons the interactions between order arrivals and cancellations may become the more important factor.

Nevertheless, with this approximation, we may view the conditional expected value of some observable $O$ given the state of the order book at time $t_0$ of stock $i$ at some future time $t >t_0$ as
\begin{align*}
{\E}_{t_0}[O_{i,t}] &= \sum_z \braket{z \mid O \mid \psi(t_0) } + \sum_z  \bra{z} O \int_{t_0}^t H (\tau) d\tau \ket{\psi(t_0)}  \\
&=  O_{i,t_0} + \sum_z \bra{z} O \int_{t_0}^t \sum_q \sum_k \alpha_{M,i,j}(k,q) E_{M,i,j}  + \omega_{M,i,j}(k,q)  C_{M,i,j} d\tau \ket{\psi(t_0)}
\end{align*}
where $E$ and $C$ denote the order entry and cancellation operators laid out in \Cref{sec:time_evolution} and $O_{i,t_0}$ is the realization of the observable at time $t_0$.

We may generalize this notion for the conditional expected value to some function that depends on the intensity of events, and, thus, again on order size $q$, the price level $k$ and additionally further variables that determine arrival and cancellation rates, e.g. the spread.

Making the very crude assumption that the expected value of some observable is linear in the intensity of events, we could formulate the approximation as
\begin{align*}
 {\E}_{t_0}[O_{i,t}] &\approx \gamma_{0,i} + \gamma_{1,i} \E[ r_{M,i,j,e}(k,q \mid \Delta) ] .
\end{align*}
Decomposing $r_{M,i,j,e}(k,q \mid \Delta)$ as done in \Cref{eq:rate_decomp} and additionally assuming that the the average intensity  $\bar{r}_{0,M,i,j,a}(\Delta)$ is some function of the prevailing spread yields
\begin{align}
  {\E}_{t_0}[O_{i,t}] &\approx \gamma_{0,i} + \gamma_{1,i} \E[  (\alpha_{M,i,j}(k,q \mid \Delta) + \omega_{M,i,j}(k,q\mid \Delta))] \notag \\
  &\approx \gamma_{0,i} + \gamma_{1,i} \E[  \bar{r}_{0,M,i,j,a}(\Delta) \,\, p_{K,M}(k;\boldsymbol{\theta}_{M,a}) \,\,  p_{Q,M}(q;\boldsymbol{\phi}_{M,a}) \notag \\
  &\hspace{2.2cm}+ \bar{r}_{0,M,i,L,c}(\Delta)  \,\, p_{K,M}(k;\boldsymbol{\theta}_{M,c}) \,\, p_{Q,M}(q;\boldsymbol{\phi}_{M,c})] \label{eq:observableapprox}
\end{align}
Now, expanding each term by a Taylor series expansion around the respective expected value we have the following expansions for $p_{Q,M}(q;\boldsymbol{\phi}_{M,e})$
\begin{align*}
p_{Q,M}(q;\boldsymbol{\phi}_{M,e}) &\approx p_{Q,M}(\E[q];\boldsymbol{\phi}_{M,e}) + d p_{Q,M}(\E[q];\boldsymbol{\phi}_{M,e})(q - \E[q])\\
&\hspace{2cm}+\frac{d^2 p_{Q,M}(\E[q];\boldsymbol{\phi}_{M,e})}{2}(q - \E[q]) ^2 + \ldots
\end{align*}
Taking expectations with respect to $p_{Q,M}(\E[q];\boldsymbol{\phi}_{M,e})$ yields an approximation of the expected value in moments of order 2 and higher
\begin{align*}
 \E[p_{Q,M}(q;\boldsymbol{\phi}_{M,e})] &\approx p_{Q,M}(\E[q];\boldsymbol{\phi}_{M,e}) +\frac{d^2 p_{Q,M}(\E[q];\boldsymbol{\phi}_{M,e})}{2}\Var[q] + \ldots
\end{align*}
Using then, again, the first order Taylor approximation of $p_{Q,M}(\E[q];\boldsymbol{\phi}_{M,e})$ around $0$ reintroduces the first moment
\begin{align*}
 \E[p_{Q,M}(q;\boldsymbol{\phi}_{M,e})] &\approx   p_{Q,M}(0;\boldsymbol{\phi}_{M,e}) + d p_{Q,M}(0;\boldsymbol{\phi}_{M,e}) \E[q] +\frac{d^2 p_{Q,M}(\E[q];\boldsymbol{\phi}_{M,e})}{2}\Var[q] + \ldots
\end{align*}
Thus, we may write the expected value of $p_{Q,M}(q;\boldsymbol{\phi}_{M,e})$ as a linear function of the moments
\begin{align}
  \E[p_{Q,M}(q;\boldsymbol{\phi}_{M,e})] &\approx   \xi_{M,i,j,e,0} + \xi_{M,i,j,e,1} \E[q] + \xi_{M,i,j,e,2}  \Var[q] + \ldots \label{eq:Taylorsize}
\end{align}
Changing variables in $p_{K,M}(k;\boldsymbol{\theta}_{M,e})$ by considering the relative distance $d_l$ to the opposing best quote instead of the absolute price level $k$, the same can be done for $\E[p_{K,M}(k;\boldsymbol{\theta}_{M,e})]$
\begin{align}
  \E[p_{K,M}(k;\boldsymbol{\theta}_{M,e})] &\approx   \kappa_{M,i,j,e,0} + \kappa_{M,i,j,e,1} \E[d_l] + \kappa_{M,i,j,e,2}  \Var[d_l] + \ldots \label{eq:Taylorprice}
\end{align}

Last but not least, we may model the expected value of the event specific intensity by its expected value shifted by a event specific factor $\rho_{0,M,i,j,e}$ and the expected value of an event
unspecific function $f(\Delta)$ solely dependent on the spread
\begin{align}
 \E[\bar{r}_{0,M,i,j,e,t}(\Delta)] = {\E}_{t_0}[\bar{r}_{0,M,i,j,e,t}] \rho_{0,M,i,j,e} \E[f(\Delta)] \label{eq:intensityapprox}.
\end{align}
Approximating $\E[f(\Delta)]$ by an expansion in moments as above yields
\begin{align}
\E[f(\Delta)]  = \delta_{i,0}+\delta_{i,1} {\E}_{t_0}[\Delta_t]\,\,+ \sum_{v=2}^{4}\delta_{i,v}
 {\E}_{t_0}[(\Delta_t-\mu_{\Delta,t})^v] \label{eq:Taylorspread}.
\end{align}

Reinserting the Taylor expansions in Equations~\eqref{eq:Taylorsize}, \eqref{eq:Taylorprice}~and~\eqref{eq:Taylorspread} up to order four together with \Cref{eq:intensityapprox} in \Cref{eq:observableapprox} yields \Cref{eq:LinApproximation1}
\begin{align}
 {\E}_{t_0}[O_{i,t}] &= \gamma_{0,i} +   \bigg(\delta_{i,0}+\delta_{i,1} {\E}_{t_0}[\Delta_t]\,\,+ \sum_{v=2}^{4}\delta_{i,v}
 {\E}_{t_0}[(\Delta_t-\mu_{\Delta,t})^v]\bigg) \,\,\times \notag\\
 &\quad\sum_{M,j,e} \rho_{0,M,i,j,e}  {\E}_{t_0}[\bar{r}_{0,M,i,j,e,t}]\,\,\times \notag\\
 &\hspace{1cm}\bigg(\kappa_{M,i,j,e,0}+ \kappa_{M,i,j,e,1} {\E}_{t_0,M,i,j,e}[d_{l,t}]\,\,+   \notag\\
 &\hspace{2.5cm}\sum_{v=2}^{4} \kappa_{M,i,j,e,v} {\E}_{t_0,M,i,j,e}[(d_{l,t}-\mu_{d_l,t})^v]\bigg) \,\,\times \notag\\
  &\hspace{1cm}\bigg(\xi_{M,i,j,e,0}+ \xi_{M,i,j,e,1} {\E}_{t_0,M,i,j,e}[q_{t,M,j,e}]\,\,+  \notag\\
  &\hspace{2.5cm}\sum_{v=2}^{4}\xi_{M,i,j,e,v} {\E}_{t_0,M,i,j,e}[(q_{t,M,j,e}-\mu_{q,t})^v]\bigg)+ \varepsilon_i, \notag
\end{align}
\end{document}